\documentclass{aa}  

\usepackage{textcomp}
\usepackage{graphicx} %
\usepackage{amsmath} %
\usepackage{amssymb} %
\usepackage{bm} %
\usepackage{upgreek} %
\usepackage{IEEEtrantools} %
\usepackage{multirow}
\usepackage[colorlinks=true,allcolors=blue,urlcolor=blue]{hyperref}
\usepackage{txfonts}
\usepackage[normalem]{ulem} 
\usepackage{multirow}
\usepackage[dvipsnames]{xcolor}

\newcommand{\sect}[1]{\text{Sect.~\ref{#1}}}
\newcommand{\fig}[1]{\text{Fig.~\ref{#1}}}

\newcommand{\multitd}{\textsc{multi3d}}

\newcommand{\balder}{\textsc{balder}}
\newcommand{\blue}{\textsc{blue}}
\newcommand{\mtd}{\textlangle3D\textrangle}
\newcommand{\marcs}{\textsc{marcs}}

\newcommand{\stagger}{\textsc{stagger}}
\newcommand{\atmo}{\textsc{atmo}}

\newcommand{\kms}{\mathrm{km\,s^{-1}}}
\newcommand{\teff}{T_{\mathrm{eff}}}
\newcommand{\lgg}{\log{g}}

\newcommand{\lggf}{\log{gf}}
\newcommand{\feh}{\mathrm{\left[Fe/H\right]}}
\newcommand{\abrat}[2]{\mathrm{\left[#1/#2\right]}}
\newcommand{\lgeps}[1]{\log{\epsilon_{\mathrm{#1}}}}

\newcommand{\lgt}{\log{\tau_{500}}}
\newcommand{\dex}{\mathrm{dex}}
\newcommand{\lyalpha}{\mathrm{Ly\upalpha}}

\newcommand{\hbeta}{\mathrm{H\upbeta}}

\newcommand{\nm}{\mathrm{nm}}
\newcommand{\eV}{\mathrm{eV}}
\newcommand{\epot}{\mathrm{\chi_{exc.}}}
\newcommand{\vmic}{\mathrm{\xi_{mic}}}
\newcommand{\vmicol}{\mathrm{\xi^{1D,LTE}_{mic}}}
\newcommand{\vmicon}{\mathrm{\xi^{1D,NLTE}_{mic}}}
\newcommand{\vmac}{\mathrm{\xi_{mac}}}
\newcommand{\depcof}{\upbeta}

\newcommand{\corr}[2]{\Delta^{\text{#1}}_{\text{#2}}}
\begin{document} 

\title{Carbon, oxygen, and iron abundances in
disk and halo stars\thanks{Tables 1--7
are available in electronic form at the CDS
via anonymous ftp to 
\url{cdsarc.u-strasbg.fr (130.79.128.5)} or 
via~\url{http://cdsarc.u-strasbg.fr/viz-bin/qcat?/A+A/XXX/xxx}.}}
\subtitle{Implications of 3D non-LTE spectral line formation}
\author{A.~M.~Amarsi\inst{1}
\and
P.~E.~Nissen\inst{2}
\and
\'A.~Sk\'ulad\'ottir\inst{1}}
\institute{Max Planck Institute f\"ur Astronomy, K\"onigstuhl 17, 
D-69117 Heidelberg, Germany\\
\email{amarsi@mpia.de},\,{\ttfamily skuladottir@mpia.de}
\and
Stellar Astrophysics Centre, Department of Physics and Astronomy, Aarhus
University, Ny Munkegade 120, DK-8000 Aarhus C, Denmark\\
\email{pen@phys.au.dk}}

\abstract{The abundances of carbon, oxygen, and iron in late-type stars
are important parameters in exoplanetary and stellar physics, as well
as key tracers of stellar populations and Galactic chemical evolution.
However, standard spectroscopic abundance analyses 
can be prone to severe systematic errors, 
based on
the assumption that the stellar atmosphere
is one-dimensional (1D) and hydrostatic,
and by ignoring departures from local thermodynamic equilibrium (LTE).
In order to address this, we carried out
three-dimensional (3D) non-LTE radiative transfer calculations
for \ion{C}{I} and \ion{O}{I}, and 3D LTE radiative transfer 
calculations for \ion{Fe}{II}, across the 
\stagger-grid of 3D hydrodynamic model atmospheres.
The absolute 3D non-LTE versus 1D LTE abundance corrections
can be as severe as $-0.3\,\dex$~for \ion{C}{I} lines
in low-metallicity F dwarfs, and $-0.6\,\dex$~for \ion{O}{I} lines
in high-metallicity F dwarfs. The 3D LTE versus 1D LTE abundance 
corrections for \ion{Fe}{II} lines are less severe, typically 
less than $+0.15\,\dex$. 
We used the corrections in a re-analysis of 
carbon, oxygen, and iron in $187$~F~and G~dwarfs in the Galactic disk and halo.
Applying the differential 3D non-LTE corrections 
to 1D LTE abundances visibly reduces
the scatter in the abundance plots.
The thick disk and high-$\upalpha$~halo population
rise in carbon and oxygen with decreasing metallicity,
and reach a maximum of $\abrat{C}{Fe}\approx0.2$~and a plateau of 
$\abrat{O}{Fe}\approx0.6$~at $\feh\approx-1.0$.
The low-$\upalpha$~halo population is qualitatively similar,
albeit offset towards lower metallicities and with larger scatter.
Nevertheless, these populations overlap in the
$\abrat{C}{O}$~versus $\abrat{O}{H}$~plane, 
decreasing to a plateau
of $\abrat{C}{O}\approx-0.6$~below $\abrat{O}{H}\approx-1.0$.
In the thin-disk, stars having confirmed planet detections tend
to have higher values of $\mathrm{C/O}$~at given
$\abrat{O}{H}$; this potential signature of
planet formation is only apparent after 
applying the abundance corrections to the 1D LTE results.
Our grids of line-by-line abundance corrections are publicly available
and can be readily used to improve the accuracy
of spectroscopic analyses of late-type stars.}


\maketitle

\section{Introduction}
\label{introduction}

Carbon, oxygen, and iron are 
among the most interesting
elements in astrophysics.
They are two of the most important sources of opacity
in stellar interiors, while carbon and oxygen are also catalysts in the CNO
cycle and affect energy generation.
Hence, they have a large influence on
stellar structure (e.g.~\citealt{2008PhR...457..217B})
and stellar evolution 
(e.g.~\citealt{2012ApJ...755...15V}).
They are also important in the context of
exoplanets, providing insight into their formation properties,
compositions, and atmospheres
\citep[e.g.][]{2012ApJ...757..192J,2019ApJ...873...32M}.

The role of carbon, oxygen, and iron as 
diagnostics of stellar populations and Galactic chemical evolution
is of particular interest.
All three elements are released 
into the cosmos by core-collapse supernova of massive stars
($M\gtrsim8\,M_{\odot}$); carbon and oxygen
form through hydrostatic helium burning in their cores
and iron forms during the explosion itself
\citep[e.g.][]{2002RvMP...74.1015W}.
Carbon could also 
be released into the cosmos by massive stars before they 
explode
via metal-line driven, metallicity-dependent winds,
especially from Wolf-Rayet (WR) stars \citep[e.g.][]{2018ApJS..237...13L}.
In addition, carbon is dredged up by thermal pulses in 
asymptotic giant branch (AGB) stars
and released into the cosmos through mass loss
\citep[e.g.][]{2014PASA...31...30K}.
Lastly, significant iron is formed at later Galactic times
via the radioactive decay of $^{56}\mathrm{Ni}$~in 
Type Ia supernova~\citep[e.g.][]{2013ARA&amp;A..51..457N}.
Thus, the different abundance ratios of these three elements can
be used to probe different astrophysical phenomena, 
occurring on different timescales, and associated with stars of 
different masses
\citep[e.g.][]{2003MNRAS.339...63C,
2005ApJ...623..213C,2006ApJ...653.1145K,
2009A&amp;A...505..605C,
2016ApJ...827..126B,2019ApJ...874...93B}.

There is already literature on 
the abundances of these elements in the atmospheres of
late-type stars, both in the disk \citep[e.g.][]{2010ApJ...725.2349D,
2016ApJ...830..159N,2017A&A...599A..96S}
and metal-poor halo
\citep[e.g.][]{2004A&amp;A...414..931A,
2004A&amp;A...416.1117C,2009A&amp;A...500.1143F,
2013ApJ...762...26Y,
2019A&A...622L...4A}.
In particular, \citet{2014A&amp;A...568A..25N}~measured
carbon, oxygen, and iron abundances in $152$~F~and G~dwarfs 
with $-1.8\lesssim\feh\lesssim+0.5$\footnote{$\abrat{A}{B}\equiv
\log_{10}(N_{\mathrm{A}}/N_{\mathrm{B}})-
\log_{10}(N_{\mathrm{A}}/N_{\mathrm{B}})_{\odot}$}
in both the halo and the disk.
The introduction of that paper 
includes a review of earlier
studies of late-type stars.
Later studies have generally supported the various conclusions of
that work, concerning for example the elemental abundance separation
of the thin and thick disks, the low- and high-$\upalpha$~halo
populations \citep[see][]{2010A&amp;A...511L..10N,2011A&A...530A..15N},
and their mean Galactic chemical evolutions
\citep[e.g.][]{2015MNRAS.453..758H,2018ApJ...852...49H}.
Furthermore, different studies 
\citep[e.g.][]{2014ApJ...788...39T,
2016ApJ...831...20B,2018ApJ...865...68B,
2018A&A...614A..84S} now generally 
agree that exoplanet host stars do not typically have
high enough ratios of carbon to oxygen to form
carbon planets \citep{2005astro.ph..4214K,2014ApJ...787...81M}.

The results of \citet{2014A&amp;A...568A..25N} and similar studies
can be considered precise in two ways. 
First, the observed spectra are of high 
spectral resolution 
(resolving power $R=\lambda/\Delta\lambda\gtrsim40000$~in this case)
and have high signal-to-noise ratios 
($\mathrm{S/N}\gtrsim100$);
consequently, the measured equivalent widths have 
relatively small random errors.
Second, the sample size is large enough
for statistically-significant conclusions to be drawn.

However, systematic modelling errors can 
limit the accuracy of spectroscopic studies.
Standard spectroscopic analyses of late-type stars
are based on the assumption that stellar atmospheres
are one-dimensional (1D) and hydrostatic,
and that the atmospheric
matter satisfies Boltzmann-Saha
excitation and ionisation balance
as is implied by local thermodynamic equilibrium (LTE).
These two assumptions
can impart significant errors on the inferred stellar parameters
and elemental abundances,
that vary depending on the parameters of the star under 
investigation, and depending on the spectral 
line(s) under investigation
\citep[e.g.][]{2005ARA&amp;A..43..481A}.

Nevertheless, the outlook on stellar spectroscopy is promising.
It is now possible to 
carry out highly realistic spectroscopic analyses 
of late-type stars, owing to
advances in three-dimensional (3D)
hydrodynamic modelling of stellar atmospheres 
\citep[e.g.][]{2013A&A...557A...7T,2018MNRAS.475.3369C}
and in 3D non-LTE radiative transfer post-processing of 
these model atmospheres 
\citep[e.g.][]{2010A&amp;A...522A..26S,
2016MNRAS.455.3735A},
combined with progress in atomic astrophysics 
\citep[e.g.][]{2016A&amp;ARv..24....9B}
not least in the area of ab initio calculations 
for inelastic collisions with electrons
\citep[e.g.][]{2017A&A...606A..11B} and with atomic hydrogen
\citep[e.g.][]{2016PhRvA..93d2705B,2018ApJ...867...87B}.
A short review about the state-of-the-art
can be found in Sect.~2.4 of \citet{2018A&ARv..26....6N}.

There has been a particularly rapid development of 3D non-LTE methods
for analysing carbon and oxygen abundances.
In late-type stars, carbon and 
oxygen abundances can be measured using
atomic lines (although carbon abundances are 
more commonly measured using lines of CH).
Improved atomic models 
have recently been developed
for \ion{C}{I} \citep{2019A&A...624A.111A}
and \ion{O}{I} \citep{2018A&A...616A..89A},
that utilise ab initio calculations
for inelastic collisions with atomic hydrogen
\citep{2018A&A...610A..57B,2019A&A...625A..78A}
such data typically being the largest 
source of uncertainty in non-LTE 
modelling \citep[e.g.][]{2011A&amp;A...530A..94B}.
It was shown that 3D non-LTE synthetic spectra
based on these atomic models successfully reproduces
the observed solar centre-to-limb variations
of various \ion{C}{I}~and \ion{O}{I}~lines.
This is a sensitive test of the atomic models and especially of
the reliability of the data 
for inelastic collisions with atomic hydrogen
\citep[e.g.][]{2004A&amp;A...423.1109A,2015A&amp;A...583A..57S}.

Recent studies of iron in late-type stars
have demonstrated the 
potentially large impact of
3D non-LTE effects on \ion{Fe}{I} lines 
\citep{2016MNRAS.463.1518A,2017MNRAS.468.4311L,2017A&amp;A...597A...6N}.
However, such calculations are still prohibitively expensive
for large samples of stars.
Fortunately, progress can be made by focusing 
on \ion{Fe}{II} lines instead.
The departures from LTE in this majority species are thought to be
insignificant in late-type stars,
at least for \ion{Fe}{II} lines of low excitation potential
($\epot\lesssim8\,\eV$) and at 
metallicities $\feh\gtrsim-3.0$~\citep{2012MNRAS.427...50L}.
The impact of non-LTE abundance errors in \ion{Fe}{II} lines
are even smaller when the lines are measured differentially
with respect to the Sun or to a standard star
\citep[e.g.][]{2017A&A...608A.112N}.
Assuming that these 1D non-LTE results for \ion{Fe}{II}
are also applicable in 3D hydrodynamic model atmospheres,
and that 3D hydrodynamic simulations better represent 
real stellar atmospheres than
1D hydrostatic ones, it follows that 
3D LTE models of \ion{Fe}{II} lines should give
iron abundances that are more reliable
than those based on 1D (non-)LTE models.

Our goal here is to derive carbon,
oxygen, and iron abundances in Milky Way disk and halo stars,
that are both of high precision, and of improved accuracy.
We present detailed 3D non-LTE radiative transfer calculations 
for \ion{C}{I} and \ion{O}{I}, and 
3D LTE radiative transfer calculations for \ion{Fe}{II},
using the code \balder~(\sect{method}).
We explain the 3D non-LTE effects across stellar parameter
space, and present extensive grids of 3D non-LTE 
and 1D non-LTE versus 1D LTE abundance corrections 
for \ion{C}{I} and \ion{O}{I},
and 3D LTE versus 1D LTE abundance corrections for 
\ion{Fe}{II} (\sect{results}).
For \ion{C}{I} and \ion{Fe}{II}, these are the first grids
of their type to be presented in the literature;
for \ion{O}{I}, this updates the grids of 
\citet{2016MNRAS.455.3735A}, benefiting from the 
improvements to the atomic model described above, 
as well as to the 3D non-LTE radiative transfer code.
Based on these abundance corrections,
we present a reanalysis of precise literature data 
\citep{2014A&amp;A...568A..25N,2019A&A...622L...4A},
to obtain carbon, oxygen, and iron abundances 
in a sample of $187$~disk and halo stars (\sect{obs}).
Finally, we discuss how these new measurements,
of high precision and improved accuracy, alter
our understanding of stellar populations, 
Galactic chemical evolution, 
and the formation of planets (\sect{discussion}),
before presenting a short summary and 
some closing remarks on the outlook
for precise and accurate spectroscopic analyses of late-type stars
(\sect{conclusion}).

\section{3D non-LTE method}
\label{method}

\subsection{Model atmospheres}
\label{method_atmospheres}

\begin{figure}
    \begin{center}
        \includegraphics[scale=0.33]{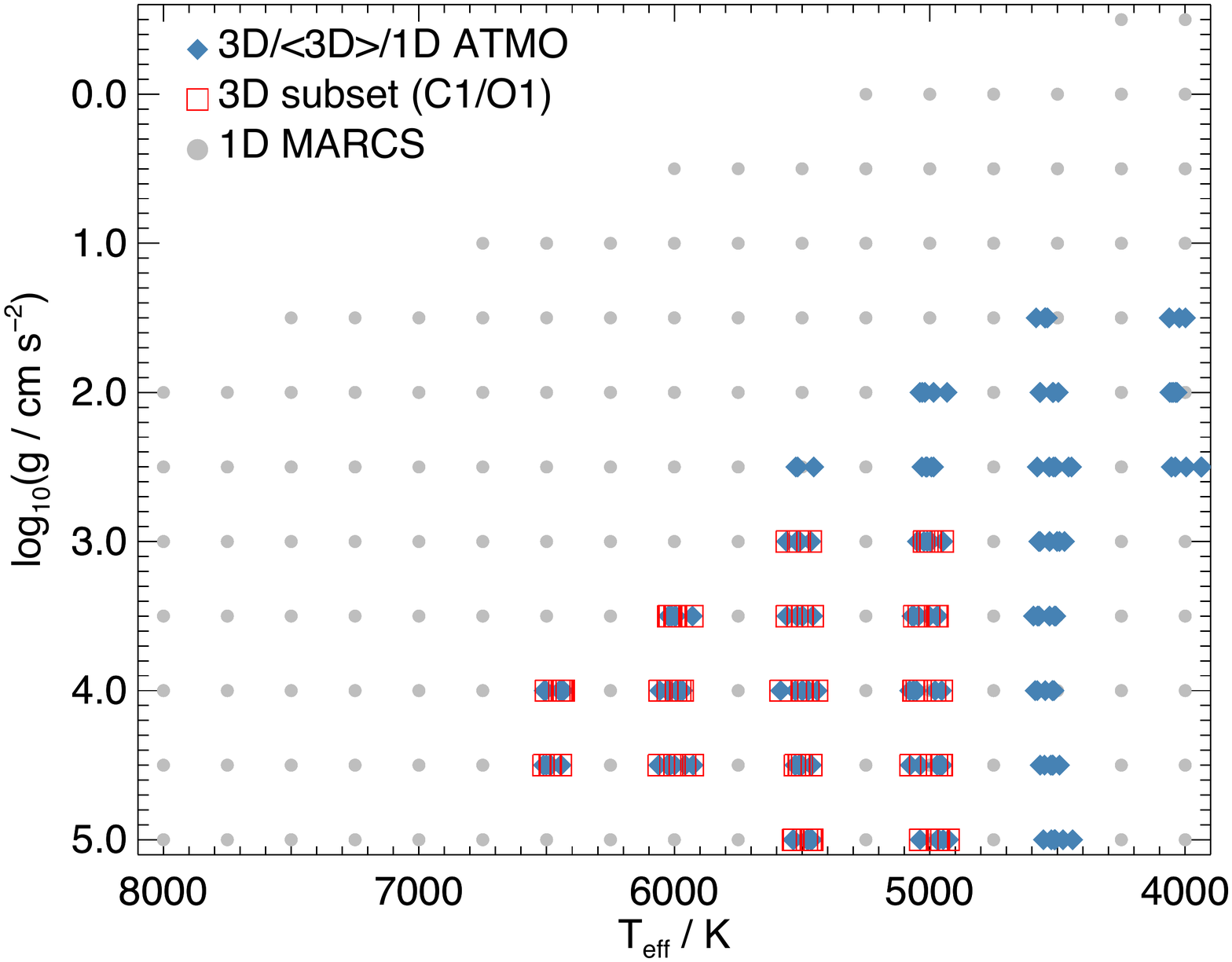}
        \caption{Kiel diagram illustrating 
        model atmospheres in
        $\lgg$---$\teff$~space.
        The $1807$~\marcs~nodes (grey circles) 
        are regularly spaced in effective temperature,
        so model atmospheres with the same $\teff$~label and $\lgg$~label
        but different $\feh$~labels overlap in this figure.
        However the $164$~\stagger~nodes (blue diamonds) are not
        regularly space in effective temperature: thus,
        model atmospheres with different $\feh$~labels
        are apparent as a horizontal scatter of the
        \stagger~nodes around nearby \marcs~nodes.
        The 3D LTE and 3D non-LTE calculations 
        for \ion{C}{I} and \ion{O}{I}
        were only performed on a subset of the full \stagger-grid
        ($74$~red squares),
        covering dwarfs and sub-giants.}
        \label{fig:kiel}
    \end{center}
\end{figure}

\begin{figure*}
    \begin{center}
        \includegraphics[scale=0.33]{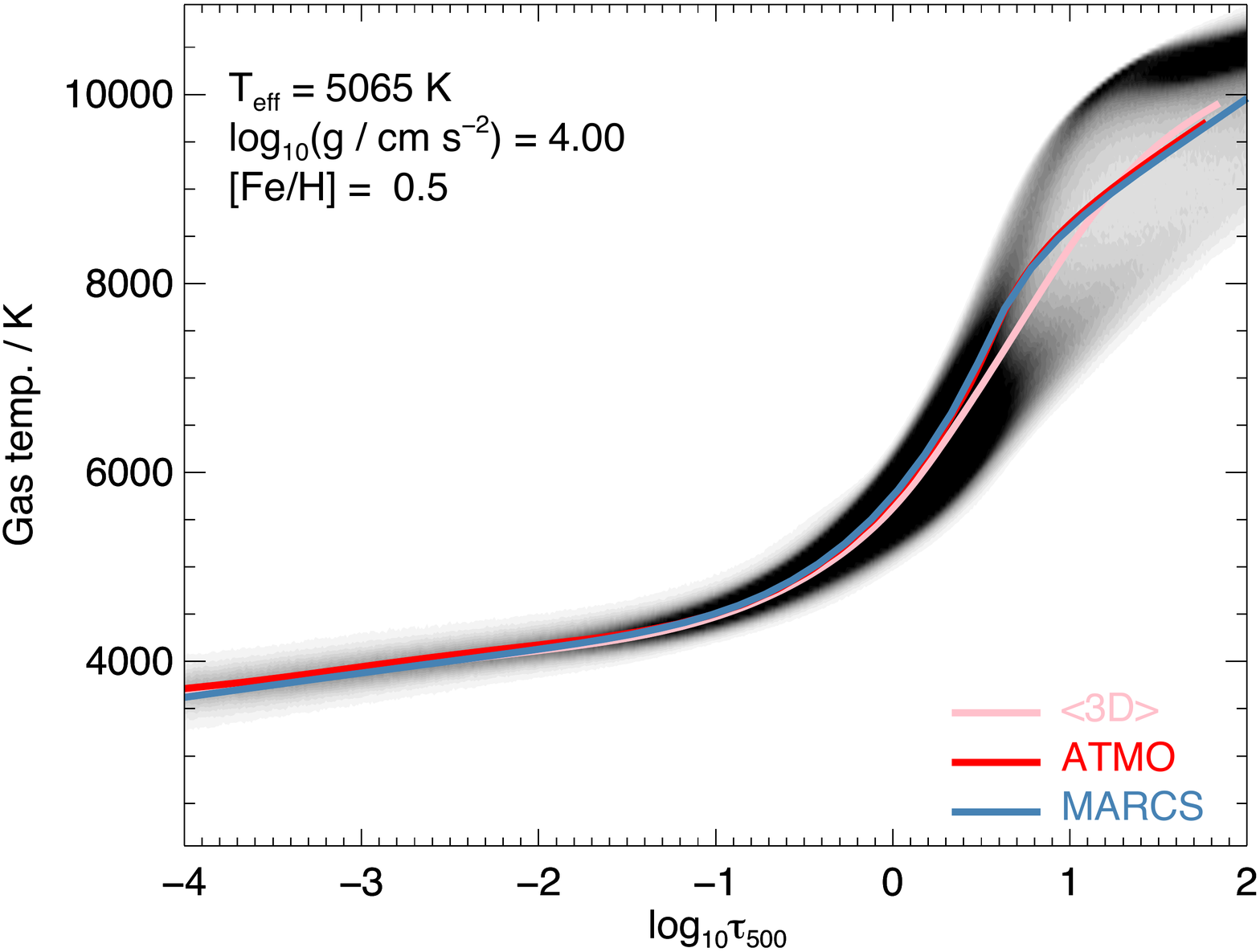}\includegraphics[scale=0.33]{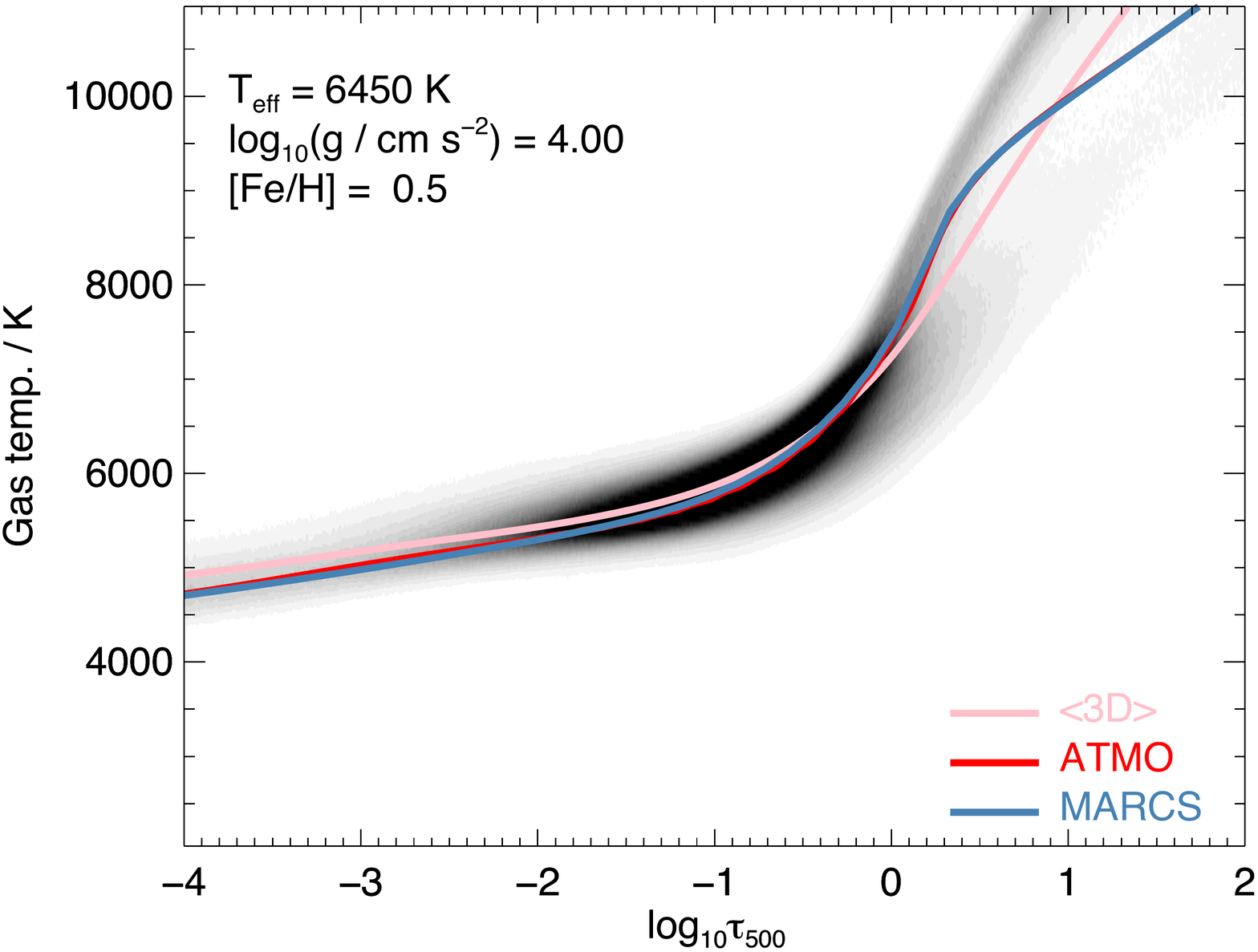}
        \includegraphics[scale=0.33]{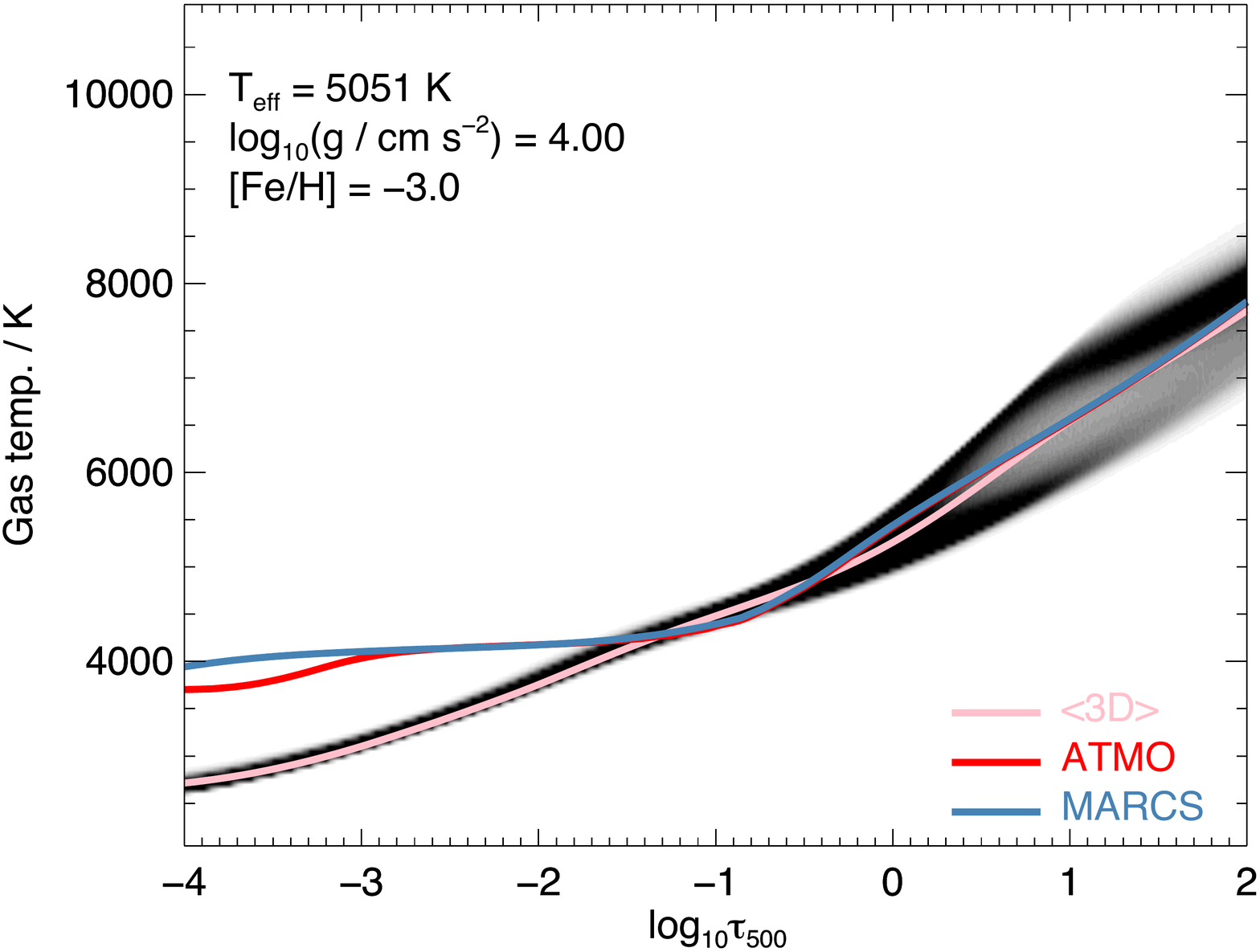}\includegraphics[scale=0.33]{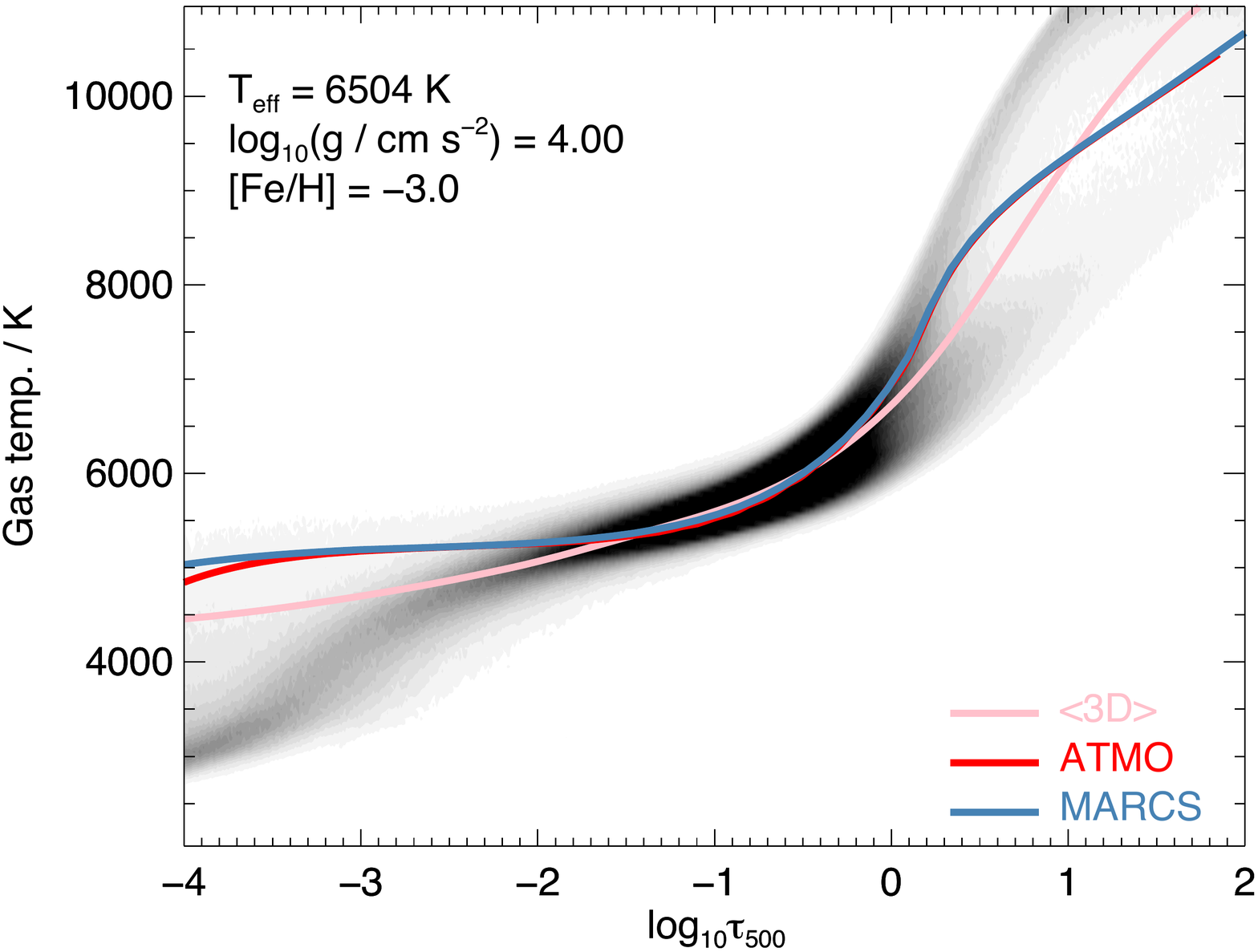}
        \caption{Gas temperature distributions in four different
        3D \stagger~model atmospheres. The \mtd~and 1D \atmo~model
        atmospheres are also plotted, as are the
        1D \marcs~model atmospheres after interpolating
        them onto the same effective temperature.
        The 1D \atmo~and \marcs~model atmospheres were computed
        using the same fixed set of MLT parameters:
        $\alpha_{\text{MLT}}=1.5$, $y=0.076$, $\varv_{\text{conv}}=8.0$.}
        \label{fig:temp}
    \end{center}
\end{figure*}

Spectrum synthesis calculations were performed on four 
different families of model atmospheres:
3D hydrodynamic model atmospheres from
the \stagger-grid \citep{2013A&amp;A...557A..26M};
1D model atmospheres determined by 
averaging the 3D \stagger~model atmospheres (henceforth
\mtd~model atmospheres; \citealt{2013A&amp;A...560A...8M});
theoretical 1D hydrostatic model atmospheres
from the \atmo-grid (the 1D equivalent of the \stagger-grid, see
Appendix A of \citealt{2013A&amp;A...557A..26M});
and theoretical 1D hydrostatic model atmospheres
from the \marcs-grid \citep{2008A&amp;A...486..951G}.
We illustrate the grids in $\lgg$---$\teff$~space in
\fig{fig:kiel}, and the temperature distributions for a few 
example models in \fig{fig:temp}.

\subsubsection{3D model atmospheres}
\label{method_atmospheres_3d}

The 3D hydrodynamic model atmospheres were adopted
from the \stagger-grid
\citep{2013A&amp;A...557A..26M},
which was constructed using the \stagger-code
\citep[e.g.][]{Nordlund:1995,2018MNRAS.475.3369C}.
The model atmospheres are labelled by their effective temperatures 
($\teff$), surface gravities 
($\lgg$),
and iron abundance with respect to that of the Sun 
($\feh$).
Their chemical compositions are that of the Sun
\citep{2009ARA&amp;A..47..481A}, scaled by $\feh$,
and with $\upalpha$-element abundances enhanced
by $+0.4\,\dex$~for $\feh\leq-1.0$, to roughly 
account for the mean Galactic chemical evolution.
The grid is not regular:
the effective temperature step size varies by
around $50\,\mathrm{K}$~across the grid~(\fig{fig:kiel}),
because the emergent flux and hence the effective temperature
is an output of a given simulation, rather than an input 
parameter.

The 3D LTE radiative transfer calculations for \ion{Fe}{II}
were performed on a set of
model atmospheres of dwarfs, sub-giants, and giants.
This set spans $29$~nodes in $\lgg$---$\teff$ space,
and up to $6$~nodes in $\feh$:
$4000\lesssim\teff/\mathrm{K}\lesssim6500$~(in steps of
roughly $500\,\mathrm{K}$), 
$1.5\leq\log\left(g / \mathrm{cm\,s^{-2}}\right)\leq5.0$~(in
steps of $0.5\,\dex$), 
and $-4.0\leq\feh\leq0.5$~(in steps of $1.0\,\dex$~for $\feh\leq0.0$).
The set
is illustrated in $\lgg$---$\teff$~space in \fig{fig:kiel}. 
Model atmospheres with the same $\lgg$~labels
but different $\feh$~labels (input parameters
in the simulations) generally have different effective temperatures.
This results in a horizontal scatter 
in \fig{fig:kiel}
about the targeted $\teff$~nodes that are $500\,\mathrm{K}$~apart.
Although the set as described above should contain
$29\times6=174$~model atmospheres, calculations were only performed
on $164$~model atmospheres; $10$~model atmospheres are missing
from this grid (mainly corresponding to models
having $\feh=+0.5$, and having lower effective
temperatures and surface gravities).

For \ion{C}{I} and \ion{O}{I}, 
the 3D non-LTE radiative calculations were performed on
a subset of model atmospheres of dwarfs and sub-giants,
as shown in \fig{fig:kiel}.
This subset spans $15$~nodes in $\lgg$---$\teff$ space,
and up to $5$~nodes in $\feh$:
$5000\lesssim\teff/\mathrm{K}\lesssim6500$,
$3.0\leq\log\left(g / \mathrm{cm\,s^{-2}}\right)\leq5.0$,
and $-3.0\leq\feh\leq0.5$.
Calculations were performed
on $74$~model atmospheres: the 
$\teff\approx5500\,\mathrm{K}$, $\lgg=3.0$,
$\feh=+0.5$~model is missing from this grid.

Prior to carrying out the 3D non-LTE calculations,
the model atmospheres were re-sampled and re-grided, from their
original Cartesian mesh having 
$240\times240\times230$~physical grid-points,
to one having $80\times80\times220$~physical grid-points
(Sect.~2.1.1 of \citealt{2018A&A...615A.139A}).  
Calculations were performed on typically
five snapshots of each model atmosphere,
equally spaced in stellar time,
from which temporally-averaged
emergent line fluxes could be determined.

\subsubsection{\mtd~model atmospheres}
\label{method_atmospheres_mean3d}

In this work, \mtd~model atmospheres were taken
from the averaged \stagger-grid presented 
in \citet{2013A&amp;A...560A...8M}.
The line formation calculations on these 
model atmospheres were only
used to study the general behaviour of departure coefficients
(\sect{results_departure});
however, \mtd~non-LTE versus 1D LTE abundance corrections
were also computed, and are available upon request.
The \mtd~model atmospheres have 1D geometry, but were determined
from the 3D \stagger~model atmospheres.
Specifically, in this work 
the \mtd~models are
horizontal- and temporal-averages
(on surfaces of constant Rosseland mean optical depth)
of the gas temperature, logarithmic gas density, and logarithmic
electron number density \citep{2013A&amp;A...560A...8M}.
For all three elements,
radiative transfer calculations were performed on the entire 
set of \mtd~model atmospheres depicted in \fig{fig:kiel}.

In principle, the effective temperature 
can change after
averaging the 3D model atmosphere.  These changes were not taken
into account here: in other words, the $\teff$~label of the 
\mtd~model atmosphere
is identical to that of the 
\stagger~model atmosphere from 
which they were constructed.
The illustration in \fig{fig:temp} suggests that 
any changes to the 
effective temperature are in any case small.

\subsubsection{1D model atmospheres}
\label{method_atmospheres_1d}

Theoretical 1D hydrostatic model atmospheres were adopted 
from the \atmo-grid (Appendix A of \citealt{2013A&amp;A...557A..26M}).
The \atmo~model atmospheres
are the 1D versions of the \stagger~model
atmospheres, using the same 
radiative transfer solver, angle quadrature, and binned opacities,
and computed to have
exactly the same effective temperatures, 
surface gravities,
and chemical compositions,
as their 3D equivalents.
Thus, the 3D~non-LTE versus 1D LTE and
3D LTE versus 1D LTE abundance corrections derived
(\sect{method_abcor}) and presented (\sect{results_abcor}) here
are based on \stagger~and \atmo~model atmospheres, to make
them as differential as possible. 
For all three elements,
radiative transfer calculations were performed on the entire 
set of \atmo~model atmospheres depicted 
in \fig{fig:kiel}.

In addition, calculations were performed on
$1807$~theoretical 1D hydrostatic model atmospheres adopted 
from the standard \marcs-grid \citep{2008A&amp;A...486..951G}.
Compared to the \stagger- and \atmo-grids, 
the \marcs-grid has the benefits of 
of using a monochromatic 
opacity-sampling treatment for the radiative transfer,
and of being finer and more extended in stellar parameter space.
For these reasons, 1D non-LTE versus 1D LTE abundance
corrections for \ion{C}{I} and \ion{O}{I}
based on \marcs~model atmospheres are also presented here.

The predicted atmospheric structure, and hence
the resultant LTE and non-LTE emergent line fluxes,
are in practice very similar between
the \atmo~and \marcs~model atmospheres,
as is evident in \fig{fig:temp}.
This is in part because the 
two grids adopt the same implementation of the Mixing Length Theory
\citep[MLT;][]{1958ZA.....46..108B,1965ApJ...142..841H} to model
the convective flux, using the same fixed set of MLT parameters:
$\alpha_{\text{MLT}}=1.5$, $y=0.076$, $\varv_{\text{conv}}=8.0$.
Both sets of theoretical 1D model atmospheres 
effectively enforce radiative equilibrium in the upper layers:
\fig{fig:temp}~demonstrates that in the metal-poor regime,
this leads them to significantly overestimating
the temperature of the upper layers, where 
in reality the temperature is set by the competing effects
of radiative heating and adiabatic cooling
\citep[e.g.][]{1999A&amp;A...346L..17A}.

As with the \stagger, \mtd, and 
\atmo~model atmospheres, the \marcs~model atmospheres
are labelled by $\teff$, $\lgg$, and $\feh$;
however, their chemical compositions are based on an 
older compilation of solar abundances \citep{2007coma.book..105G},
and the $\upalpha$-element abundances are enhanced 
by $0.1\,\dex$~for each step of $0.25\,\dex$~from $\feh=0.0$~to
$\feh=-1.0$. 
For all three elements,
radiative transfer calculations were performed on the entire 
set of \marcs~model atmospheres depicted 
in \fig{fig:kiel}.
This set spans
$4000\leq\teff/\mathrm{K}\leq8000$~(in 
steps of $250\,\mathrm{K}$),
$-0.5\leq\log\left(g / \mathrm{cm\,s^{-2}}\right)\leq5.0$~(in
steps of $0.5\,\dex$),
and $-5.0\leq\feh\leq0.5$~(in steps of $0.25$~to $1.0\,\dex$).
Following \citet{2018MNRAS.478.4513B}, for $\lgg\geq4.0$~plane-parallel 
model atmospheres 
computed with a microturbulence of $1.0\,\kms$~were adopted; 
otherwise,
solar-mass spherically-symmetric model atmospheres
computed with a microturbulence of $2.0\,\kms$~were adopted.

\subsection{Non-LTE atomic models}
\label{method_atom}

\begin{figure}
    \begin{center}
        \includegraphics[scale=0.33]{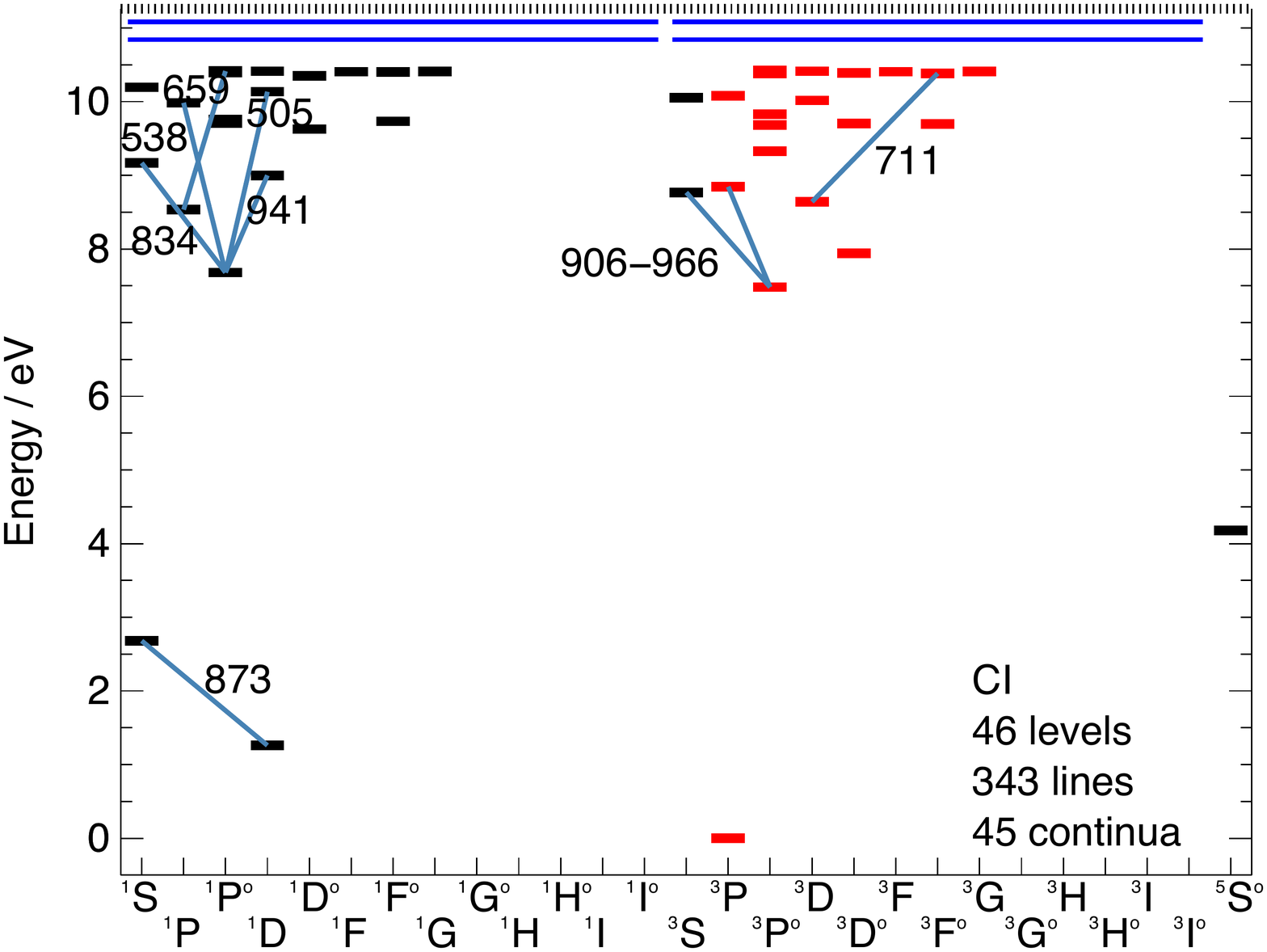}
        \includegraphics[scale=0.33]{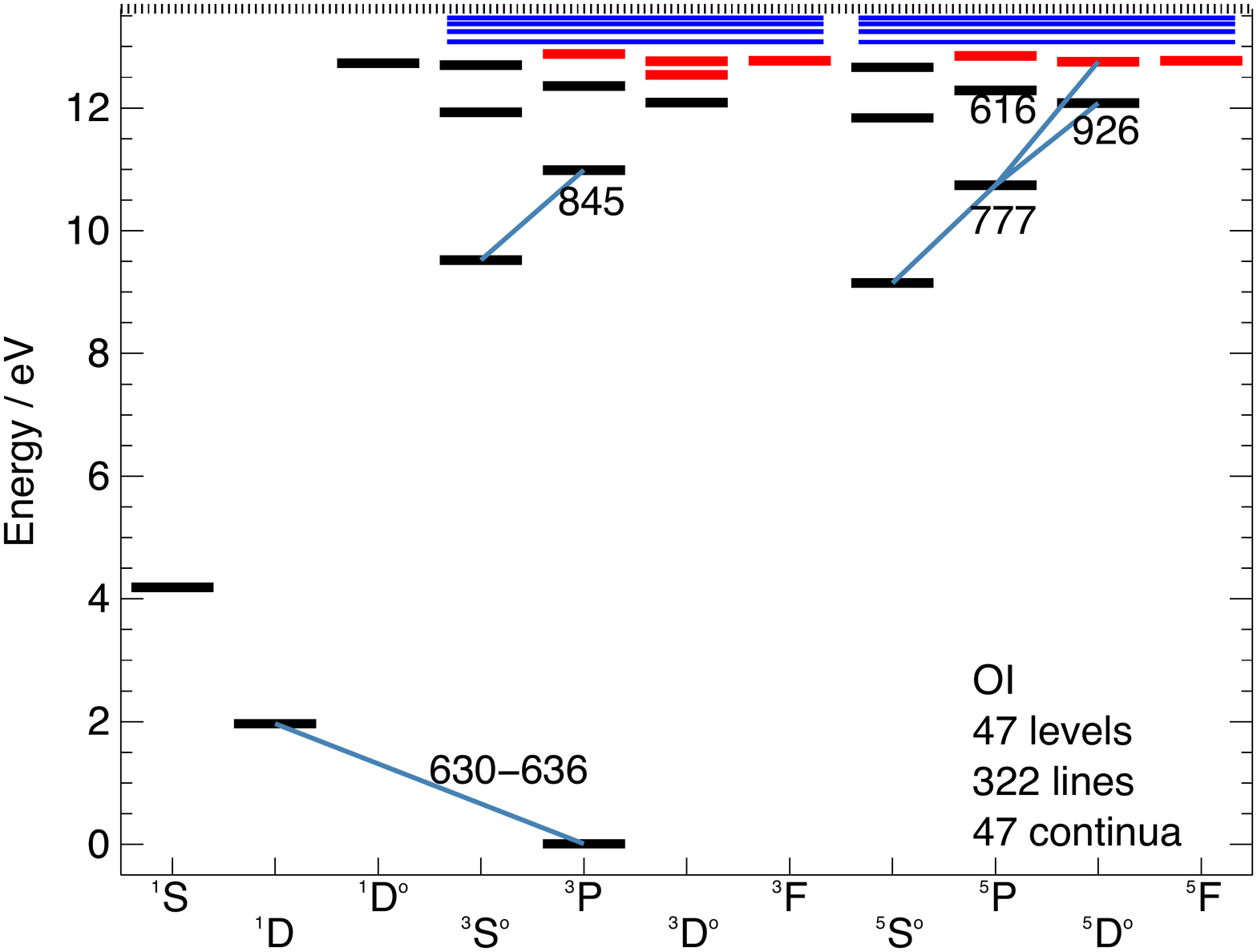}
        \includegraphics[scale=0.33]{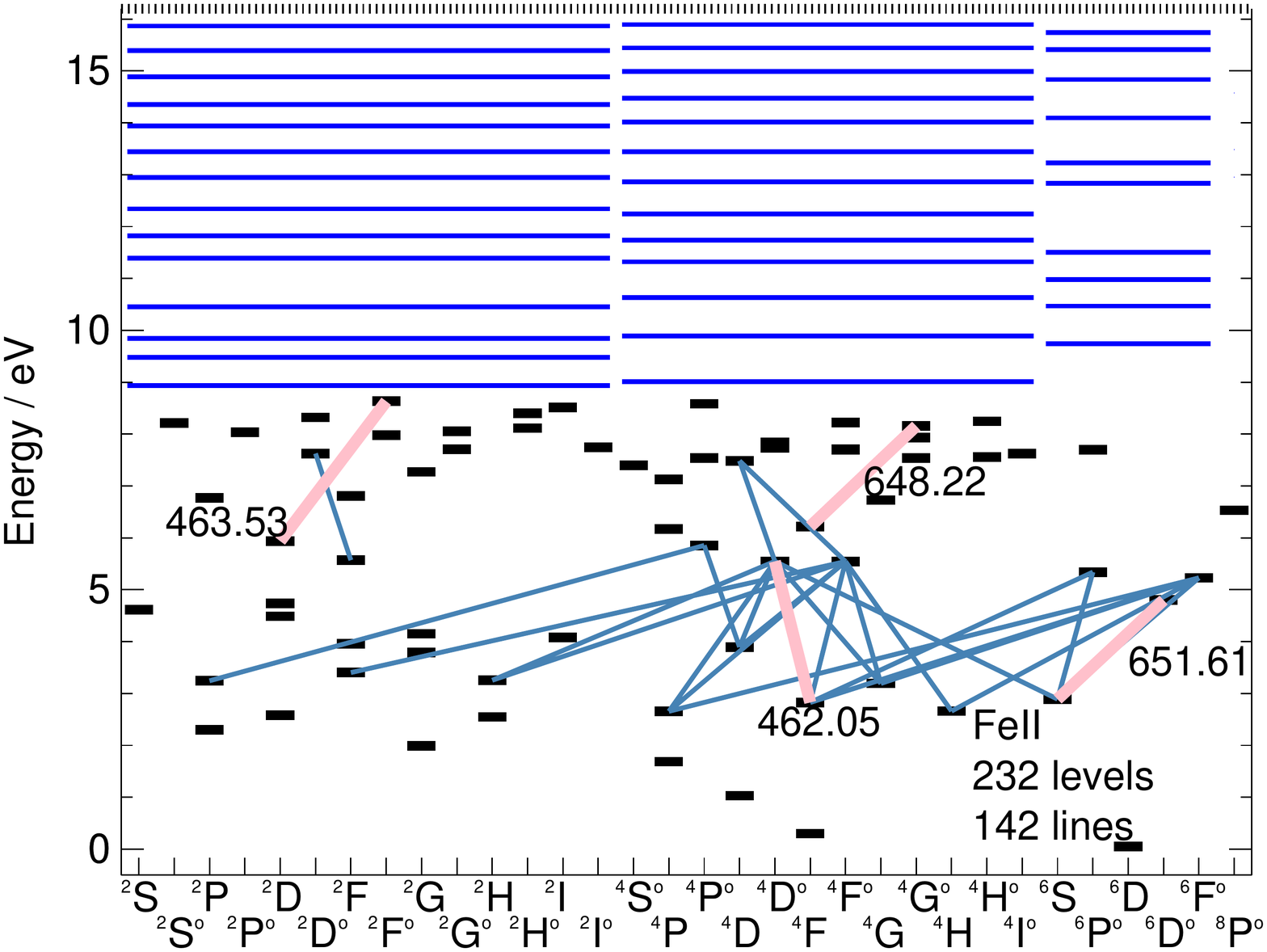}
        \caption{Grotrian term diagrams of the atomic models
        for \ion{C}{I} and \ion{O}{I} used for the non-LTE iterations.
        Also shown is a Grotrian term diagram for 
        \ion{Fe}{II}, based on which 
        LTE emergent line fluxes were calculated.
        Levels for which fine structure has been collapsed
        are shown in red, and 
        super levels are shown as long horizontal blue
        lines at the top of the figures.
        The transitions for which emergent line fluxes
        and abundance corrections were calculated are 
        shown, connecting different energy levels. 
        Their approximate wavelengths in air ($\nm$) are indicated;
        for \ion{Fe}{II} this is done only for the 
        four lines 
        for which the 3D LTE versus 1D LTE abundance corrections are
        shown in \fig{fig:abcor_strength_fe2}.}
        \label{fig:grotrian}
    \end{center}
\end{figure}

For \ion{C}{I}, the `No-FS' atomic model presented in
\citet{2019A&A...624A.111A} was adopted here.
This model is composed of
$46$~levels of \ion{C}{I} plus the ground state
of \ion{C}{II}, $343$~radiative bound-bound transitions,
and $45$~radiative bound-free transitions in total.
For \ion{O}{I}, the `reduced' atomic model presented in
\citet{2018A&A...616A..89A} was adopted here.
This model is composed of $47$~levels of \ion{O}{I} plus the
three lowest levels of \ion{O}{II}, $322$~radiative bound-bound transitions,
and $47$~radiative bound-free transitions in total.
We illustrate the atomic models used for the non-LTE iterations 
in \fig{fig:grotrian}; 
full details can be found in the above references.

A simple atomic model was also constructed for 
\ion{Fe}{II}, which we illustrate this model in
\fig{fig:grotrian}.  By using this atomic model,
the calculations for this species
were carried out in the same way as for \ion{C}{I} and \ion{O}{I},
using the exact same 3D non-LTE code that
we discuss in \sect{method_rt},
albeit without performing any non-LTE iterations
(for the reasons given in \sect{introduction}).
Although the synthetic spectra were thus calculated under LTE conditions,
completeness of the atomic model is important here
to ensure that its partition function is consistent 
with that adopted by the internal equation of state module 
of the 3D non-LTE code.  For this reason,
super levels were included in the atomic model,
as shown in \fig{fig:grotrian}.

\subsection{Radiative transfer post-processing}
\label{method_rt}

The radiative transfer post-processing of the model atmospheres
was carried out using 
\balder. This is a 3D non-LTE MPI-parallelised FORTRAN code
that both solves the equations of statistical equilibrium
and calculates the normalised, disk-integrated emergent spectrum.  It is 
based on \multitd~\citep{2009ASPC..415...87L_short},
but has various important modifications
in particular concerning the 
parallelisation scheme and the
equation of state and opacities \citep{2016MNRAS.463.1518A},
the emergent spectrum solver \citep{2018A&A...615A.139A},
and the statistical equilibrium solver \citep{2019A&A...624A.111A}.

Emergent line fluxes for \ion{Fe}{II} lines were calculated
by \balder~in LTE (without any non-LTE iterations,
for the reasons given in \sect{introduction}).
For \ion{C}{I}~and \ion{O}{I}, 
non-LTE iterations were carried out by \balder~first,
using the atomic models described in \sect{method_atom}.
After the solutions had converged, emergent
line fluxes were calculated by \balder~using 
comprehensive atomic models that include all fine structure. 
Emergent line fluxes were 
calculated only for specific lines,
as we discuss in \sect{method_lines}.

The calculations on the 3D, \mtd, and 1D model atmospheres
(\sect{method_atmospheres})
follow an approach that is very similar
to that described in \citet{2018A&A...615A.139A},
and that paper can be consulted for 
further details. Here, we just note that the main difference
between the calculations on the 3D model atmospheres,
and on the \mtd~and 1D model atmospheres,
pertains to the broadening of
disk-averaged, temporally-averaged spectral lines
effected by temperature and velocity gradients as well as oscillatory 
motions. These effects are naturally accounted for during the 
post-processing of 3D hydrodynamic model atmospheres
\citep[e.g.][]{2000A&amp;A...359..729A}.
In contrast, post-processing of  \mtd~and 1D model atmospheres
generally needs to include extra broadening parameters:
microturbulence $\vmic$~and macroturbulence 
$\vmac$~(e.g.~Chapter 17 of \citealt{2008oasp.book.....G}).
Here, the calculations on \mtd~and 1D model atmospheres
were performed for three different values of 
depth-independent microturbulence: 
$\xi=0.0$, $1.0$, and $2.0\,\kms$.
The current study relies primarily on
abundance corrections based on equivalent widths
(\sect{method_abcor}, \sect{results_abcor}),
rather than on spectrum fitting, and consequently macroturbulence,
which by definition conserves the equivalent width, is 
not considered here.

The equation of state and background opacities
--- all line (bound-bound) and continuous 
(bound-free and free-free) opacities not already included in 
the atomic model --- were computed by the \blue~module within \balder,
once given the temperature, density, and chemical composition
of the model atmosphere.
As in previous work 
(see Sect.~2.1.2 of \citealt{2016MNRAS.463.1518A}),
the background continuous opacities were computed on the fly,
whereas the background line opacities were precomputed 
on regular grids of temperature, density, and chemical composition
(labelled by $\feh$), and interpolated onto the model atmosphere
at runtime.
The elemental abundances adopted by \blue~were generally
set to the values that the model atmosphere was computed with.
For the \ion{Fe}{II} calculations, the iron abundance
was always fixed to that of the model atmosphere.
However for the \ion{C}{I} and \ion{O}{I} calculations,
the carbon and oxygen abundances were respectively varied
between $-0.4\leq\abrat{X}{Fe}\leq1.2$, 
in steps of $0.4\,\dex$~for the 3D calculations,
and in steps of $0.2\,\dex$~for the \mtd~and 1D calculations,
independent of the atmospheric chemical composition
(labelled by $\feh$).

In all cases,
natural (radiative) broadening coefficients were estimated from
the available line and level data.
Pressure broadening due to 
elastic collisions with atomic hydrogen were generally based on
the theory of Anstee, Barklem, and O'Mara
\citep[ABO;][]{1995MNRAS.276..859A,1997MNRAS.290..102B,1998MNRAS.296.1057B}.
For lines outside of these tables, 
the theory of \citet{1955psmb.book.....U} was used instead,
with an enhancement factor of $2.0$~for \ion{C}{I} and
\ion{O}{I}, and $1.5$~for \ion{Fe}{II}.

\subsection{Line selection}
\label{method_lines}

Emergent line fluxes (and subsequently equivalent widths and
abundance corrections) were only determined for
the \ion{C}{I}, \ion{O}{I}, and \ion{Fe}{II}~lines
of most relevance to spectroscopic studies of late-type stars.
This set of lines is larger than, and includes all of, the lines
used in the subsequent re-analysis of literature data (\sect{obs}).
We illustrate the lines in \fig{fig:grotrian} and
provide a brief overview here; we provide a complete list of the adopted
line parameters in the online Table 1.

The \ion{C}{I} lines are listed in the first
$17$~rows of Table 1 in \citet{2019A&A...624A.111A}.
They are all in the optical and near infra-red,
spanning wavelengths from $505\,\nm$~to $966\,\nm$.
The $16$~permitted lines
are all of high excitation potential,
$\epot\gtrsim7.5\,\eV$, and with oscillator strengths
$-1.6<\lggf<+0.3$; 
the forbidden, weak ($\lggf=-8.165$), low-excitation 
($\epot=1.264\,\eV$) 
[\ion{C}{I}] $872.7\,\nm$~line is also included.

The set of \ion{O}{I} lines include those studied by
\citet{2016MNRAS.455.3735A}, namely the
permitted \ion{O}{I} $616\,\nm$~and $777\,\nm$~multiplets
and the forbidden, low-excitation
[\ion{O}{I}] $630.0\,\nm$~and $636.4\,\nm$~lines.
In this study this set is extended 
to include the permitted 
\ion{O}{I} $844\,\nm$~and $926\,\nm$~multiplets.
In contrast to \citet{2016MNRAS.455.3735A}, 
the abundance corrections presented here do not 
take into account the \ion{Ni}{I}~line that blends
the [\ion{O}{I}] $630.0\,\nm$~line \citep[e.g.][]{2001ApJ...556L..63A}.

The set of \ion{Fe}{II}~lines
includes the same $142$~lines studied by 
\citet{2009A&amp;A...497..611M}.
They are all in the optical and near infra-red,
spanning wavelengths from $409\,\nm$~to $771\,\nm$,
of low to intermediate excitation potentials,
$2.6\lesssim\epot\lesssim6.2\,\eV$, and with oscillator strengths
$-5.1<\lggf<-1.0$.
This set of atomic data is perhaps the
best currently available for \ion{Fe}{II} lines, in the absence
of more complete laboratory investigations \citep{2019arXiv190711760D}.
Combined with 3D abundance corrections, these lines
offer a promising way to obtain highly accurate
iron abundances in late-type stars.

\subsection{Definition of abundance corrections}
\label{method_abcor}

The (absolute)
abundance corrections presented in this work are based on 
equivalent widths, rather than on spectrum fitting.
For a given atmospheric model (3D \stagger, 1D \atmo,
1D \marcs), and a given radiative transfer
post-processing approach (non-LTE, LTE),
the equivalent widths of the 
lines listed in \sect{method_lines} were determined
by directly integrating across the normalised emergent line fluxes.

The 3D non-LTE versus 1D LTE abundance correction
for a given \ion{C}{I} or \ion{O}{I} line was then calculated as
the difference between the absolute 3D non-LTE abundance
$\log\epsilon^{\mathrm{3D,NLTE}}$~and 
the absolute 1D LTE abundance~$\log\epsilon^{\mathrm{1D,LTE}}$,
corresponding to the same, measured equivalent width:
\phantomsection\begin{IEEEeqnarray}{rCl}
\label{abundancecorrection}
    \corr{3N}{1L}&=&
    \log\epsilon^{\mathrm{3D,NLTE}}-\log\epsilon^{\mathrm{1D,LTE}}\,.
\end{IEEEeqnarray}
These are generally five-dimensional functions:
\phantomsection\begin{IEEEeqnarray}{rCl}
    \label{abcor1}
    \corr{3N}{1L}
    &=&
    \corr{3N}{1L}\left(\log\epsilon^{\mathrm{1D,LTE}},
    \teff,\lgg,\feh,\vmic\right)\,.
\end{IEEEeqnarray}
Here, $\teff$, $\lgg$, and $\feh$~are the atmospheric parameters
(\sect{method_atmospheres}), and $\vmic$~is
the microturbulence with which the 1D LTE line synthesis
was performed (\sect{method_rt}).

The 3D LTE versus 1D LTE abundance correction
for a given \ion{Fe}{II} line, $\corr{3L}{1L}$,
was calculated in a similar fashion.
Here, the iron abundance is always consistent
with the atmospheric chemical composition as labelled
by $\feh$~(\sect{method_rt}).
Consequently $\corr{3N}{1L}$~is a four-dimensional function
in this case.

The 1D non-LTE versus 1D~LTE abundance corrections
for \ion{C}{I}~and \ion{O}{I}, $\corr{1N}{1L}$,
are also generally five-dimensional functions, and
are defined in an analogous way to $\corr{3N}{1L}$.
For these abundance corrections between two sets of 1D calculations,
the same microturbulence was assumed in both sets
($\vmicol=\vmicon=\vmic$).
As discussed in \sect{method_atmospheres}, 
although the 3D (non-)LTE versus 1D LTE abundance corrections
presented in this work are based on the
3D \stagger~and 1D \atmo~model atmospheres,
the 1D non-LTE versus 1D LTE abundance corrections 
work are based only on 1D \marcs~model atmospheres.

For the \ion{Fe}{II}~lines, 
prior to calculating the abundance corrections,
the equivalent widths were first
interpolated onto a grid that is regularly spaced in effective temperature.  
This had to be done for \ion{Fe}{II},
because the iron abundance was always forced to be consistent 
with the atmospheric chemical composition, labelled by $\feh$,
and because the \stagger-grid nodes are irregular in effective temperature
(\sect{method_atmospheres}), with nodes having the same 
$\lgg$~label and $\feh$~label generally having different effective
temperatures.  
The resulting 3D LTE versus 1D LTE abundance corrections for \ion{Fe}{II}
are thus presented on a regular grid of effective temperatures.
In contrast, this was not necessary for
\ion{C}{I} and \ion{O}{I}, 
because the carbon and oxygen abundances were varied independently of $\feh$.
The resulting 3D non-LTE versus 1D LTE abundance corrections for \ion{C}{I}
and \ion{O}{I} are thus presented on an irregular grid
of effective temperatures, corresponding to the nodes
shown in \fig{fig:kiel}.

\section{3D non-LTE effects across parameter space}
\label{results}

\subsection{Departure coefficients}
\label{results_departure}

\begin{figure*}
    \begin{center}
        \includegraphics[scale=0.33]{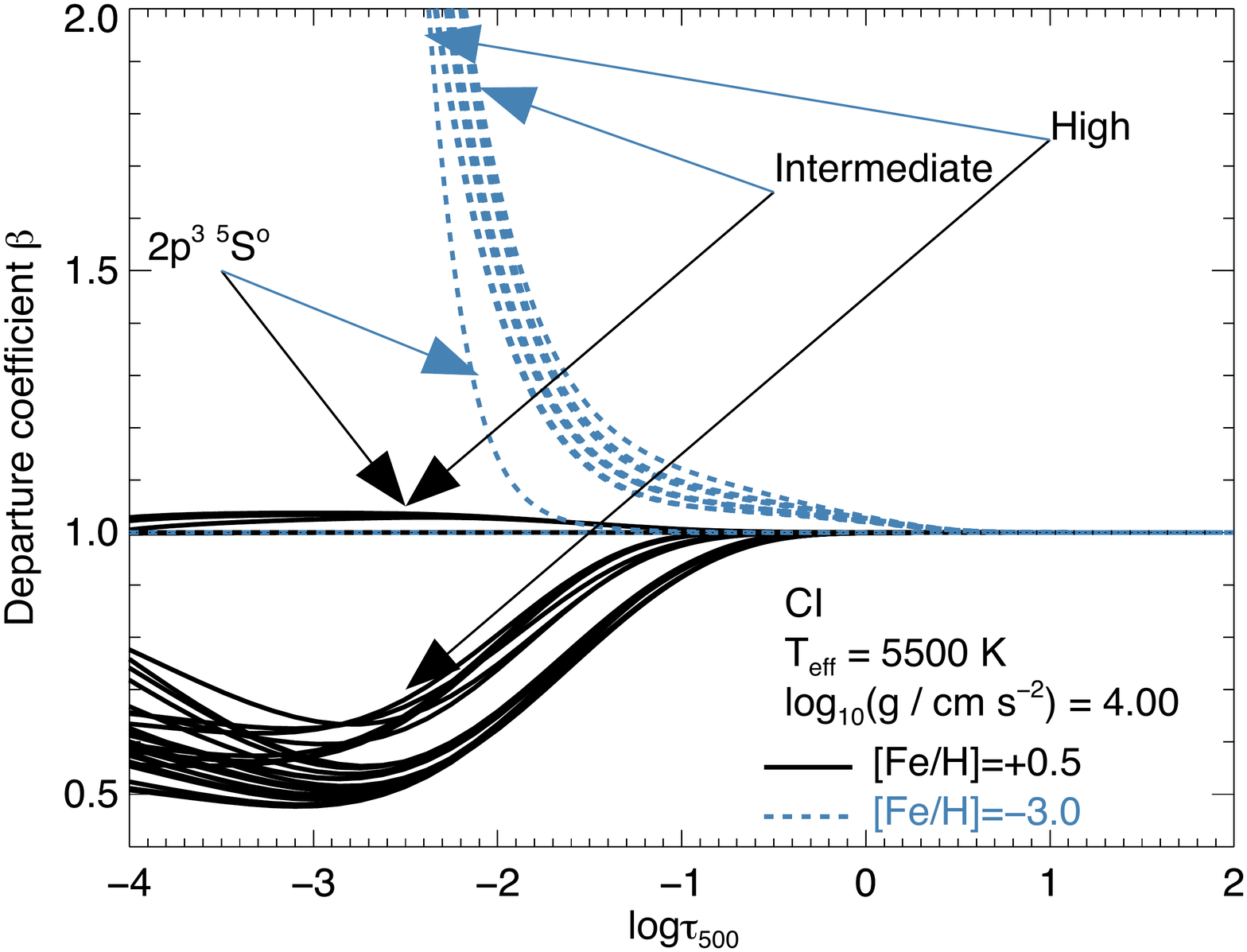}\includegraphics[scale=0.33]{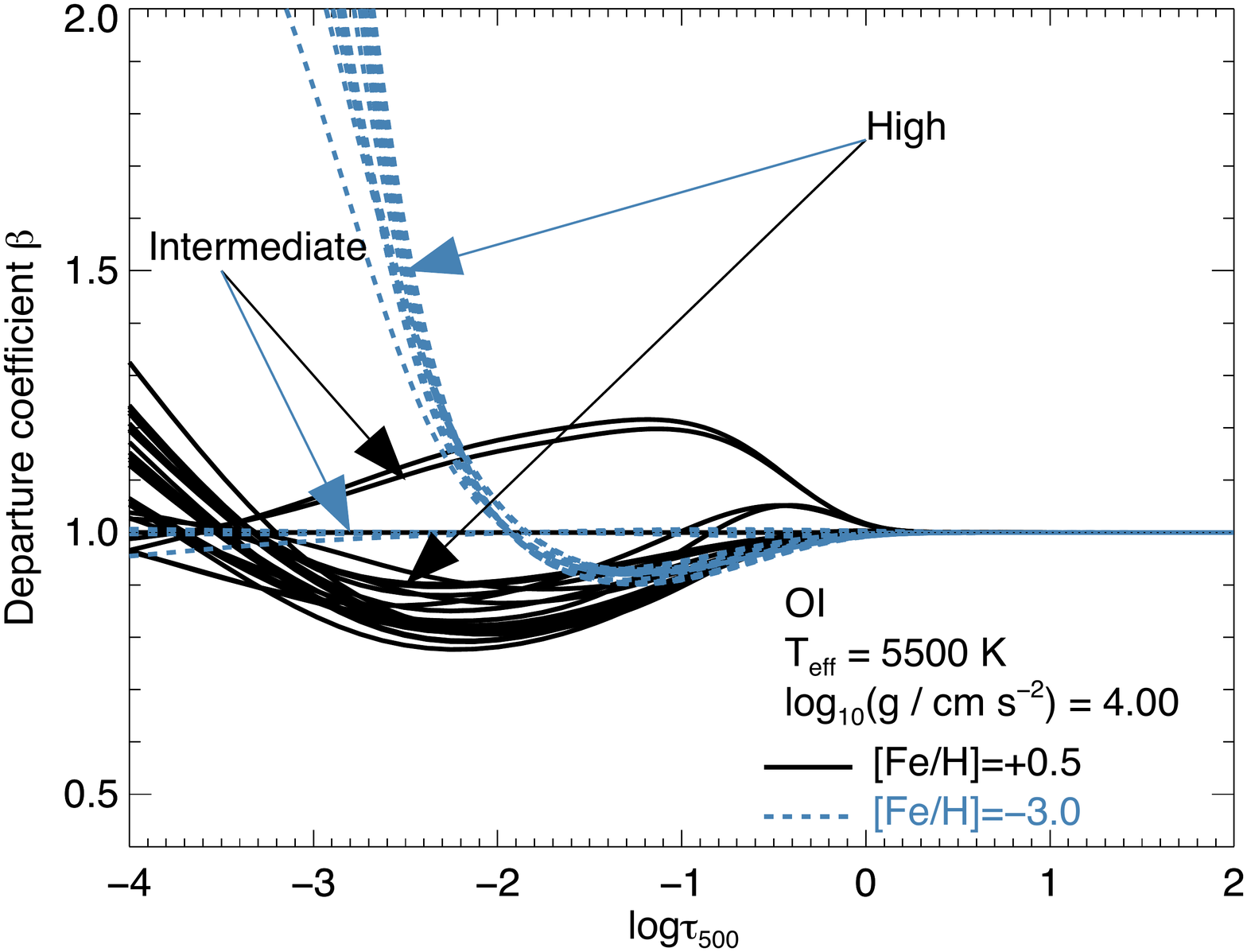}
        \caption{Departure coefficients for levels of \ion{C}{I} (left panel),
        and for levels of \ion{O}{I} (right panel), in different
        \mtd~model atmospheres assuming
        $\abrat{C}{Fe}=\abrat{O}{Fe}=0.0$~and $\vmic=1.0\,\kms$.
        The levels of intermediate excitation indicated here are,
        for \ion{C}{I},
        $\mathrm{2p.3s\,^{3}P^{o}}$~($7.49\,\mathrm{eV}$),
        $\mathrm{2p.3s\,^{1}P^{o}}$~($7.68\,\mathrm{eV}$),
        and $\mathrm{2p^{3}\,^{3}D^{o}}$~($7.95\,\mathrm{eV}$);
        and for \ion{O}{I}, 
        $\mathrm{2p^{3}.3s\,^{5}S^{o}}$~($9.15\,\mathrm{eV}$)
        and 
        $\mathrm{2p^{3}.3s\,^{3}S^{o}}$~($9.52\,\mathrm{eV}$).
        The \ion{C}{I} $\mathrm{2p^{3}\,^{5}S^{o}}$~($4.18\,\mathrm{eV}$)
        is indicated separately.
        The levels of high excitation indicated here are the ones above
        these levels.
        In both panels, the levels of low excitation potential
        remain close to unity throughout the atmospheres shown.}
        \label{fig:depcof}
    \end{center}
\end{figure*}

To understand the behaviour of the
abundance corrections for \ion{C}{I} and 
\ion{O}{I} lines, it is helpful to first 
consider how the level populations deviate from their 
LTE predictions in the atmospheres of different late-type stars.
In \fig{fig:depcof} we plot the departure coefficients
$\depcof\equiv n_{\text{NLTE}}/n_{\text{LTE}}$~for
the different \ion{C}{I} and \ion{O}{I} levels in the atomic 
models (\sect{method_atom}).
For clarity we only show the departure coefficients in the
\mtd~model atmospheres, which tend to 
follow the distributions of departure coefficients
in the 3D model atmospheres
(see for example Fig.~4 of \citealt{2017MNRAS.464..264A}).
We also only show results for the models with
$\teff\approx5500\,\mathrm{K}$, $\lgg=4.00$,
and two different chemical compositions corresponding to
the labels $\feh=+0.5$~and $-3.0$,
with $\abrat{C}{Fe}=\abrat{O}{Fe}=0.0$, and $\vmic=1.0\,\kms$.
The departure coefficients in stars with different
effective temperatures and surface gravities
show trends that are at least
qualitatively similar to those discussed here.

Before discussing the behaviour of
levels of intermediate and high excitation potential
for \ion{C}{I} and \ion{O}{I} in detail,
we note that for both species, 
the levels of low excitation potential (below around $3\,\eV$; see
\fig{fig:grotrian}) have the same behaviour.
Namely, their populations
do not significantly deviate from their LTE predictions,
across the entire grid of stellar parameters.
Since \ion{C}{I} and \ion{O}{I} are majority species 
in late-type stellar atmospheres, these levels
are very highly populated and are thus relatively insensitive
to changes in the populations of the sparsely-populated 
levels of higher excitation potential.
Consequently, the current model 
predicts negligible
non-LTE effects for the [\ion{C}{I}]
$872.7\,\nm$~line, and for the
[\ion{O}{I}] $630.0\,\nm$~and $636.4\,\nm$~lines.

\subsubsection{Levels of atomic carbon}
\label{results_departure_c1}

\fig{fig:depcof} shows that, in the high
metallicity case ($\feh=+0.5$), the populations of the 
levels of intermediate
excitation potential (the lower levels of the 
near infra-red \ion{C}{I} lines 
around $906\,\nm$~to $966\,\nm$
as well as of the \ion{C}{I} $505.2\,\nm$~and $538.0\,\nm$~lines)
are close to unity
(in the main line-forming regions $-1.0\lesssim\lgt\lesssim0.5$; see
Fig.~1 of \citealt{2019A&A...624A.111A}).
The populations of the levels of high excitation potential
(the upper levels of the aforementioned lines)
drop slightly below unity.
This behaviour is driven by photon losses in the network 
of intermediate- and high-excitation \ion{C}{I} lines, causing a population
cascade downwards.

This implies a slight source function effect on the
intermediate- and high-excitation \ion{C}{I} lines
in the high metallicity case.
A similar effect was recently discussed
for the Sun \citep{2019A&A...624A.111A}.
The line source function follows
$\depcof_{\text{upper level}}/\depcof_{\text{lower level}}$~\citep{2003rtsa.book.....R},
and drops below the Planck function, thus
the lines are strengthened with respect to LTE.

Towards lower metallicities however, 
photon losses in the \ion{C}{I} lines become less significant.
\fig{fig:depcof} shows that 
in the low metallicity case ($\feh=-3.0$) the populations of the 
levels of intermediate and high excitation potential 
rise above unity in the \mtd~model atmospheres,
Non-thermal UV photons pump the various
\ion{C}{I} lines around $160$~to $250\,\nm$~that connect 
the levels of low excitation potential 
to the levels of intermediate excitation potential, enhancing
the populations of the latter. This overpopulation
is then communicated to the rest of the levels,
primarily via inelastic collisions with neutral hydrogen. 
Consequently, the departure coefficients are largest for
the levels of intermediate excitation potential,
and become slightly closer to unity with increasing excitation
potential. 
The exception to this general trend with excitation potential is the 
$\mathrm{2p^{3}\,^{5}S^{o}}$~level, as can be seen in 
\fig{fig:depcof}: this is because, being
in the quintet system, 
it is only weakly coupled to all of the other levels in the network,
which are in the singlet and triplet systems (\fig{fig:grotrian}).

The overexcitation effect at low metallicities
is slightly enhanced in the 3D model atmospheres 
where steeper temperature gradients 
results in a larger non-thermal UV radiation field
\citep[e.g.][]{1999A&amp;A...346L..17A}.
At higher metallicities, background metal 
lines and continua block the UV \ion{C}{I} lines
so as to make this mechanism inefficient
relative to the photon loss mechanism discussed above.

Thus, at low metallicities there is an opacity effect on the 
intermediate- and high-excitation \ion{C}{I} lines.
The line opacity follows
$\depcof_{\text{lower level}}$~\citep{2003rtsa.book.....R},
and the lines are again strengthened with respect to LTE.

This picture of the non-LTE effects
agrees well with that presented in 
Sect.~3.2 of \citet{2006A&amp;A...458..899F}.
More details of the relative importance of different
radiative and collisional transitions can be found
in that study.


\subsubsection{Levels of atomic oxygen}
\label{results_departure_o1}

At first glance the departure coefficients for
\ion{O}{I} resemble those of \ion{C}{I} discussed in
\sect{results_departure_c1}. However, there are differences
in the details that result in 
non-LTE effects on the \ion{O}{I} lines that 
are typically much more severe 
in the high metallicity case 
($\feh=+0.5$) and much less severe in the
low metallicity case ($\feh=-3.0$),
compared to those on \ion{C}{I} lines.

\fig{fig:depcof} shows that, in the high
metallicity case ($\feh=+0.5$), the populations of the 
levels of intermediate
excitation potential (the lower levels of the 
\ion{O}{I} $777\,\nm$~and $844\,\nm$~multiplets, namely
the $\mathrm{3s\,^{5}S^{o}}$~and $\mathrm{3s\,^{3}S^{o}}$~levels)
rise above unity 
(in the main line-forming regions $-1.0\lesssim\lgt\lesssim0.5$; see
Fig.~1 of \citealt{2018A&A...616A..89A}).
The populations of the 
levels of high excitation potential typically drop slightly below unity.
The exceptions seen in the plot are the 
$\mathrm{3p\,^{5}P}$~and
$\mathrm{3p\,^{3}P}$~levels 
(the upper levels of the 
\ion{O}{I} $777\,\nm$~and $844\,\nm$~multiplets,
and the lower levels of the 
\ion{O}{I} $616\,\nm$~and $926\,\nm$~multiplets):
these rise above unity due to efficient collisional 
coupling with the levels of intermediate excitation potential.
Photon losses in intermediate- and high-excitation
\ion{O}{I} lines drive a population cascade downwards,
similar to \ion{C}{I}.  Unlike \ion{C}{I} however,
this population flow stops at the metastable 
$\mathrm{3s\,^{5}S^{o}}$~level, which is efficiently 
coupled to the $\mathrm{3s\,^{3}S^{o}}$~level of slightly
higher excitation potential
\citep[e.g.][]{2009A&amp;A...500.1221F}.
This also means that
the strength of the \ion{O}{I} $777\,\nm$~multiplet can dictate
the overall statistical equilibrium \citep{2016MNRAS.455.3735A}.

This implies a
strong opacity effect 
on the \ion{O}{I} $777\,\nm$~and $844\,\nm$~multiplets
in the high metallicity case
that strengthens the lines with respect to LTE.
The more highly excited
\ion{O}{I} $616\,\nm$~and $926\,\nm$~multiplets
(\fig{fig:grotrian})
suffer from both an opacity effect as well as a source function effect.
The latter also strengthens the lines with respect to LTE,
as the line source function drops below the Planck function.

Towards lower metallicities, as with 
\ion{C}{I} (\sect{results_departure_c1}), photon
losses in the \ion{O}{I} lines become less significant.
\fig{fig:depcof} shows that 
in the low metallicity case ($\feh=-3.0$),
and at least in the main line-forming regions, 
levels of intermediate excitation potential stay close to unity,
while the populations of the 
levels of high excitation potential drop slightly below unity.
This is consistent with the same picture as in the high metallicity
case, namely of photon losses driving a population cascade downwards,
albeit with less impact owing to the 
\ion{O}{I} lines being weaker.

This implies
a source function effect on the 
\ion{O}{I} $777\,\nm$~and $844\,\nm$~multiplets
in the low metallicity case.  The source function
drops below the Planck function so as to strengthen
the lines with respect to LTE. 
On the other hand, the more highly excited 
\ion{O}{I} $616\,\nm$~and $926\,\nm$~multiplets
form even deeper in metal-poor stellar atmospheres,
where conditions are closer to LTE.
These lines thus
tend to suffer only very mildly from non-LTE effects towards lower
metallicities.

There are no strong \ion{O}{I} lines in the mid-UV region
that connect the levels of low excitation potential 
to the levels of intermediate excitation potential.
Consequently there is apparently no low-metallicity 
overexcitation effect driven by photon pumping,
in contrast to \ion{C}{I} 
as discussed in \sect{results_departure_c1}.
Photon pumping through 
the \ion{O}{I} $130\,\nm$~line would be possible
\citep{2009A&amp;A...500.1221F},
if not for the large \ion{H}{I}~$\lyalpha$~opacity
in the atmospheres of these stars \citep{2015MNRAS.454L..11A}.

\subsection{Abundance corrections}
\label{results_abcor}

\begin{figure*}
    \begin{center}
        \includegraphics[scale=0.33]{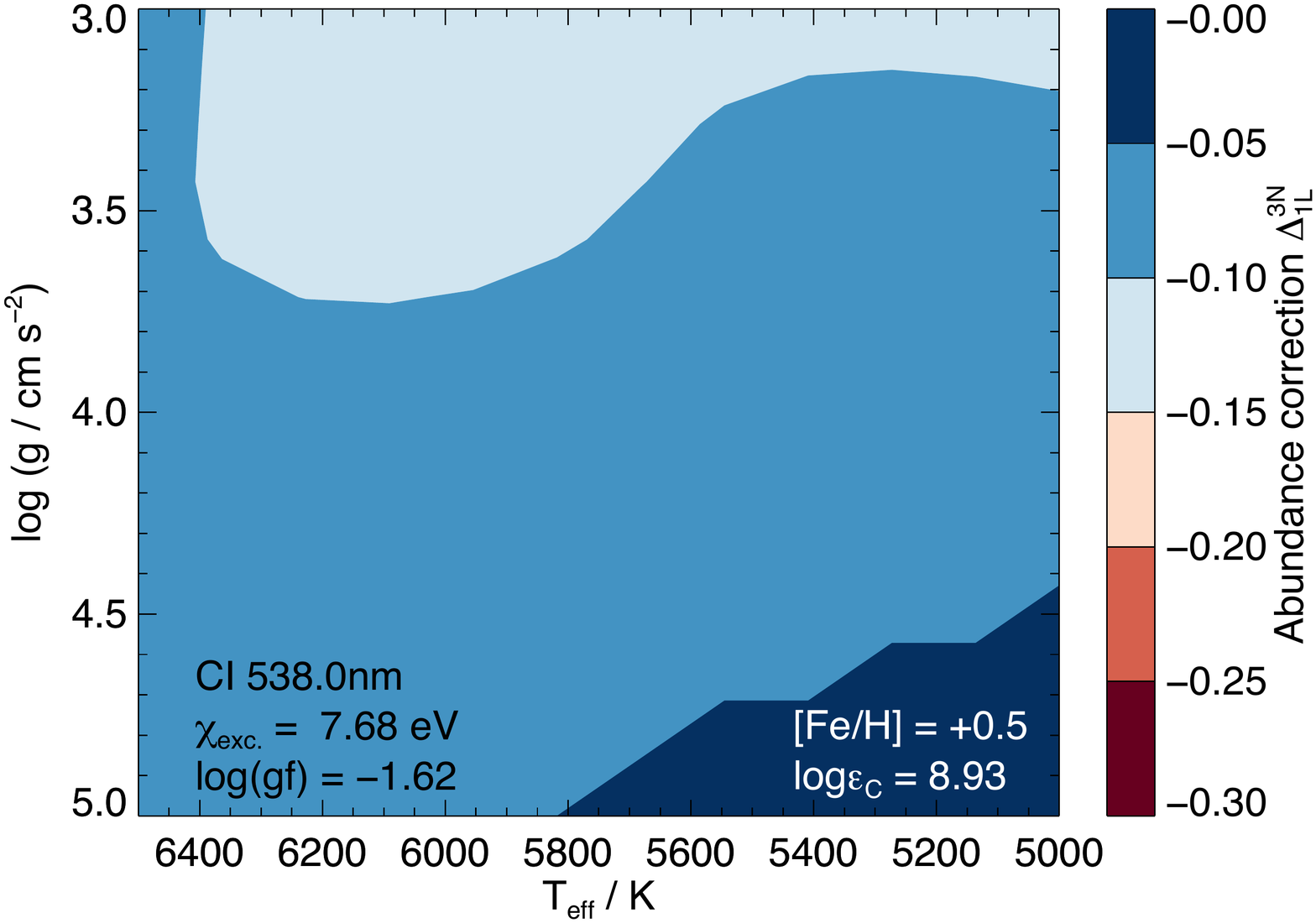}\includegraphics[scale=0.33]{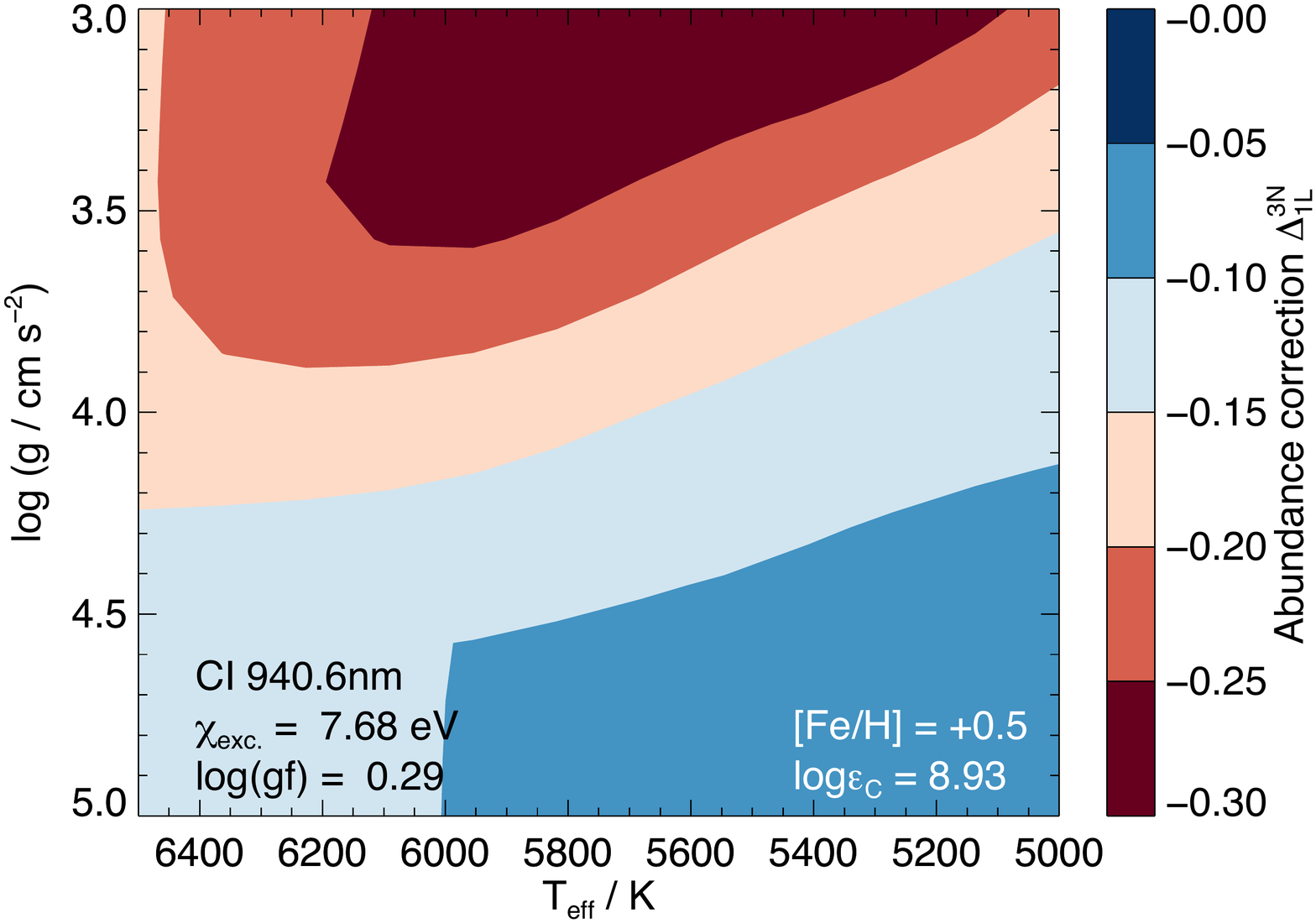}
        \includegraphics[scale=0.33]{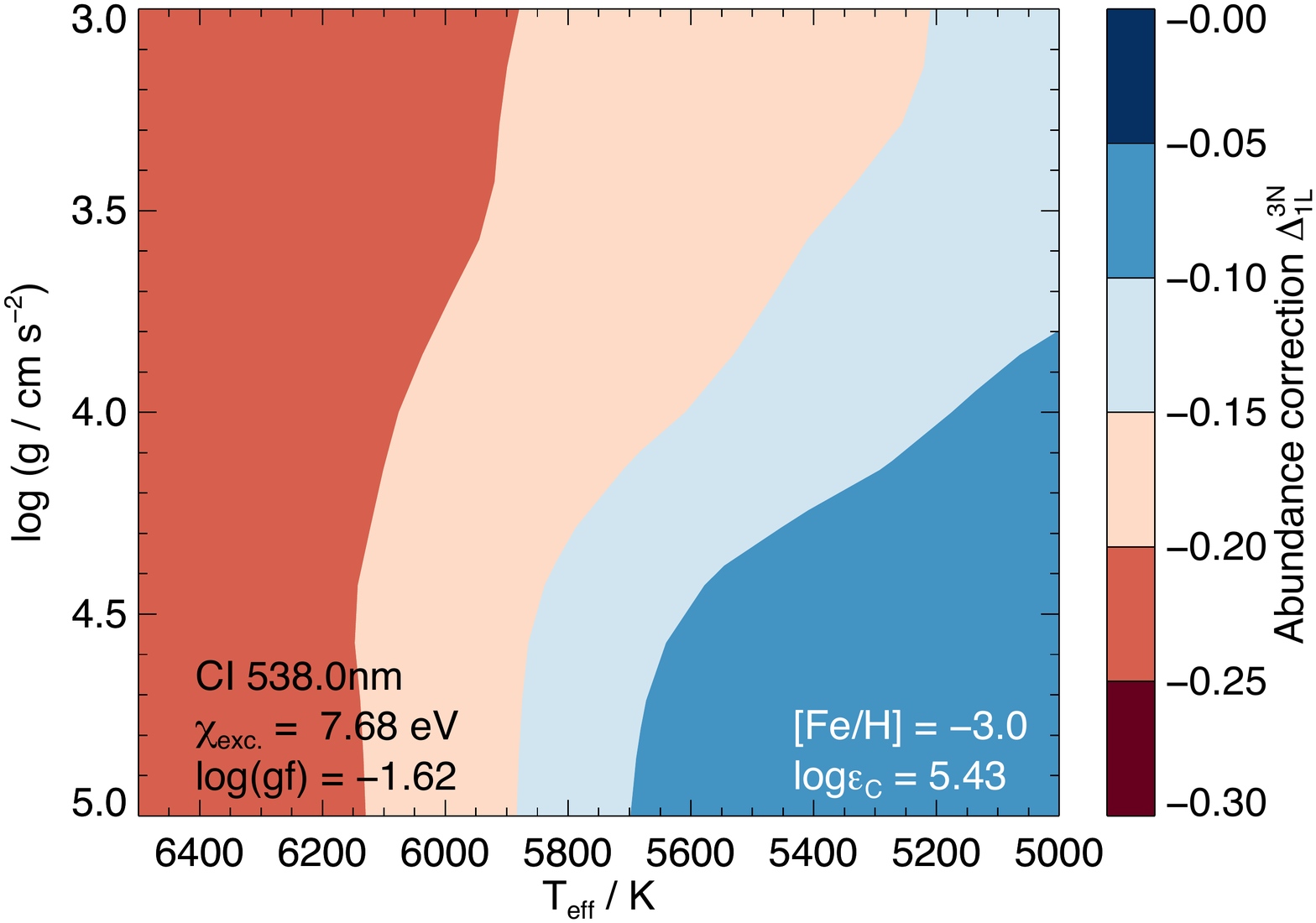}\includegraphics[scale=0.33]{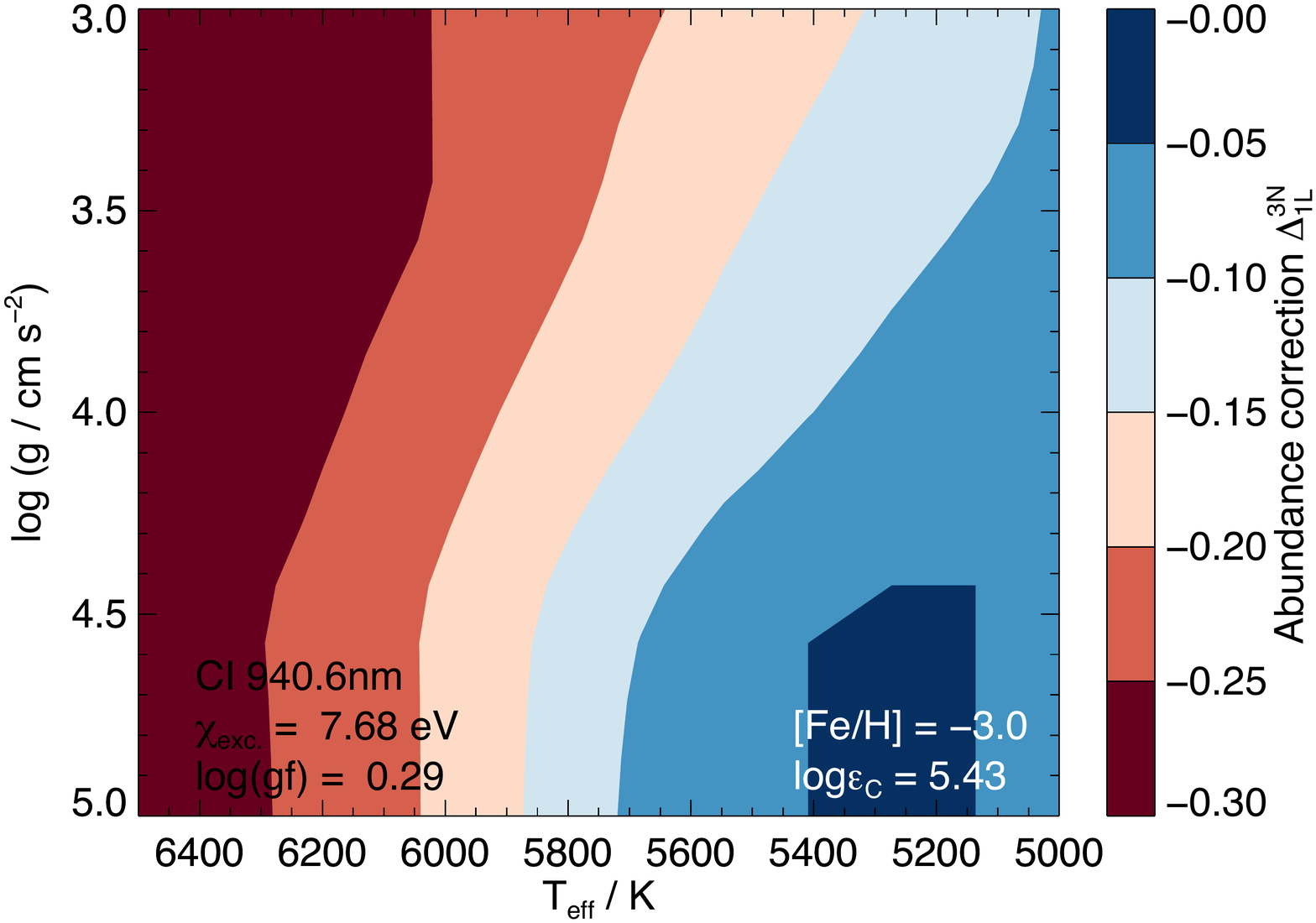}
        \caption{Kiel diagram of 3D non-LTE versus 1D LTE abundance
        corrections for different \ion{C}{I} lines (columns),
        at different metallicities and carbon abundances
        (rows).
        The 1D microturbulence was fixed to $\vmic=1.0\,\kms$.}
        \label{fig:abcor_c1}
    \end{center}
\end{figure*}

\begin{figure*}
    \begin{center}
        \includegraphics[scale=0.33]{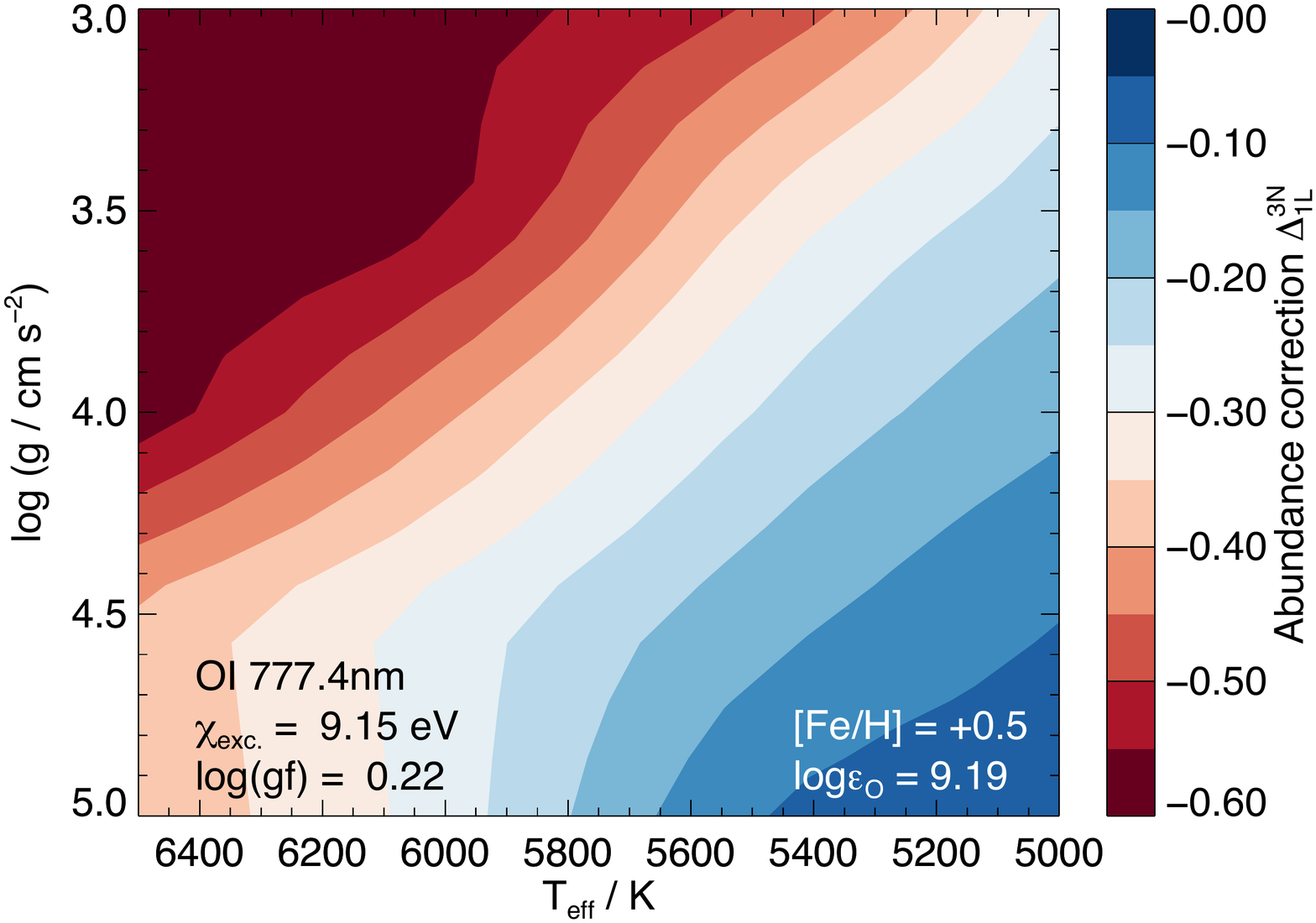}\includegraphics[scale=0.33]{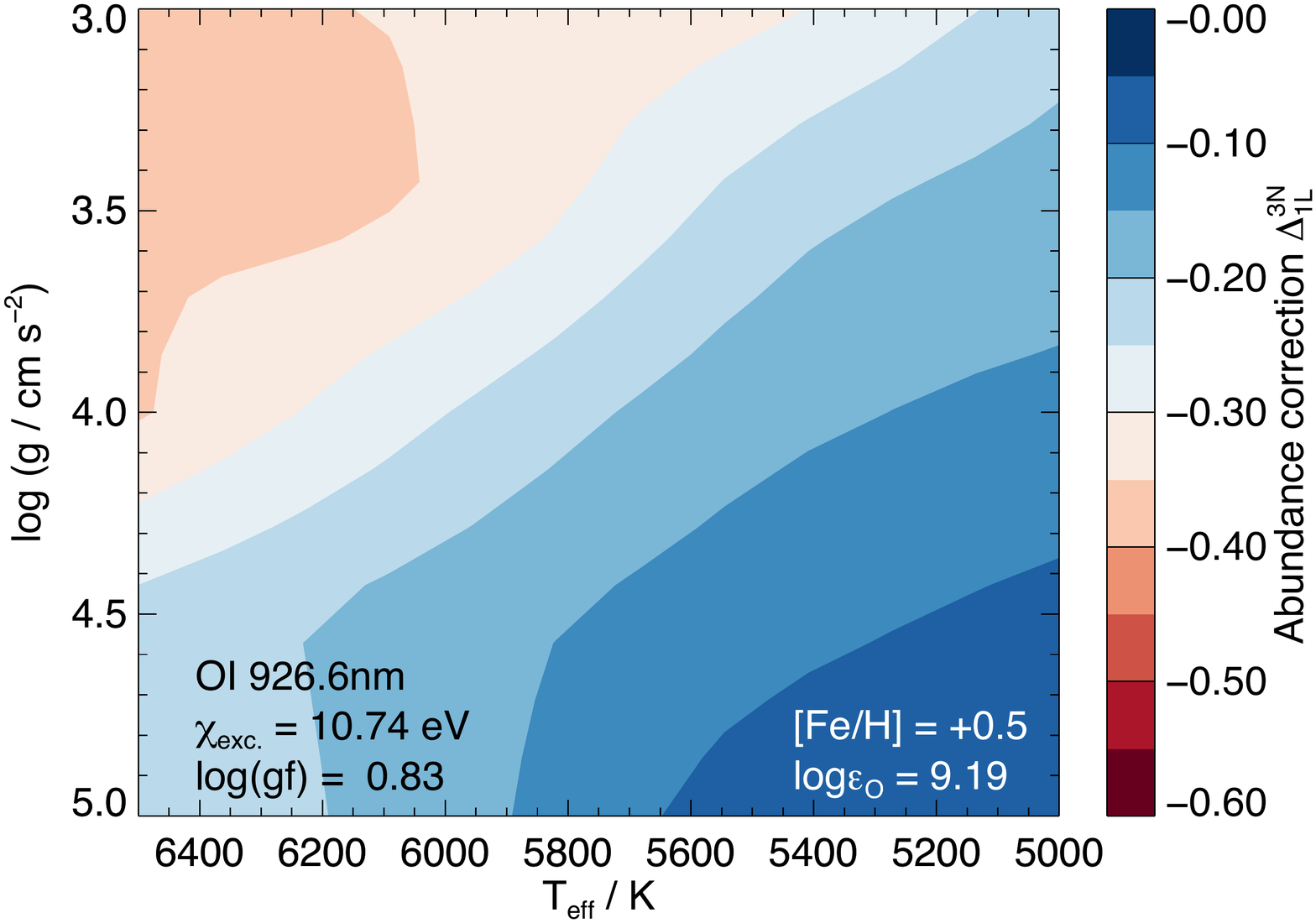}
        \includegraphics[scale=0.33]{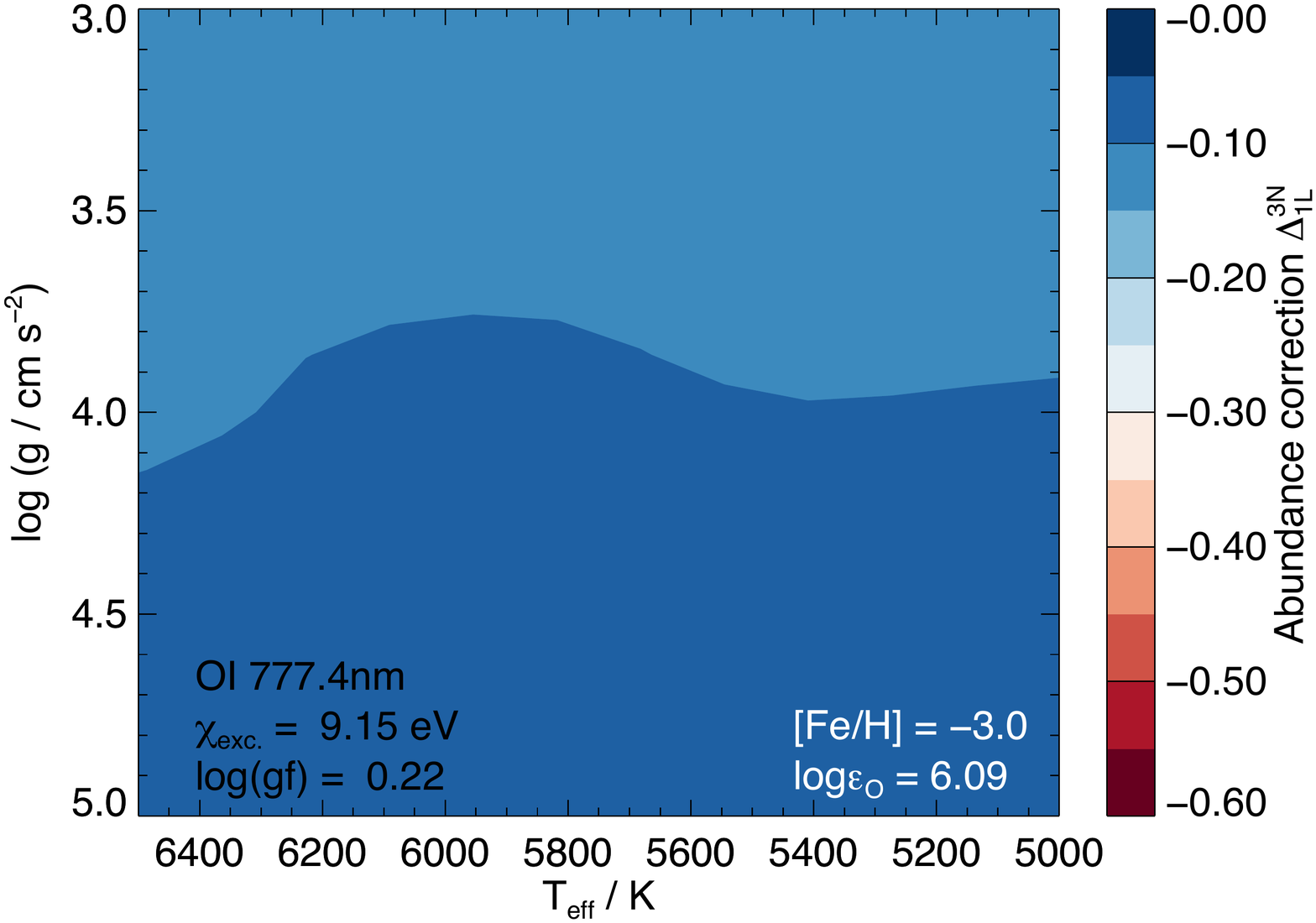}\includegraphics[scale=0.33]{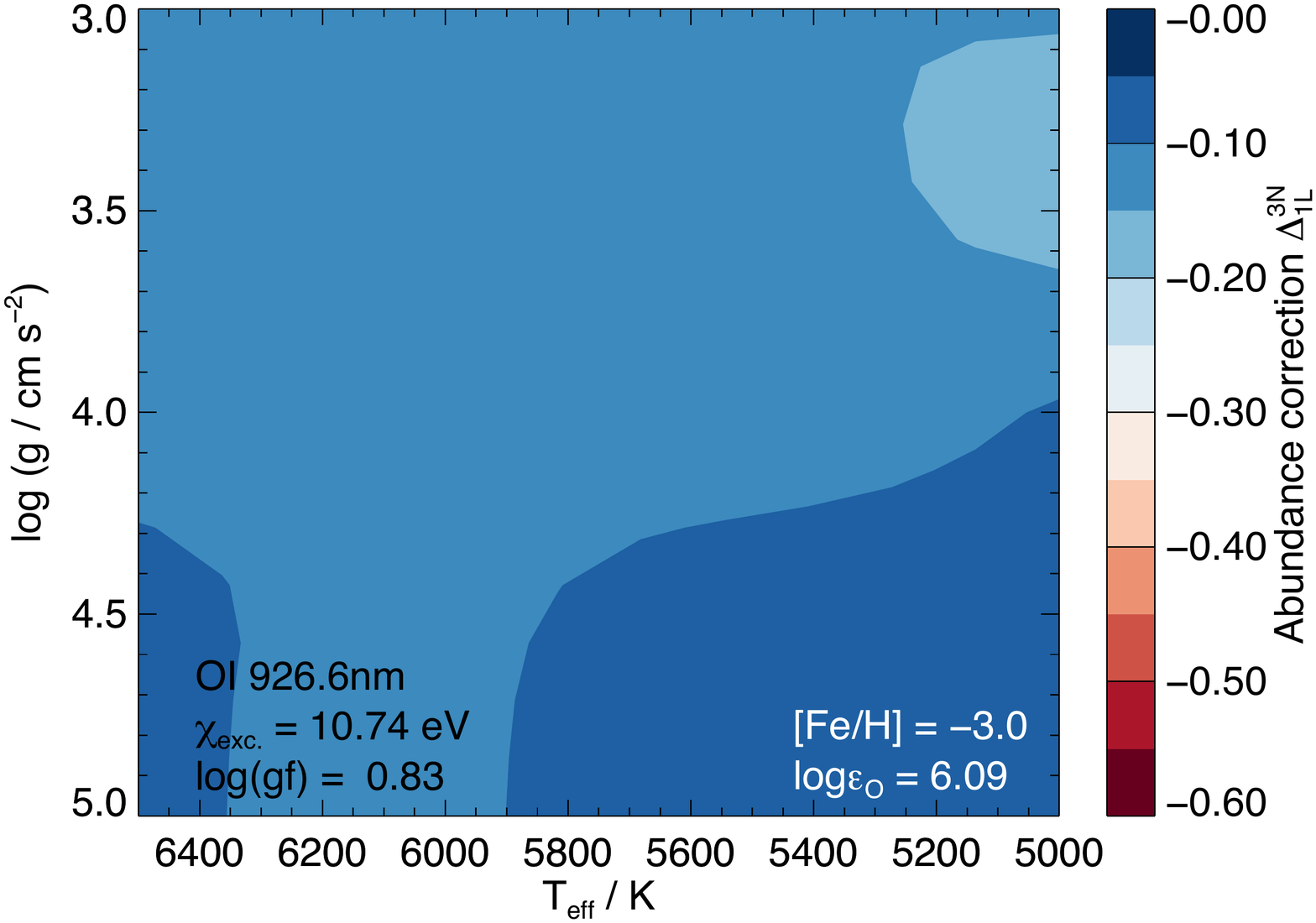}
        \caption{Kiel diagram of 3D non-LTE versus 1D LTE abundance
        corrections for different \ion{O}{I} lines (columns),
        at different metallicities and oxygen abundances
        (rows).
        The 1D microturbulence was fixed to $\vmic=1.0\,\kms$.}
        \label{fig:abcor_o1}
    \end{center}
\end{figure*}

\begin{figure*}
    \begin{center}
        \includegraphics[scale=0.33]{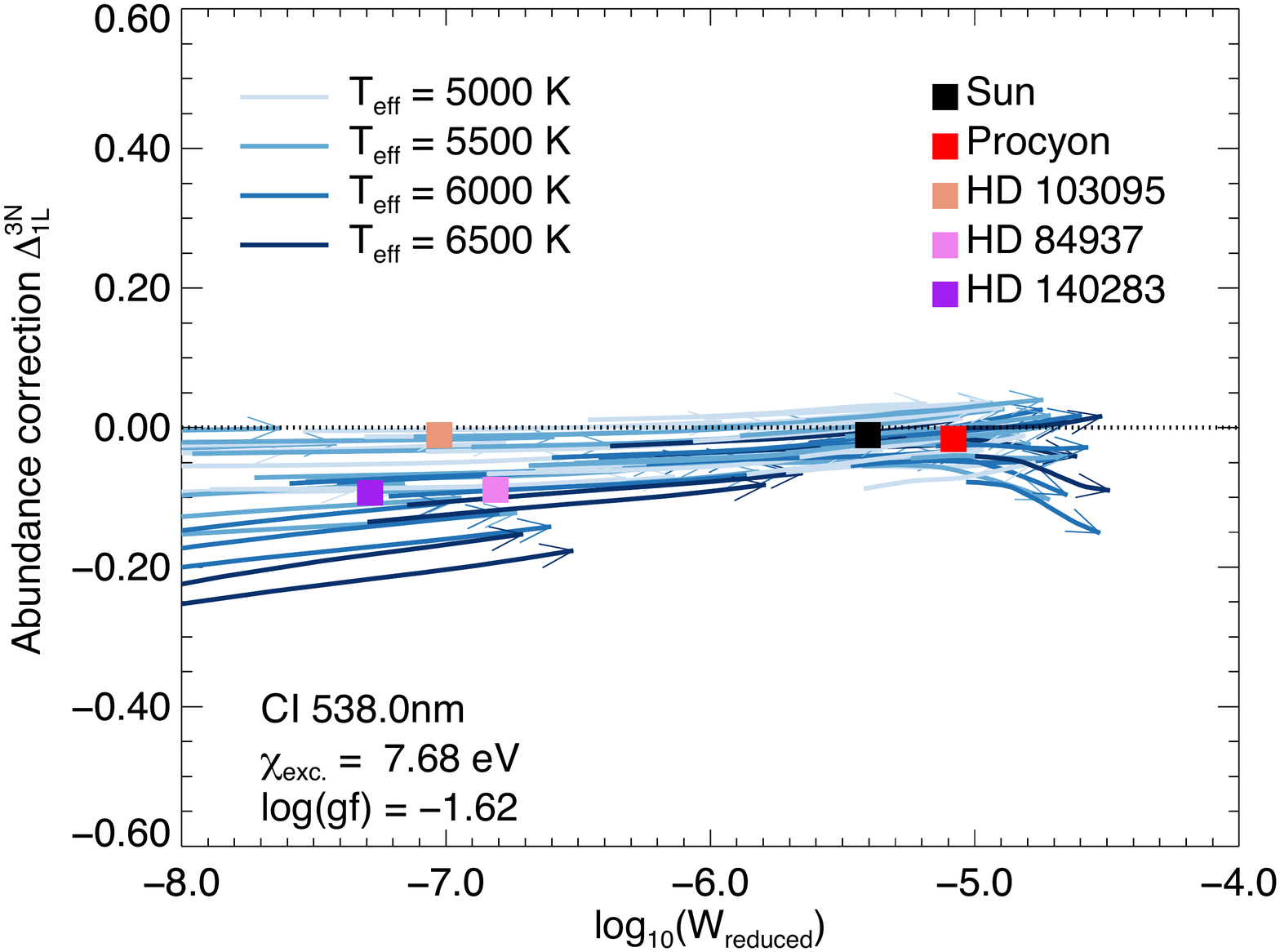}\includegraphics[scale=0.33]{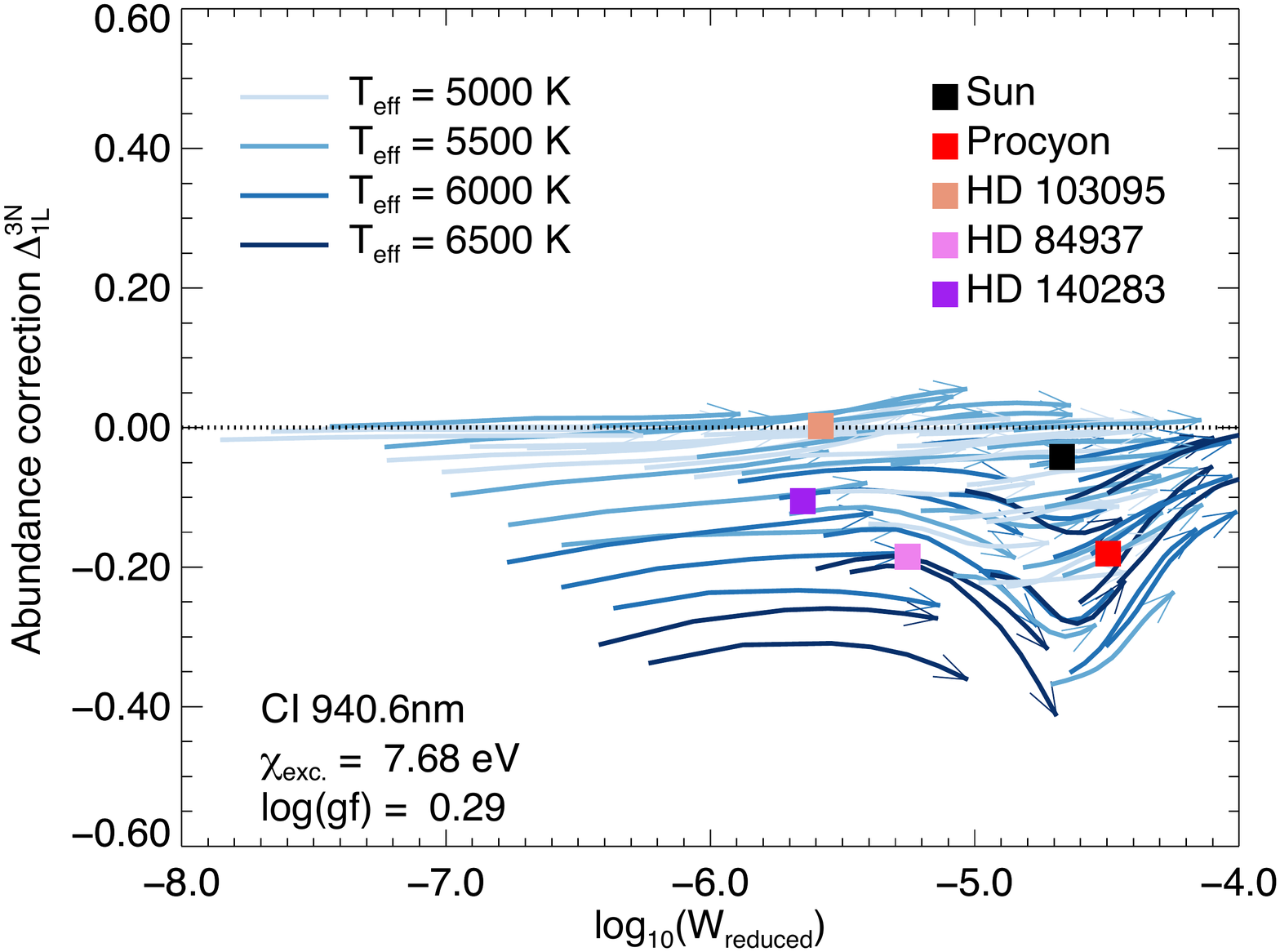}
        \includegraphics[scale=0.33]{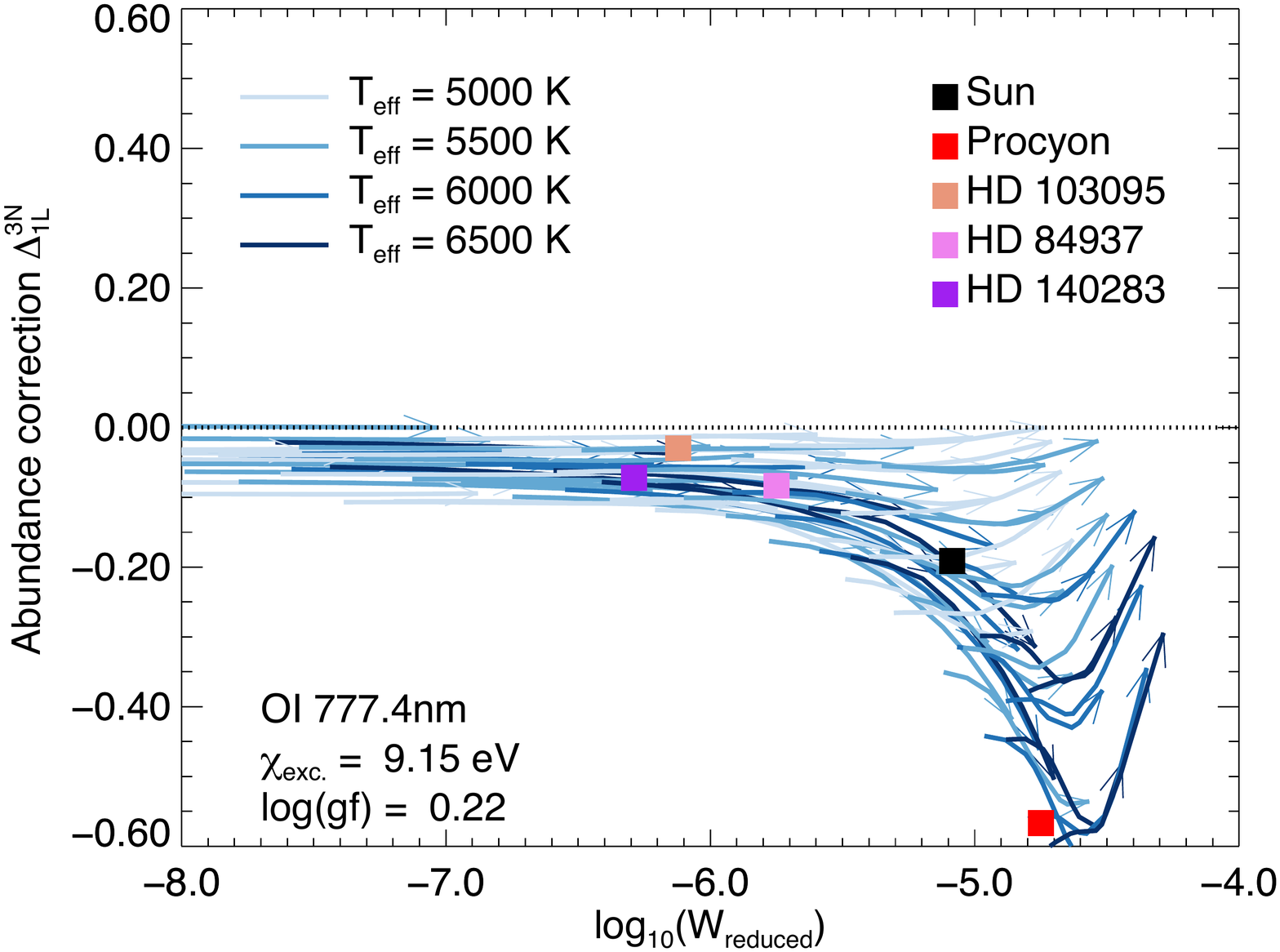}\includegraphics[scale=0.33]{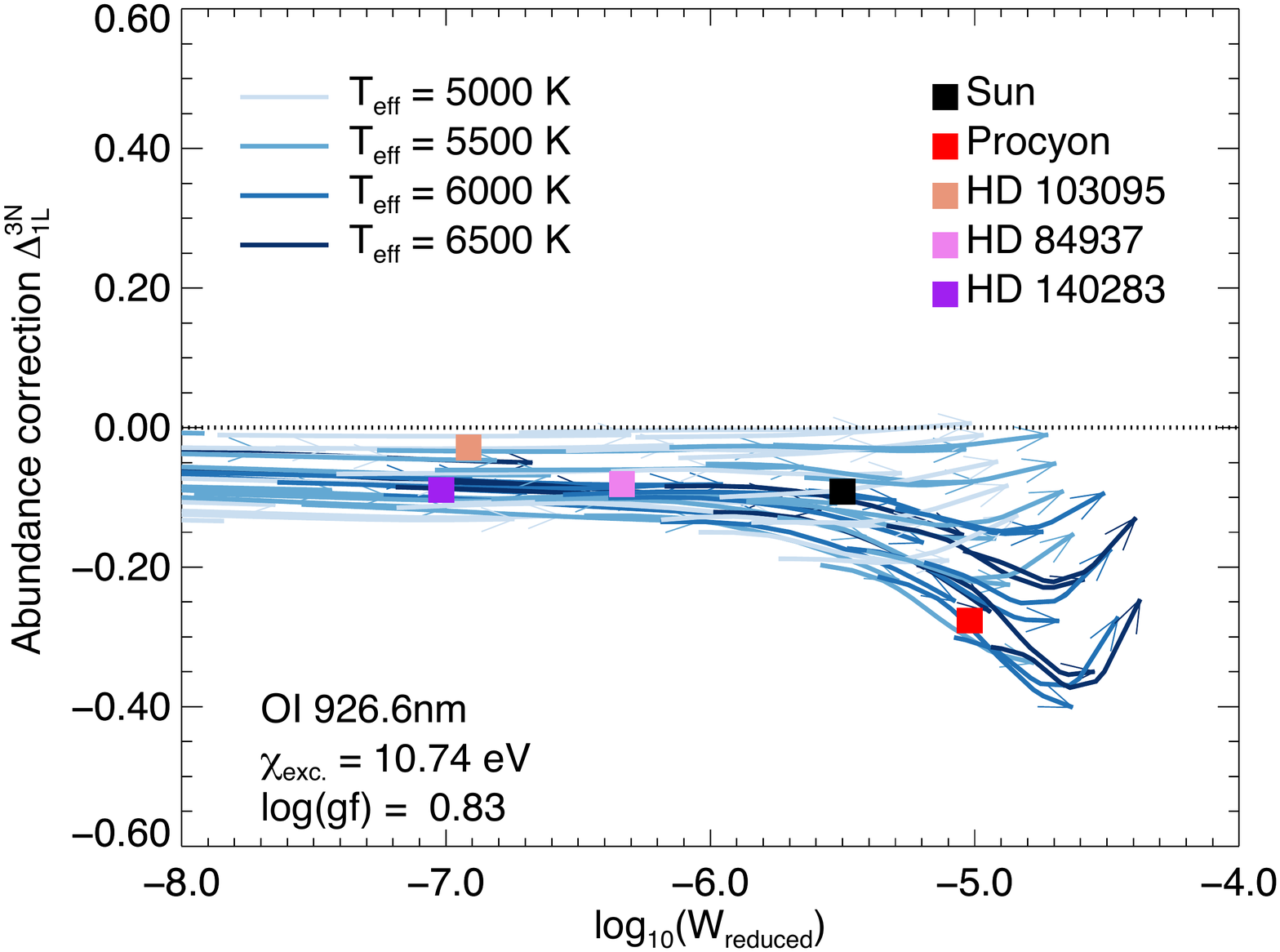}
        \caption{3D non-LTE versus 1D LTE abundance
        corrections as functions of
        3D non-LTE reduced equivalent widths,
        $W_{\text{reduced}}=W/\lambda$,
        for different
        \ion{C}{I} and \ion{O}{I} lines.
        The arrows show how the results change
        by increasing $\lgeps{C}$~and $\lgeps{O}$~respectively,
        and the curves are for
        fixed effective temperatures, surface gravities
        and metallicities, roughly corresponding to the
        nodes of the 3D \stagger~grid 
        in \fig{fig:kiel}.
        The 1D microturbulence was fixed to $\vmic=1.0\,\kms$.
        Also plotted are approximate abundance corrections
        for reference dwarfs, adopting the stellar parameters
        listed in Table 1 of \citet{2018A&A...615A.139A}
        and assuming 3D non-LTE abundance ratios
        of $\abrat{C}{Fe}=\abrat{O}{Fe}=0.0$~for the Sun 
        and Procyon, and of
        $\abrat{C}{Fe}=0.0$~and $\abrat{O}{Fe}=0.4$~for the metal-poor
        stars.
        The effective temperature of 
        Procyon ($\teff\approx6556\,\mathrm{K}$) lies slightly 
        outside of the grid of 3D models and the abundance corrections
        are extrapolated here.}
        \label{fig:abcor_strength_c1o1}
    \end{center}
\end{figure*}

\begin{figure*}
    \begin{center}
        \includegraphics[scale=0.33]{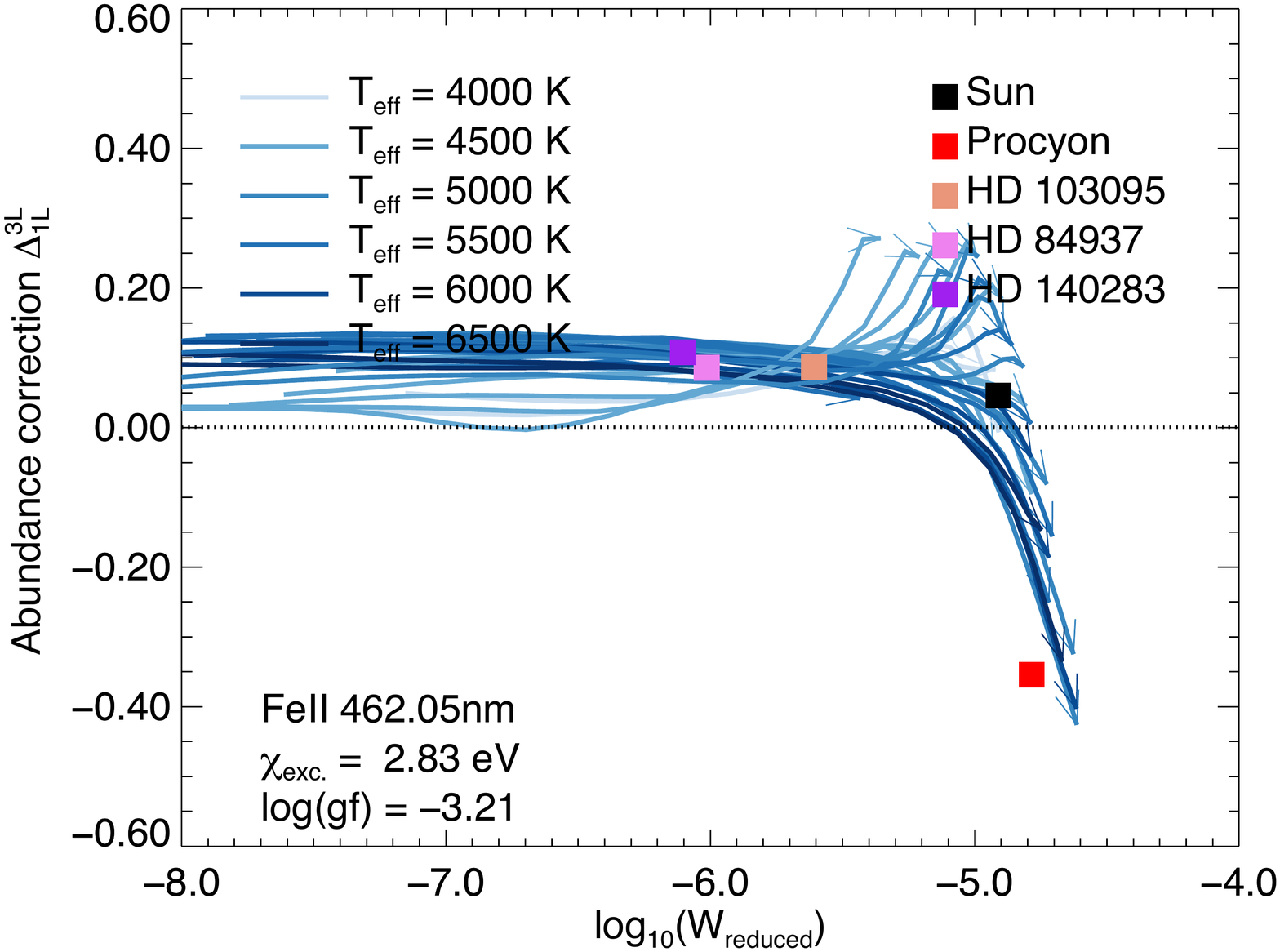}\includegraphics[scale=0.33]{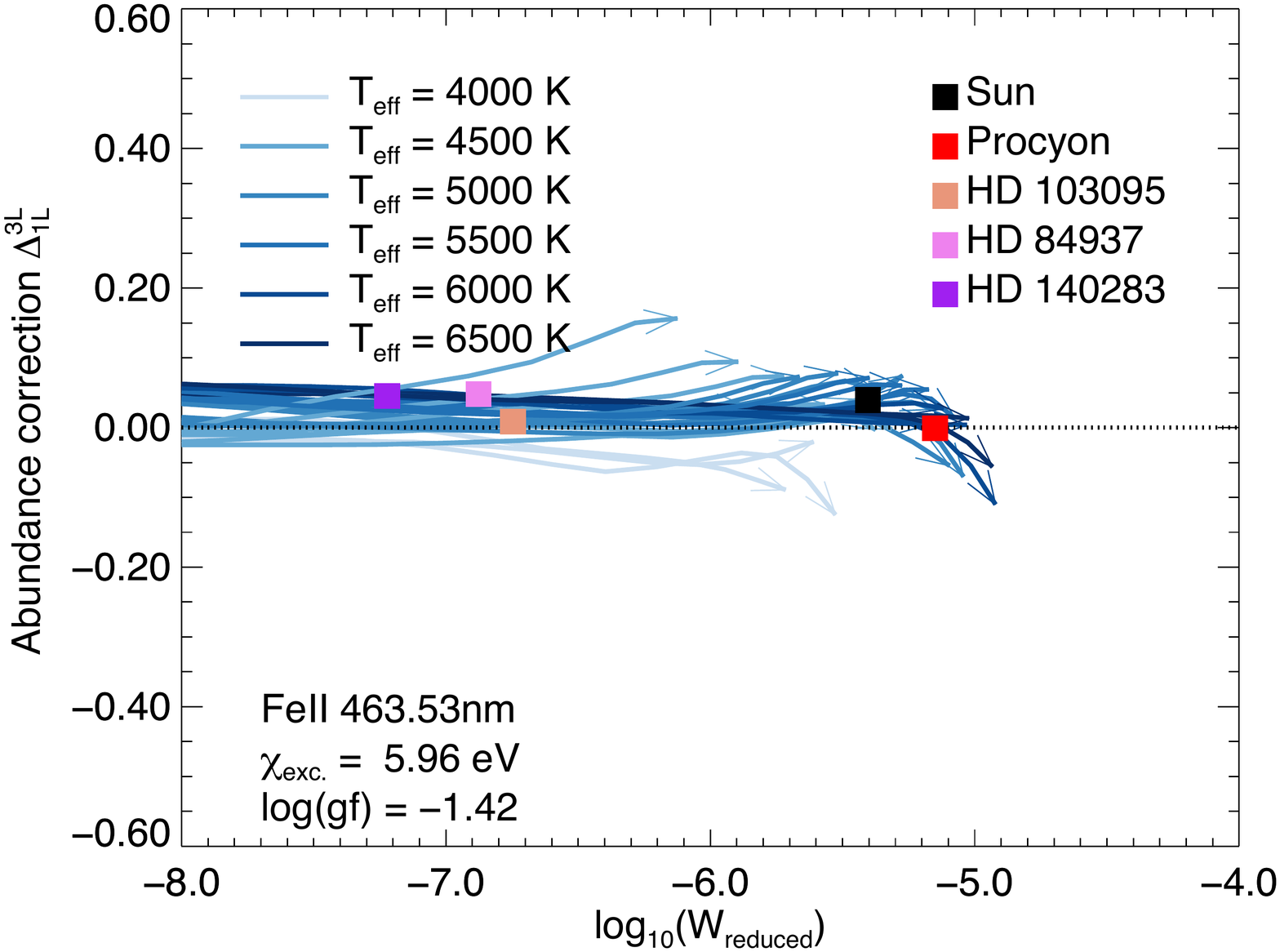}
        \includegraphics[scale=0.33]{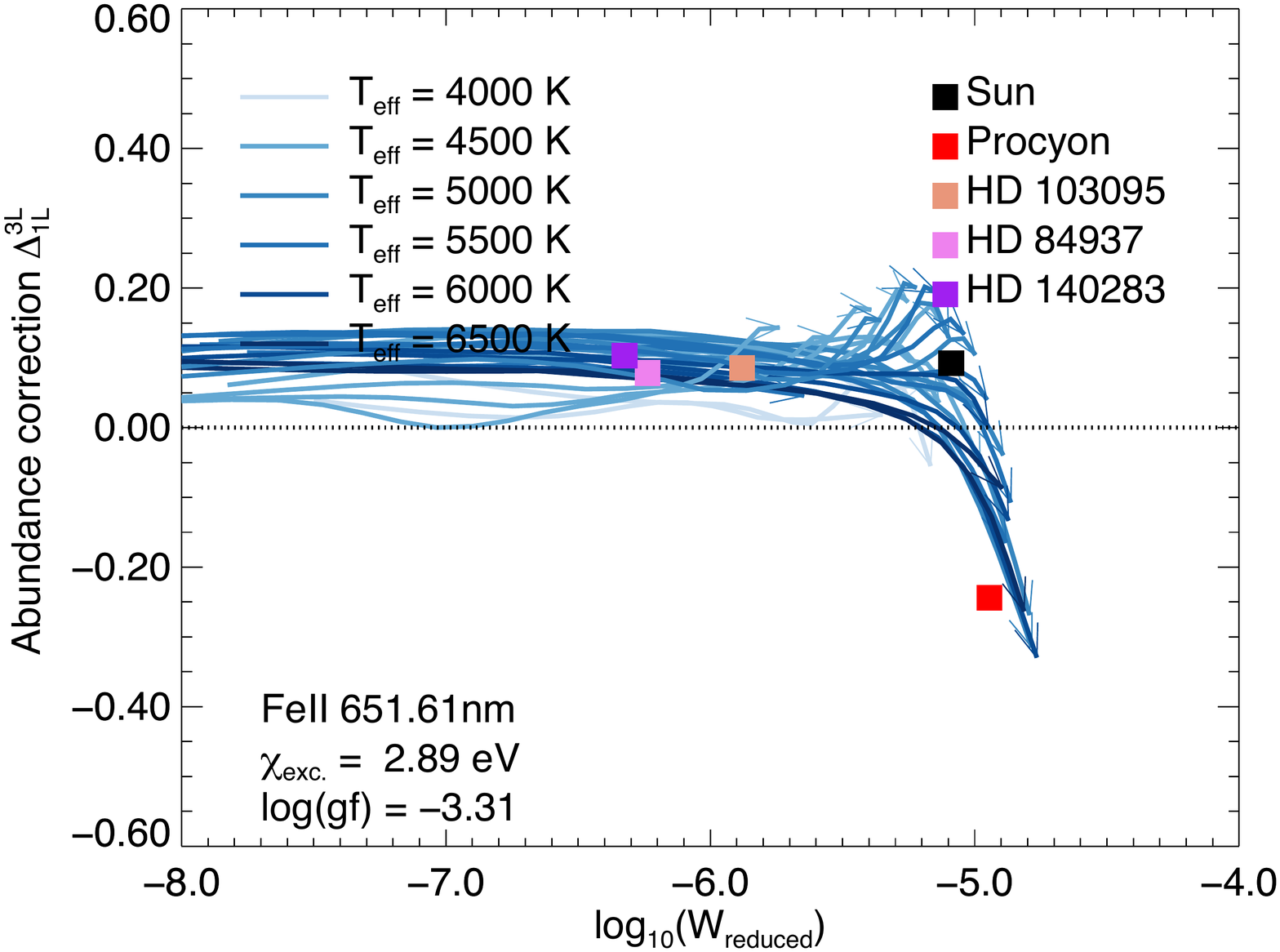}\includegraphics[scale=0.33]{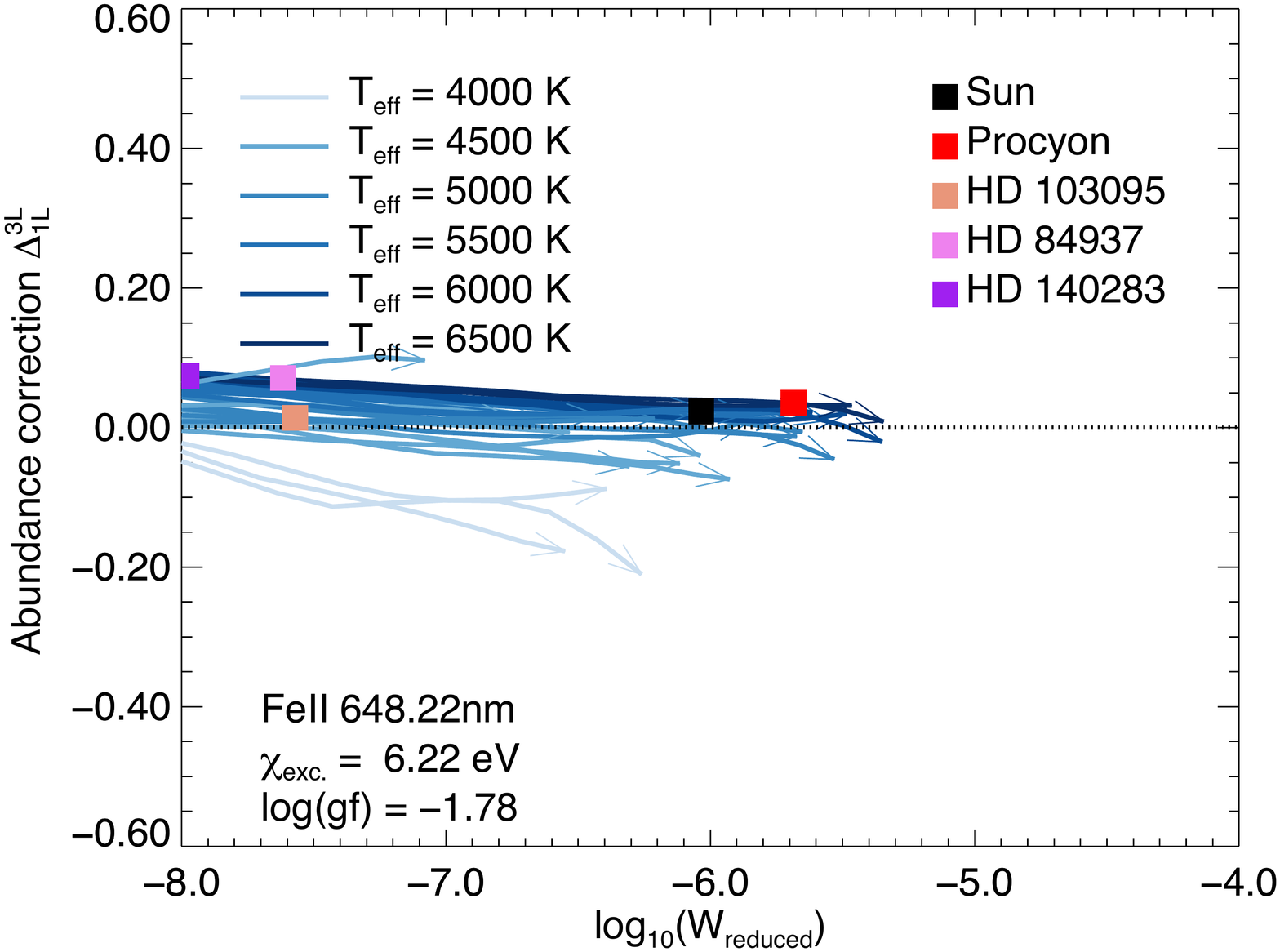}
        \caption{3D LTE versus 1D LTE abundance
        corrections as functions of
        3D LTE reduced equivalent widths, 
        $W_{\text{reduced}}=W/\lambda$,
        for different
        \ion{Fe}{II}~lines. 
        The arrows show how the results change
        by increasing the metallicity 
        $\feh$~(noting that the iron abundance is always kept
        consistent with the atmospheric chemical composition as labelled
        by $\feh$; \sect{method_rt}). Thus
        the curves are for
        fixed effective temperatures and surface gravities
        roughly corresponding to the
        nodes of the 3D \stagger~grid 
        in \fig{fig:kiel}.
        The 1D microturbulence was fixed to $\vmic=1.0\,\kms$.
        Also plotted are approximate abundance corrections
        for some reference dwarfs, adopting the stellar parameters
        listed in Table 1 of \citet{2018A&A...615A.139A}.
        The effective temperature of 
        Procyon ($\teff\approx6556\,\mathrm{K}$) lies slightly 
        outside of the grid of 3D models and the abundance corrections
        are extrapolated here.}
        \label{fig:abcor_strength_fe2}
    \end{center}
\end{figure*}

In the online Tables 2 and 3
we provide grids of 3D non-LTE versus 1D LTE 
abundance corrections $\corr{3N}{1L}$~based on the
\stagger~and \atmo~grids of model atmospheres
(\fig{fig:kiel})
for \ion{C}{I} and \ion{O}{I}~lines.
In the online Table 4, we provide
3D LTE versus 1D LTE 
abundance corrections $\corr{3L}{1L}$~for \ion{Fe}{II} lines.
In addition, in the online Tables 5 and 6
we provide grids of 1D non-LTE versus 1D LTE 
abundance corrections $\corr{1N}{1L}$~based on the
more extensive \marcs~grid of model atmospheres (\fig{fig:kiel})
for \ion{C}{I} and \ion{O}{I} lines.
These abundance corrections 
can be added directly to line-by-line 1D LTE inferred abundances,
to immediately improve their accuracy.
Tools for interpolating these data, as well as other data
(spectra, equivalent widths, and other abundance corrections)
may be acquired by contacting the authors directly.

To aid intuition, in Figs~\ref{fig:abcor_c1}---\ref{fig:abcor_strength_fe2}~we 
illustrate the 3D non-LTE abundance corrections
across stellar parameter space.
For brevity, we only show results for
representative lines, namely the
\ion{C}{I} $538.0\,\nm$~and $940.6\,\nm$~lines,
the \ion{O}{I} $777.4\,\nm$~and $926.6\,\nm$~lines,
the low-excitation
\ion{Fe}{II} $462.05\,\nm$ and $651.61\,\nm$~lines,
and the intermediate-excitation
\ion{Fe}{II} $463.53\,\nm$ and $648.22\,\nm$~lines.
The trends in the abundance corrections with
line strength and stellar parameters are at least qualitatively similar
for the other lines.

It is important to note that the data and plots are of
absolute abundance corrections
to inferred 1D LTE values of $\log\epsilon$.
In practice it is often the case that one works differentially
with respect to a reference star, usually the Sun.
The differential abundance correction is then
usually less severe, provided that the studied star and 
the reference star are not too far separated in stellar parameter
space.

\subsubsection{Forbidden \ion{C}{I} and \ion{O}{I} lines}
\label{results_abcor_forb}

The forbidden \ion{C}{I} and \ion{O}{I} lines do not
suffer from non-LTE effects, as discussed in
\sect{results_departure}.
They do however suffer from 3D effects, albeit only slightly.
These effects are typically of the order 
$0.05\,\dex$, as reported elsewhere: for 
the [\ion{O}{I}] $630.0\,\nm$~and $636.4\,\nm$~lines
see \citet{2016MNRAS.455.3735A},
and for the [\ion{C}{I}] $872.7\,\nm$~line
in the Sun see \citet{2019A&A...624A.111A}.

\subsubsection{Permitted \ion{C}{I} lines}
\label{results_abcor_c1}

In general, 1D LTE modelling of
permitted \ion{C}{I} lines
leads to overestimated carbon abundances,
across the entire parameter space under consideration here,
as shown in \fig{fig:abcor_c1}.
This is mainly driven by significant departures from LTE.
The nature of the non-LTE effects is such that
they always act to strengthen the permitted lines,
as discussed in \sect{results_departure}.

At high metallicities
\fig{fig:abcor_c1} shows that
the 3D non-LTE versus 1D LTE abundance 
corrections become more severe
towards higher effective temperatures and lower surface gravities.
This is because at high metallicities
the non-LTE effects are driven by photon losses 
in \ion{C}{I} lines of intermediate and high
excitation potential (\sect{results_departure}),
and these lines become stronger 
towards higher effective temperatures and lower surface gravities.

Towards lower metallicities,
\fig{fig:abcor_c1} shows that
the 3D non-LTE versus 1D LTE abundance corrections become more
severe, where the nature
of the non-LTE effect is different: namely, 
the effect is one of overexcitation
driven by photon pumping through \ion{C}{I} lines in the UV
(\sect{results_departure}).  This 
effect is dependent on the non-thermal UV radiation, which
increases towards higher effective temperatures.
Therefore at low metallicities
the abundance corrections become more
severe towards higher effective temperatures.

Consequently, for \ion{C}{I} lines
the most severe (absolute) 3D non-LTE versus
1D LTE abundance corrections are for
metal-poor stars of high effective temperature.
\fig{fig:abcor_c1} shows that for such stars,
the abundance corrections
can be in excess of $-0.3\,\dex$~for \ion{C}{I} lines
in the near infra-red.
However, the abundance corrections are somewhat less
severe for the \ion{C}{I} lines in the optical,
which form deeper in the atmosphere.
(see for example the contribution functions in
Fig.~1 of \citealt{2019A&A...624A.111A}).

In \fig{fig:abcor_strength_c1o1} the 3D non-LTE versus 1D LTE abundance
corrections are plotted against (reduced) equivalent widths,
for fixed stellar parameters ($\teff$, $\lgg$, $\feh$)
but varying carbon abundance.
At lower line strength, corresponding to lower metallicity, 
the trend of more severe abundance corrections
towards increasing effective temperature, as discussed above,
is immediately apparent. At higher metallicities, 
the abundance corrections increase with increasing
line strength, and there is some indication that
they turn over once the line is saturated. 
This signature is even clearer for 
\ion{O}{I}, and we discuss it further in
\sect{results_abcor_o1}.

The absolute 3D non-LTE versus
1D LTE abundance corrections can be
large at low metallicities. This means that typically,
the differential abundance
corrections with respect to the Sun are also quite large.
For the \ion{C}{I} $940.6\,\nm$~line the abundance
correction is around $-0.05\,\dex$~for the Sun
(\fig{fig:abcor_strength_c1o1}), which means that
if the absolute abundance correction for a given metal-poor star is
$-0.3\,\dex$, the differential abundance correction
is still as severe as $-0.25\,\dex$.

In light of the severity of the non-LTE effects, and
in the absence of full 3D non-LTE modelling,
1D non-LTE modelling should be used
for permitted \ion{C}{I} lines.
However,
it should be kept in mind that
when the non-LTE effects are strong they tend to 
be enhanced by the steeper temperature gradients
present in the 3D model atmospheres. This has previously been
discussed in the context of overionisation of \ion{Fe}{I}
in \citet{2016MNRAS.463.1518A} and
\citet{2017A&amp;A...597A...6N},
and is 
apparent in \citet{2019A&A...622L...4A}, where 1D non-LTE 
modelling still overestimates carbon abundances by
around $0.05$~to $0.10\,\dex$~at low metallicities.

\subsubsection{Permitted \ion{O}{I} lines}
\label{results_abcor_o1}

In general, 1D LTE modelling of
the permitted \ion{O}{I} lines
leads to overestimated oxygen abundances,
across the entire parameter space under consideration here,
as shown in \fig{fig:abcor_o1}.
As with \ion{C}{I} (\sect{results_abcor_c1}),
this is mainly driven by significant departures from LTE.
The nature of the non-LTE effects is such that
they always act to strengthen the permitted lines,
as discussed in \sect{results_departure}.

\fig{fig:abcor_o1}~shows that at all metallicities,
the 3D non-LTE versus 1D LTE abundance corrections are
more severe towards
higher effective temperatures and lower surface gravities.
This is because the non-LTE effects are driven by photon losses
in \ion{O}{I} lines of intermediate and high
excitation potential (\sect{results_departure}),
and these lines become stronger
towards higher effective temperatures and lower surface gravities.
Unlike for \ion{C}{I}, there is no 
low-metallicity overexcitation effect driven by photon pumping; 
thus
\fig{fig:abcor_o1} shows that
the abundance corrections
become less severe towards lower metallicities.

Consequently, for \ion{O}{I} lines the most severe 
(absolute) 3D non-LTE versus 1D LTE abundance corrections
are for metal-rich stars of high effective temperature and 
low surface gravity.
\fig{fig:abcor_o1} illustrates that
for such stars, the abundance corrections 
can be in excess of $-0.6\,\dex$~for 
the \ion{O}{I} $777\,\nm$~multiplet;
thus, comparing with \sect{results_abcor_c1},
the abundance corrections are more
severe in \ion{O}{I} than in \ion{C}{I} in the worst cases.
The abundance corrections are typically less severe for the
\ion{O}{I} $844\,\nm$,
$926\,\nm$, and $616\,\nm$~multiplets,
in order of decreasing severity;
the latter line is highly excited and relatively weak,
and forms very deep in the atmosphere where the departure coefficients
are much closer to unity
(\sect{results_departure}).

In \fig{fig:abcor_strength_c1o1} the 3D non-LTE versus 1D LTE abundance
corrections are plotted against (reduced) equivalent widths,
for fixed stellar parameters ($\teff$, $\lgg$, $\feh$)
but varying oxygen abundance. At higher metallicities,
the abundance corrections increase with line strength,
and turn over for reduced equivalent widths below
around $-4.8\,\dex$.
This phenomenon has previously been explained,
for example in 
Sect.~3.1 of \citet{2011A&amp;A...528A.103L} in the context 
of \ion{Na}{I} lines.
The abundance corrections rapidly grow when the 
stronger (3D) non-LTE lines
enter the damping part of the curve-of-growth and develop broad wings.
When this happens, efficient photon losses in the 
\ion{O}{I} $777\,\nm$~multiplet actively drive the non-LTE effects
in the system. The minimum corresponds to saturation of
the \ion{O}{I} $777\,\nm$~multiplet. A similar signature
can be seen for the \ion{O}{I} $926.6\,\nm$~line,
as well as for the \ion{C}{I} $940.6\,\nm$~line as pointed out
in \sect{results_abcor_c1}, however for \ion{C}{I}
it is less clear owing to a large number of \ion{C}{I} lines
influencing the statistical equilibrium.

Although the absolute 3D non-LTE versus
1D LTE abundance corrections can be
very large at high metallicities, the differential abundance
corrections with respect to the Sun can be more moderate.
For the \ion{O}{I} $777.4\,\nm$~line the abundance
correction is around $-0.2\,\dex$~for the Sun
(\fig{fig:abcor_strength_c1o1}), which means that
if the absolute abundance correction for a given metal-rich star is
$-0.4\,\dex$, the differential abundance correction is only $-0.2\,\dex$.  
Similarly, if the absolute abundance correction for a given metal-poor star is
close to zero, the differential abundance correction
becomes $+0.2\,\dex$: oxygen abundances in the metal-poor regime
are susceptible to 3D non-LTE effects via the solar reference abundance.

In the absence of full 3D non-LTE modelling,
1D non-LTE modelling should be used
for permitted \ion{O}{I} lines.
As discussed for \ion{C}{I} (\sect{results_abcor_c1}),
when the non-LTE effects are strong, they tend to 
be enhanced by the 
3D effects. For the \ion{O}{I} $777\,\nm$~multiplet,
in the metal-rich regime,
1D non-LTE modelling can systematically overestimate
oxygen abundances by of the order $0.1\,\dex$~\citep{2016MNRAS.455.3735A}.

\subsubsection{\ion{Fe}{II} lines}
\label{results_abcor_fe2}

By assumption, the \ion{Fe}{II} lines 
do not suffer from significant non-LTE effects
(see \sect{introduction}).
\fig{fig:abcor_strength_fe2} illustrates that the
3D LTE versus 1D LTE abundance corrections are 
positive for \ion{Fe}{II} lines,  at least 
while the lines are unsaturated
(see \sect{results_abcor_o1}).
This means that 1D LTE modelling of \ion{Fe}{II} lines
results in underestimated iron abundances.
The 3D effects are caused by both differences in the mean 
atmospheric structure, the presence
of atmospheric inhomogeneities, and stellar granulation
\citep{2016MNRAS.463.1518A}.

The (absolute) 3D LTE versus 1D LTE abundance corrections
for \ion{Fe}{II}~lines, at least while they are unsaturated,
are typically in the range $-0.05$~to $+0.10\,\dex$~for
lines of intermediate excitation potential,
and $0.00$~to~$+0.15\,\dex$~for
lines of low excitation potential.
They are evidently 
more severe for lines of low excitation potential
that form higher up in the atmosphere
and are sensitive to differences
in the mean temperature stratification 
and the atmospheric inhomogeneities present in
the upper layers.  In contrast 
the lines of intermediate excitation potential
form deeper and are biased towards 
the hot temperature upflows associated with stellar granulation.

In the absence of 
spectrum synthesis based on 3D model atmospheres,
spectrum synthesis based on
\mtd~model atmospheres should be preferred 
over 1D model atmospheres, when modelling
low-excitation \ion{Fe}{II} lines,
as well as when modelling lines that form higher up in the 
atmosphere in general (including the low-excitation
forbidden \ion{C}{I} and \ion{O}{I} lines):
the 3D LTE versus \mtd~LTE abundance corrections
are closer to zero for such lines.
For lines of 
higher excitation potentials
($\epot\gtrsim4\,\eV$),
in the absence of 
spectrum synthesis based on 3D model atmospheres,
spectrum synthesis based on
1D model atmospheres may be more appropriate.
This can be understood by considering
\fig{fig:temp}: in the deep atmosphere
the 1D model atmospheres have a steeper
temperature gradient than the \mtd~model atmospheres,
and thus
the 1D models more closely follow the hot upflows in the 
3D model atmosphere.

The 3D LTE versus 1D LTE
abundance corrections can become more severe towards larger line
strengths as the line core becomes saturated. 
This can be seen for the 
\ion{Fe}{II} $462.05\,\nm$~and
$651.61\,\nm$~lines of low excitation potential,
in \fig{fig:abcor_strength_fe2}.
It is possible that this reflects the failure of the LTE assumption, 
and that, at the highest metallicities, strong \ion{Fe}{II} lines
of low excitation potential are susceptible to photon losses.
For these reasons, at the highest metallicities
it is better to avoid using strong \ion{Fe}{II} lines
of low excitation potential in spectroscopic analyses.

The absolute 3D LTE versus
1D LTE abundance corrections discussed above
are slightly more severe than the 
differential abundance corrections
relative to the Sun.  The latter
are typically only around $\pm0.05\,\dex$.
This can be seen by comparing the location of the Sun
with other reference stars in \fig{fig:abcor_strength_fe2}.

\section{Re-analysis of literature data}
\label{obs}

\begin{figure}
    \begin{center}
        \includegraphics[scale=0.33]{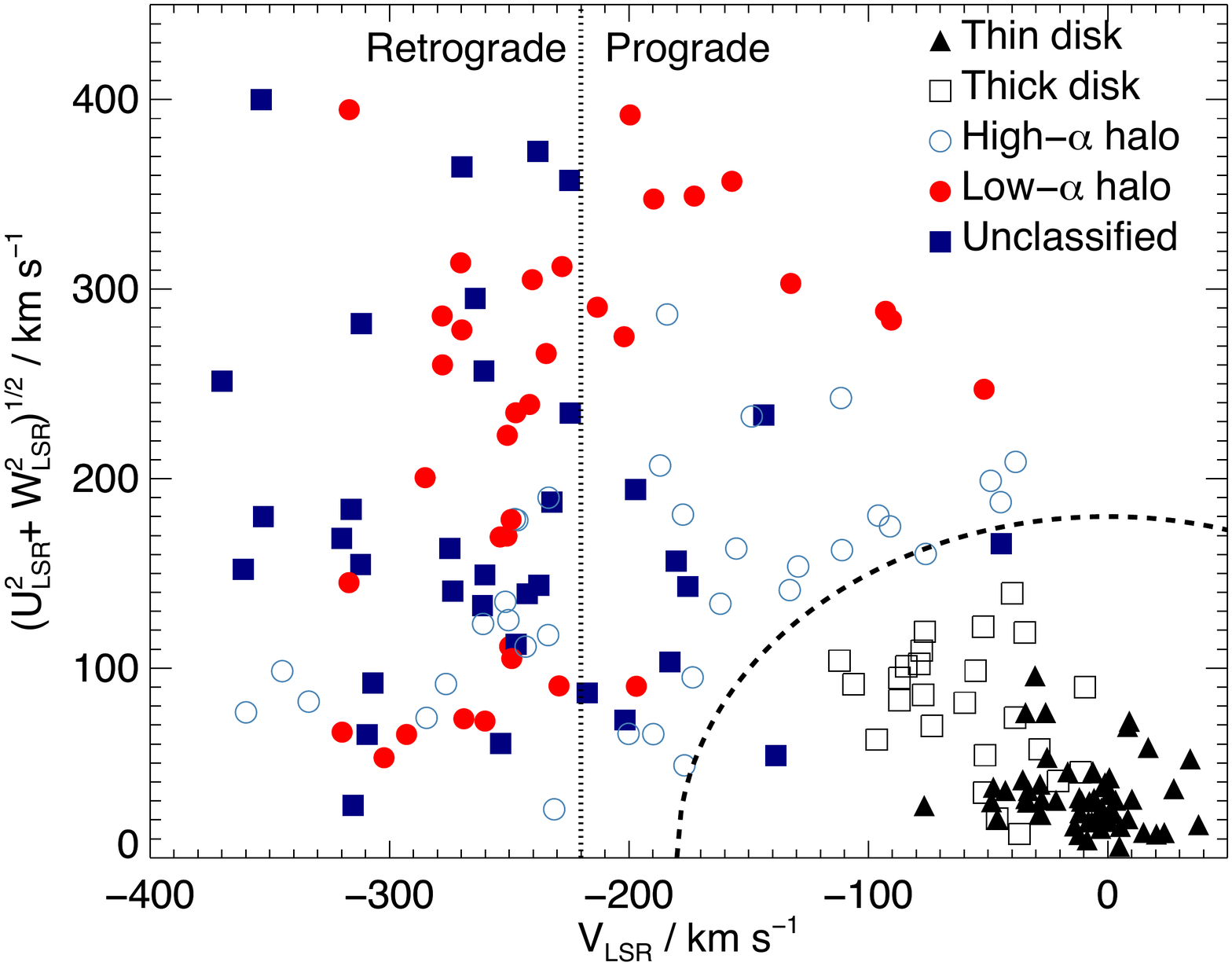}
        \caption{Toomre diagram for the entire stellar sample,
        corrected for the peculiar solar motion using
        \citet{2011MNRAS.412.1237C}.
        The kinematic information are primarily from Gaia DR2
        \citep{2018A&A...616A...1G}, but for some stars
        where these data were missing
        the kinematic information come from other sources
        \citep{1997A&A...323L..49P,
        2010A&amp;A...511L..10N,2012AstL...38..331A,
        2014ApJ...788..180C,2017AJ....153...75K}.
        The unclassified stars are from the VLT/UVES sample.
        The vertical dashed line delineates prograde and retrograde orbits,
        while the curved dashed line indicates a total velocity
        of $180\,\kms$.}
        \label{fig:toomre}
    \end{center}
\end{figure}

\subsection{Stellar sample} 
\label{obs_sample}

The sample consists of three different data sets
of F~and G~dwarfs:
the $67$~disk stars (mainly of the thin disk, and
including the Sun) in the HARPS-FEROS sample of
\citet{2014A&amp;A...568A..25N};
the $85$~thick-disk and halo stars in the UVES-FIES sample of 
\citet{2014A&amp;A...568A..25N};
and the $40$~halo stars in the VLT/UVES sample of
\citet{2007A&amp;A...469..319N} which were recently 
re-analysed by \citet{2019A&A...622L...4A},
including the carbon-poor blue straggler G~66-30.
There are five stars in common between the VLT/UVES
and the UVES-FIES samples:
in the subsequent analysis, the results of these common stars
are presented as an error-weighted average.
Consequently, the sample consists
of $187$~unique stars in total, including the Sun.
Full details about the observations,
in particular concerning the spectral resolutions and signal-to-noise
ratios of the different data, can be found in the references above.

For the HARPS-FEROS and 
UVES-FIES samples, stars were assigned to the same stellar 
populations as in \citet{2014A&amp;A...568A..25N}.
To summarise that work, stars were identified as belonging to the 
disk or halo based on whether
their total velocities with respect to the 
local standard of rest (LSR) were less than
or greater than $180\,\kms$~(the usual 
discriminant between the disk and halo populations;
e.g.~\citealt{2019A&A...624A..19B}).
The disk stars were categorised as thin- or thick-disk stars
based on the 1D LTE $\upalpha$-abundance plot
in Fig.~1 of \citet{2013A&amp;A...554A..44A},
while the 1D LTE abundances of 
magnesium, silicon, calcium, and titanium were measured
and used to separate the halo stars
into low- and high-$\upalpha$~halo populations
\citep{2010A&amp;A...511L..10N}.

We do not attempt to assign the VLT/UVES sample stars 
to specific stellar populations, with the exception
of the five stars in common with the 
UVES-FIES sample. Nevertheless, \fig{fig:toomre} 
shows that the majority of stars in this sample (labelled as unclassified)
have a similar distribution in the Toomre diagram
as those in the low-$\upalpha$~halo population, 
namely with a tendency towards slightly retrograde orbits and
a large range in $\sqrt{\mathrm{U_{\mathrm{LSR}}^{2}+W_{\mathrm{LSR}}^{2}}}$.
The majority of these stars may thus belong 
to the low-metallicity tail of the
low-$\upalpha$~halo population; future work combining
this kinematic information with information
from elemental abundance ratios
(see for example Figs~3, 4, and 5 of \citealt{2018ApJ...852...49H}), 
will be needed to confirm this.

\subsection{Stellar parameters}
\label{obs_stellarparams}

\begin{figure}
    \begin{center}
        \includegraphics[scale=0.33]{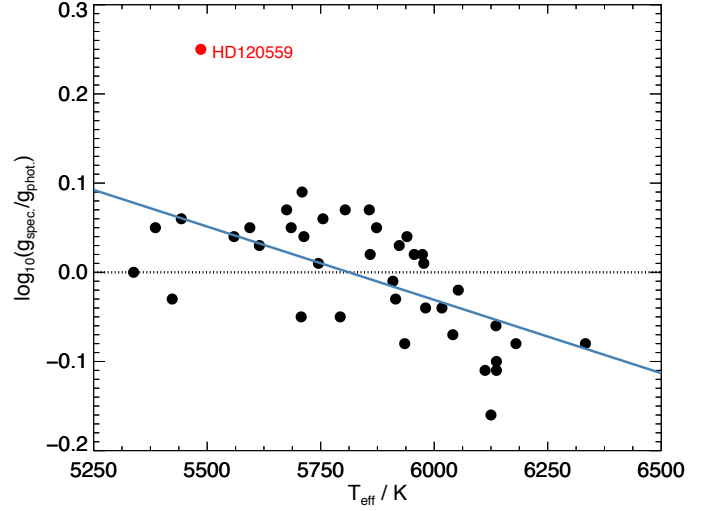}
        \includegraphics[scale=0.33]{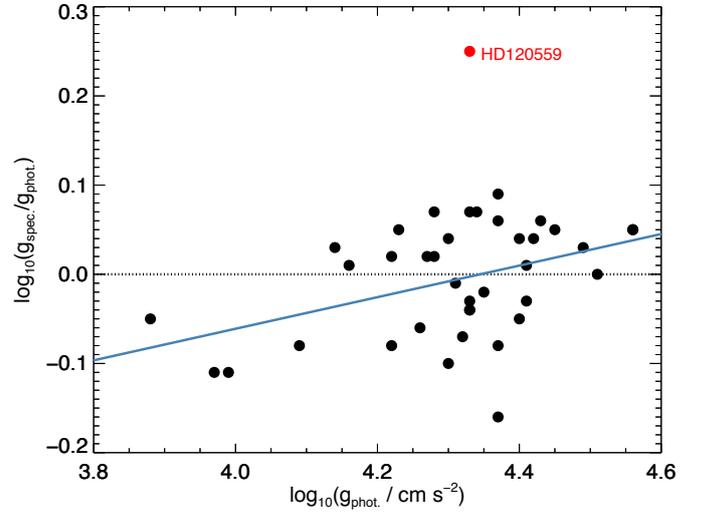}
        \includegraphics[scale=0.33]{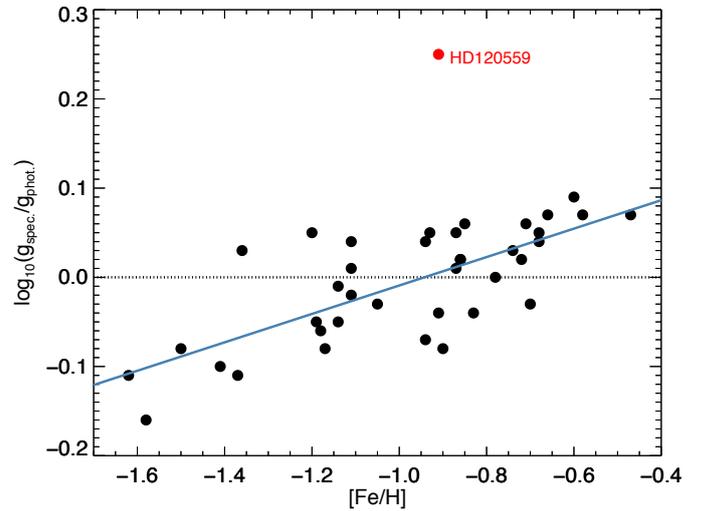}
        \caption{Differences between the spectroscopic
            surface gravities 
            adopted here, and photometric surface gravities 
            for $39$~stars in the UVES-FIES sample
            for which the photometry are not significantly
            affected by interstellar absorption and for which precise 
            parallaxes are available from Gaia DR2.
            A line of best fit is overdrawn, that excludes the
            obvious outlier HD~120559
            ($\lgg_{\text{phot.}}=4.33$),
            for which the measured Gaia DR2 parallax,
            $\uppi=29.19\pm0.18\,\mathrm{mas}$~may be in error;
            the Hipparcos parallax, 
            $\uppi=39.42\pm0.97\,\mathrm{mas}$, 
            gives $\lgg_{\text{phot.}}=4.64$, in much
            better agreement with the spectroscopic value 
            ($\lgg_{\text{spec}}=4.58$).}
        \label{fig:gaia}
    \end{center}
\end{figure}

The effective temperatures and surface gravities adopted here
are from
\citet{2014A&amp;A...568A..25N} for the 
HARPS-FEROS and UVES-FIES samples,
and those derived and used in 
\citet{2019A&A...622L...4A} for the VLT/UVES sample.
For the adopted 1D LTE abundances (\sect{obs_abundances}),
microturbulence is also relevant.
For the HARPS-FEROS and UVES-FIES samples,
these were also taken from \citet{2014A&amp;A...568A..25N},
while for the VLT/UVES sample
these were originally presented in \citet{2007A&amp;A...469..319N}.
Errors in the choice of 1D LTE microturbulence do not severely affect
the 3D non-LTE analysis presented here, because these are largely 
corrected for after applying the
3D non-LTE versus 1D LTE abundance corrections
(which are functions of $\vmic$; \sect{method_abcor}).
We present an overview of these stellar parameters here;
full details about their derivation
can be found in the references above.
In all of the above papers, 
the stellar parameters, including the chemical composition,
were iterated until consistency was achieved.

For the HARPS-FEROS sample, 
effective temperatures were derived 
by \citet{2014A&amp;A...568A..25N} using
$(b-y)$~and $(V-K)$~colours, and the 
calibration of \citet{2010A&amp;A...512A..54C} based on the infra-red
flux method. Given effective temperatures,
surface gravities were derived from the fundamental relation,
with absolute magnitudes derived 
via Hipparcos parallaxes \citep{2007A&amp;A...474..653V},
bolometric corrections from \citet{2010A&amp;A...512A..54C},
and stellar masses inferred via Yonsei-Yale evolutionary tracks
\citep{2003ApJS..144..259Y}.
Microturbulences were inferred from a 1D LTE analysis
of \ion{Fe}{II} lines in the standard way,
namely on the condition that the inferred
iron abundance should not depend on line strength.
For the Sun, the standard values were adopted here:
$\teff=5772\,\mathrm{K}$~and $\lgg=4.438$~\citep{2016AJ....152...41P},
and $\vmic=1.0\,\kms$~\citep[e.g.][]{1974SoPh...39...19H}.

For the UVES-FIES sample, 
the effective temperatures and surface gravities
were determined by
\citet{2014A&amp;A...568A..25N} through differential
1D LTE spectroscopic analyses of weak \ion{Fe}{I}
and \ion{Fe}{II}~lines, with respect to the two standard stars
HD~22879 and HD~76932.
The effective temperatures and surface gravities of the standard
stars were determined as per the HARPS-FEROS sample above, namely
using photometry and Hipparcos parallaxes.
Microturbulences were here inferred from a 1D LTE analysis
of \ion{Fe}{I} and \ion{Fe}{II} lines.

Lastly, for the VLT/UVES sample,
the effective temperatures were determined via 3D non-LTE fitting of
the $\hbeta$~line in echelle spectra,
as described in \citet{2019A&A...622L...4A}.
The surface gravities were mainly derived using the fundamental relation
as per the HARPS-FEROS sample above, except using the more precise
Gaia DR2 parallaxes \citep{2018A&A...616A...1G} instead of Hipparcos ones.
The \ion{C}{I}, \ion{O}{I}, and
\ion{Fe}{II} lines are usually sufficiently weak in these
metal-poor stars that the inferred abundances are not sensitive
to the choice of microturbulence.
Therefore $\vmic=1.5\,\kms$~was assumed for most of the stars
in this sample.

\subsection{Implications of new Gaia DR2 parallaxes}
\label{obs_stellarparams_gaia}

Newly available precise parallaxes from Gaia DR2
\citep{2018A&A...616A...1G} 
could impact the stellar parameters
derived for the HARPS-FEROS and UVES-FIES samples
(the VLT/UVES sample having already being re-analysed 
using Gaia DR2 parallaxes
as discussed in \citealt{2019A&A...622L...4A}).
For the HARPS-FEROS sample,
the Hipparcos parallax errors are sufficiently small
(corresponding to surface gravity errors of 
around $0.05\,\dex$, as discussed in \citealt{2014A&amp;A...568A..25N}),
that there would not be a significant impact on the results
if the Gaia DR2 parallaxes were used instead.

For the UVES-FIES sample,
\fig{fig:gaia} illustrates slight systematic trends
in the differences between the
spectroscopic surface gravities adopted here
and photometric surface gravities based on Gaia DR2 parallaxes,
with effective temperature, surface gravity, and metallicity.
The errors are at most $0.1\,\dex$, and largest at 
higher effective temperatures and lower metallicities.
These trends were not apparent in
Fig.~7 of \citet{2014A&amp;A...568A..25N},
partly because of the larger uncertainty in the corresponding
Hipparcos parallaxes adopted there,
and partly because of the smaller number of stars (24)
in that work with precise Hipparcos parallaxes compared to
the number of stars (39) here with precise Gaia DR2 parallaxes.

The systematic trends in \fig{fig:gaia}
likely arise from 3D non-LTE effects
in the \ion{Fe}{I} lines that were used to infer 
the spectroscopic surface gravities in 
\citet{2014A&amp;A...568A..25N}.
This was verified for the star CD~-33~3337~($\teff\approx6100\,\mathrm{K}$,
$\lgg\approx3.9$, $\feh\approx-1.4$), for which the spectroscopic
surface gravity is $0.11\,\dex$~lower than the photometric surface gravity.
The 1D non-LTE 
versus 1D LTE abundance corrections from \citet{2012MNRAS.427...50L}
amount to around $+0.06\,\dex$~for representative \ion{Fe}{I} lines,
while the abundance corrections for the standard star HD~22879 amount
to around $+0.03\,\dex$.
Noting that 
the differential 3D non-LTE effects on \ion{Fe}{II} lines are small
between these two stars ($0.003\,\dex$),
the differential increase in iron abundance from the \ion{Fe}{I} lines
of $+0.03\,\dex$~acts to increase 
the measured spectroscopic surface gravity of CD~-33~3337 by around
$0.08\,\dex$, and thus brings it into closer agreement with
the photometric surface gravity.
The residual $0.03\,\dex$~difference
could well be due to 3D effects enhancing the non-LTE effects,
as discussed in \citet{2016MNRAS.463.1518A}.

It is not possible to obtain precise photometric surface gravities
for the entire UVES-FIES sample owing to interstellar absorption.
Nevertheless, errors of $0.1\,\dex$~in the surface gravity
have only a small impact on
$\feh$~(around $0.04\,\dex$), and a negligible
impact on $\abrat{C}{Fe}$, $\abrat{O}{Fe}$, and 
$\abrat{C}{O}$~(see Table 7 of \citealt{2014A&amp;A...568A..25N}),
and thus do not affect the main conclusions of this study.

\subsection{Carbon, oxygen, and iron abundances}
\label{obs_abundances}

Iron abundances were determined from \ion{Fe}{II} lines.
For the HARPS-FEROS sample, these were the
$12$~lines listed in Table 1 of
\citet{2014A&amp;A...568A..25N}.
For the UVES-FIES and VLT/UVES samples,
up to around $16$~different optical lines were 
adopted, for which equivalent widths have previously been
presented in the literature
\citep{2002A&amp;A...390..235N,
2004A&amp;A...415..993N,2007A&amp;A...469..319N,
2011A&A...530A..15N}.

For the HARPS-FEROS and UVES-FIES samples, carbon abundances were
inferred from the \ion{C}{I} $505.2\,\nm$~and $538.0\,\nm$~lines.
These lines become very weak towards lower metallicities, and
for the VLT/UVES sample, the near infra-red \ion{C}{I} 
lines were used instead.
These lines are listed in the main text of
\citet{2019A&A...622L...4A}.

For the entire sample, oxygen abundances were based on the 
\ion{O}{I} $777.2\,\nm$, $777.4\,\nm$,
and $777.5\,\nm$~lines. While \citet{2014A&amp;A...568A..25N}
also measured and used the [\ion{O}{I}] $630.0\,\nm$~line,
we choose not to use it here, for two reasons.
First, at high metallicity the [\ion{O}{I}] $630.0\,\nm$~line is
severely blended by 
a \ion{Ni}{I} line \citep[e.g.][]{2001ApJ...556L..63A}
which constitutes more than $30\%$~of the equivalent width
in the solar spectrum, and an even higher percentage 
towards higher metallicities and effective temperatures.
Second, at low metallicity
the [\ion{O}{I}] $630.0\,\nm$~line is weak
(the equivalent width in this sample is typically 
less than $0.5\,\mathrm{pm}$), and thus difficult 
to measure reliably. 
Thus we do not consider 
the [\ion{O}{I}] $630.0\,\nm$~line
a reliable oxygen abundance indicator 
for this sample.

Given the stellar parameters (\sect{method_abcor}),
the final elemental abundances were determined by applying 
the 3D non-LTE versus 1D LTE abundance corrections 
to absolute, line-by-line 1D LTE abundances
of carbon and oxygen, 
and the 3D LTE versus 1D LTE abundance corrections 
to absolute, line-by-line 1D LTE abundances of iron.
Where possible, this was repeated for the Sun
(included in the HARPS-FEROS sample), to derive
line-by-line differential abundances that
were then averaged over the different lines to obtain
final estimates for $\abrat{C}{H}$, $\abrat{O}{H}$, and $\feh$.
For carbon and iron and for some stars, 
line-by-line absolute abundances
had to be derived and averaged instead, owing to the lines
being too blended or saturated to measure reliably in the solar spectrum.
For the HARPS-FEROS and UVES-FIES samples,
the 1D LTE abundances were taken from
\citet{2014A&amp;A...568A..25N};
for the VLT/UVES sample, they were
re-derived here using our preferred stellar parameters,
and our line formation calculations 
across the fine and extensive \marcs~grid.

The uncertainties in the abundances were calculated
from the line-by-line variations
in the inferred abundances. 
For stars and elements for which only a single line 
could be detected, the uncertainty was estimated as
$\sqrt{2}$~times the largest uncertainty determined for that
element in that population of stars.
Finally, symmetrical uncertainties were taken from
\citet{2019A&A...622L...4A} for $\abrat{O}{H}$~in 
CD~-24~17504 ($0.20\,\dex$)
and for $\abrat{C}{H}$~in G~64-12 ($0.065\,\dex$).

We present the stellar parameters and 3D non-LTE 
elemental abundances in the online Table 7.
For completeness, we also present the 
1D non-LTE, 3D LTE, and 1D LTE results,
in this online table.
We plot 3D non-LTE and 1D LTE abundance ratios
in Figs~\ref{fig:abundances1}---\ref{fig:planets1},
and discuss the results in \sect{discussion} below.

\section{Discussion}
\label{discussion}

\begin{figure*}
    \begin{center}
        \includegraphics[scale=0.33]{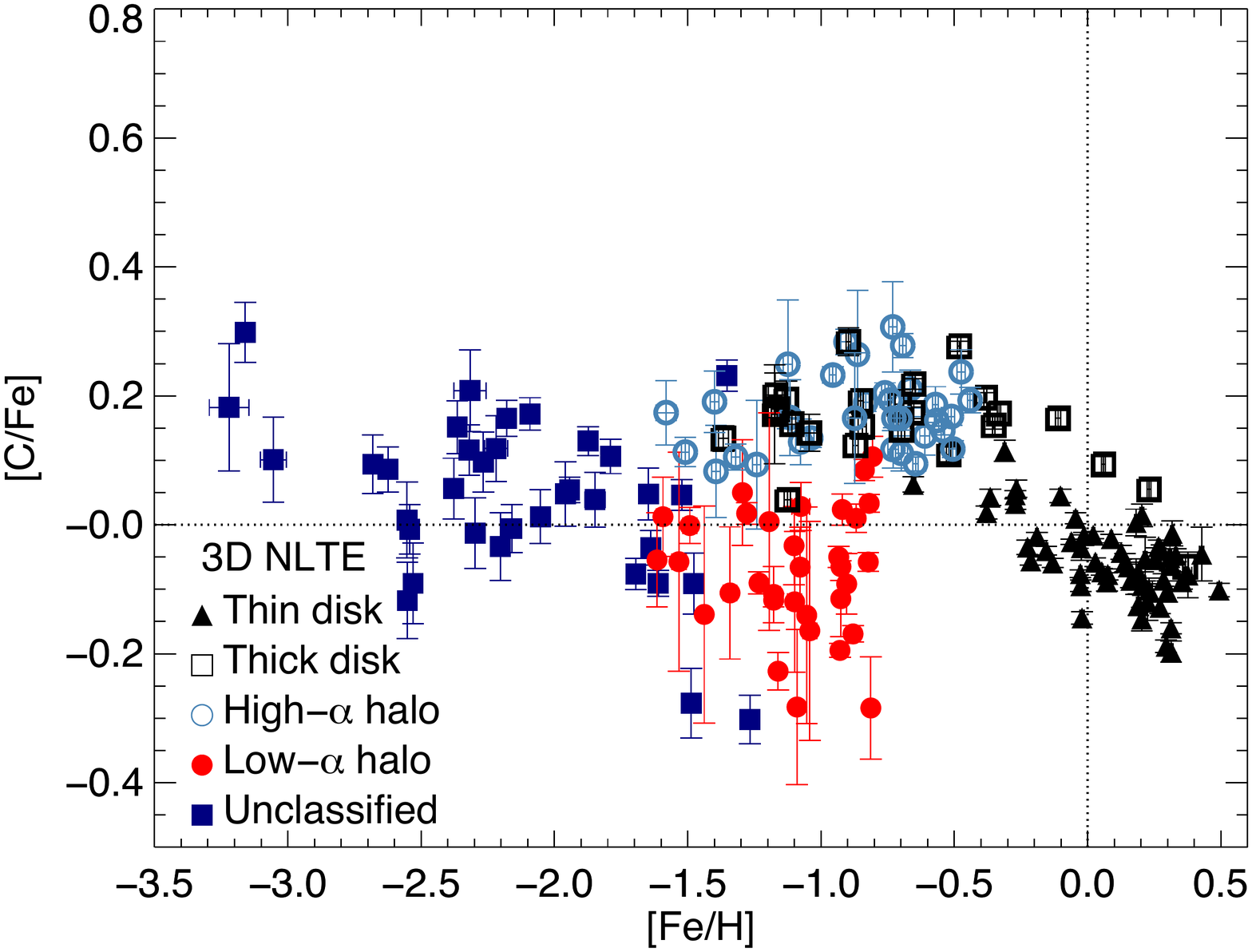}\includegraphics[scale=0.33]{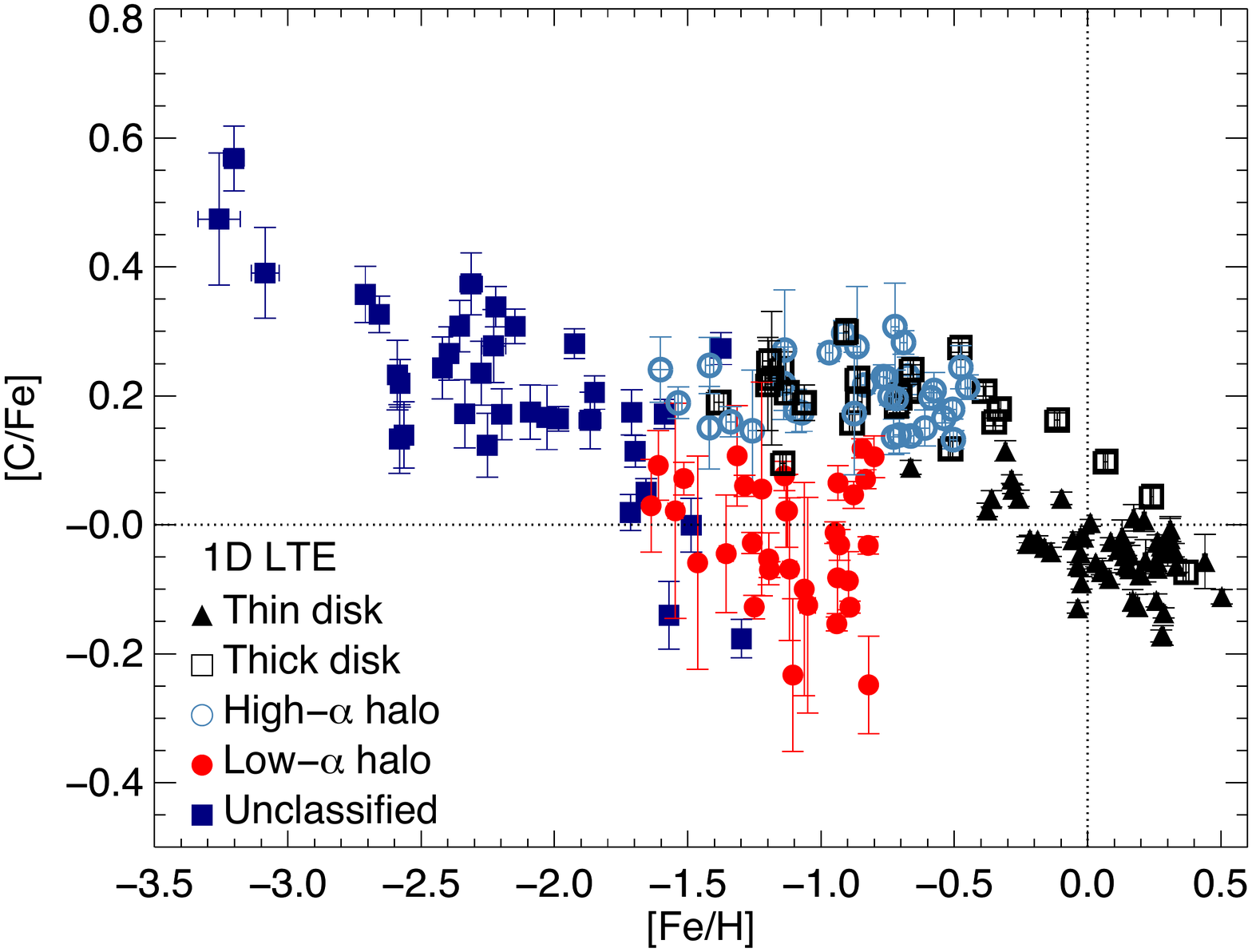}
        \caption{Carbon to iron abundance ratios 
        for the entire stellar sample. 
        The unclassified stars are from the VLT/UVES sample.
        The panels show 
        results based on different line formation models, the 3D 
        non-LTE model being preferred.}
        \label{fig:abundances1}
    \end{center}
\end{figure*}

\begin{figure*}
    \begin{center}
        \includegraphics[scale=0.33]{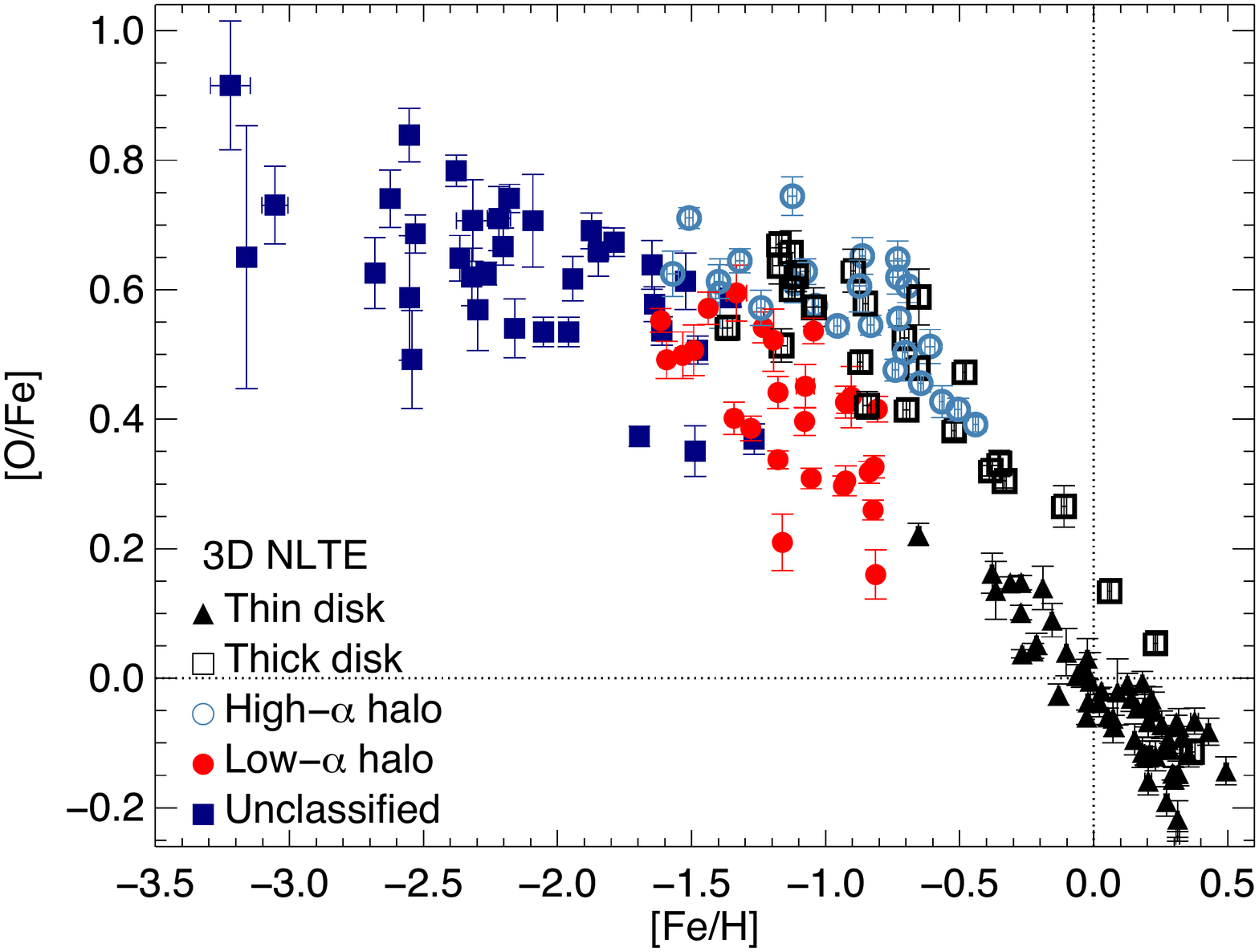}\includegraphics[scale=0.33]{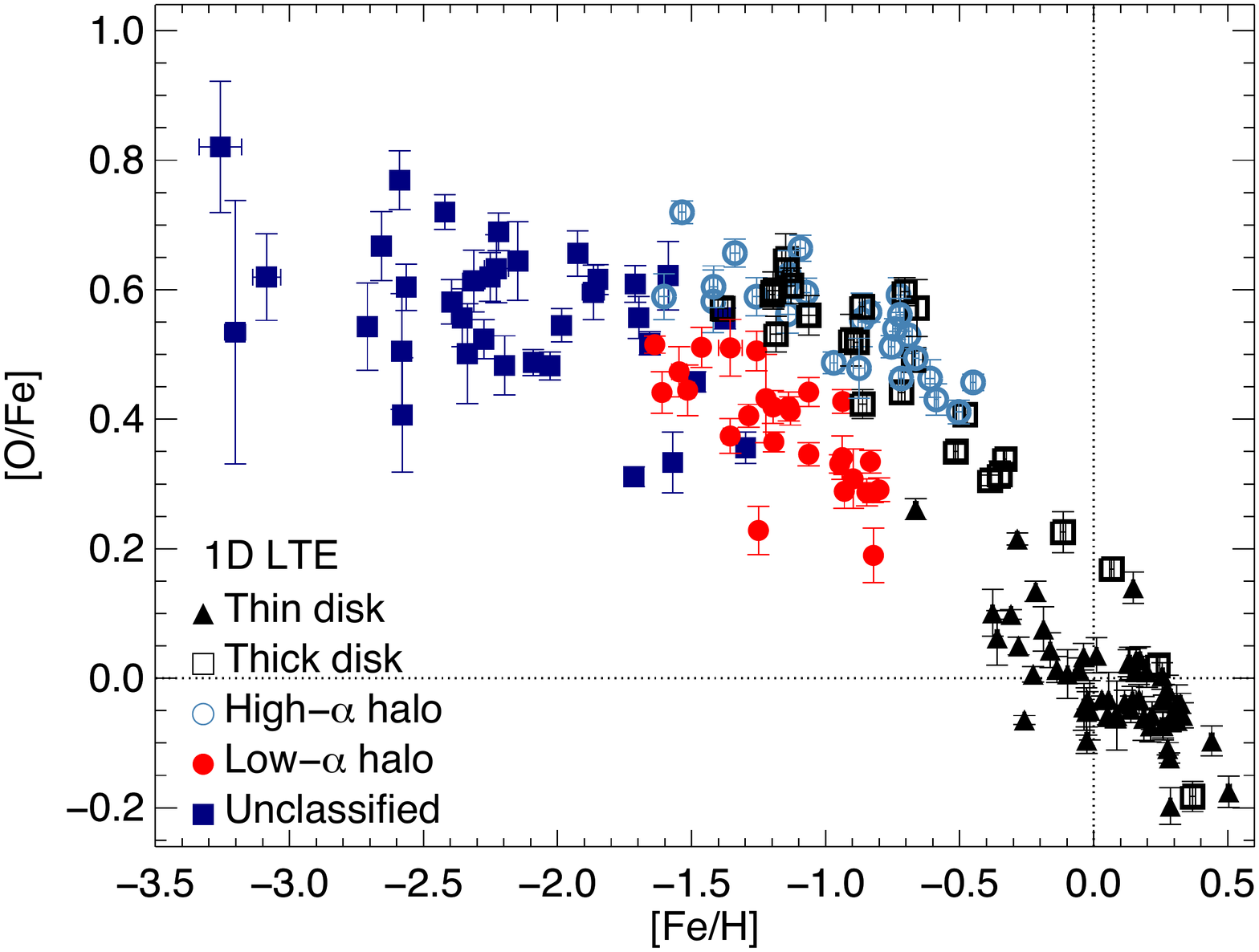}
        \caption{Oxygen to iron abundance ratios
        for the entire stellar sample. 
        The unclassified stars are from the VLT/UVES sample.
        The panels show
        results based on different line formation models, the 3D 
        non-LTE model being preferred.}
        \label{fig:abundances2}
    \end{center}
\end{figure*}

\begin{figure*}
    \begin{center}
        \includegraphics[scale=0.33]{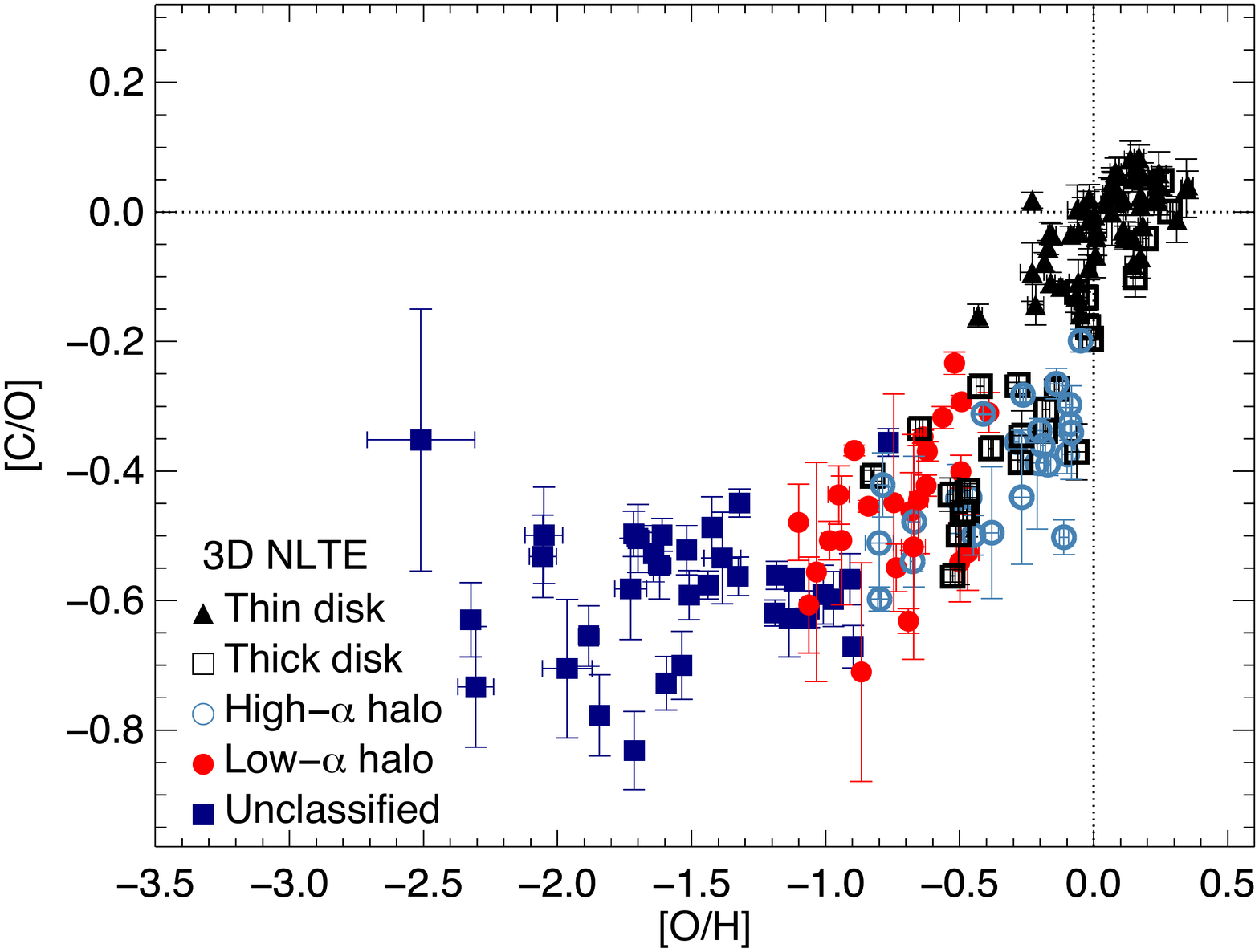}\includegraphics[scale=0.33]{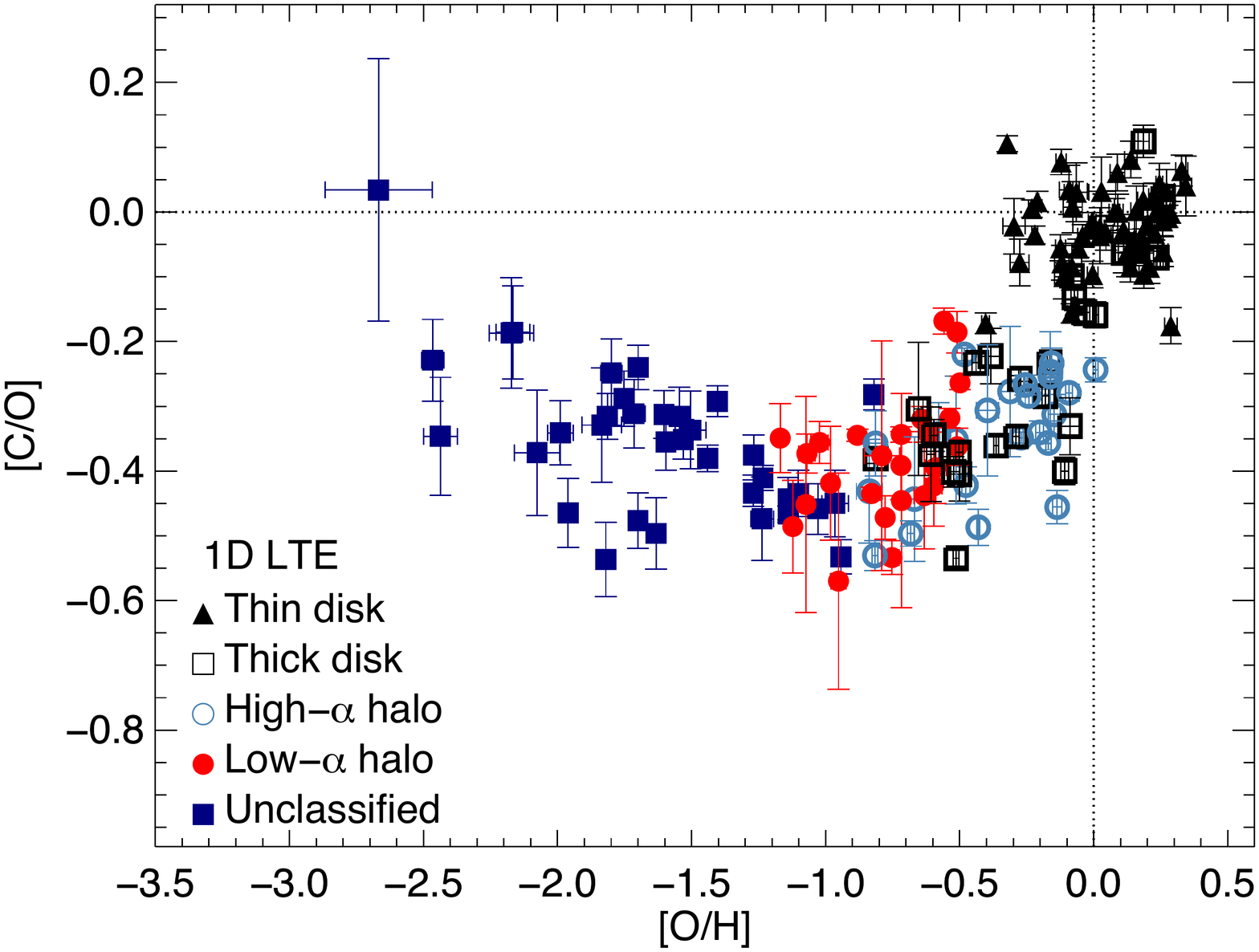}
        \caption{Carbon to oxygen abundance ratios
        for the entire stellar sample. 
        The unclassified stars are from the VLT/UVES sample.
        The panels show
        results based on different line formation models, the 3D 
        non-LTE model being preferred.}
        \label{fig:abundances3}
    \end{center}
\end{figure*}

\begin{figure*}
    \begin{center}
        \includegraphics[scale=0.33]{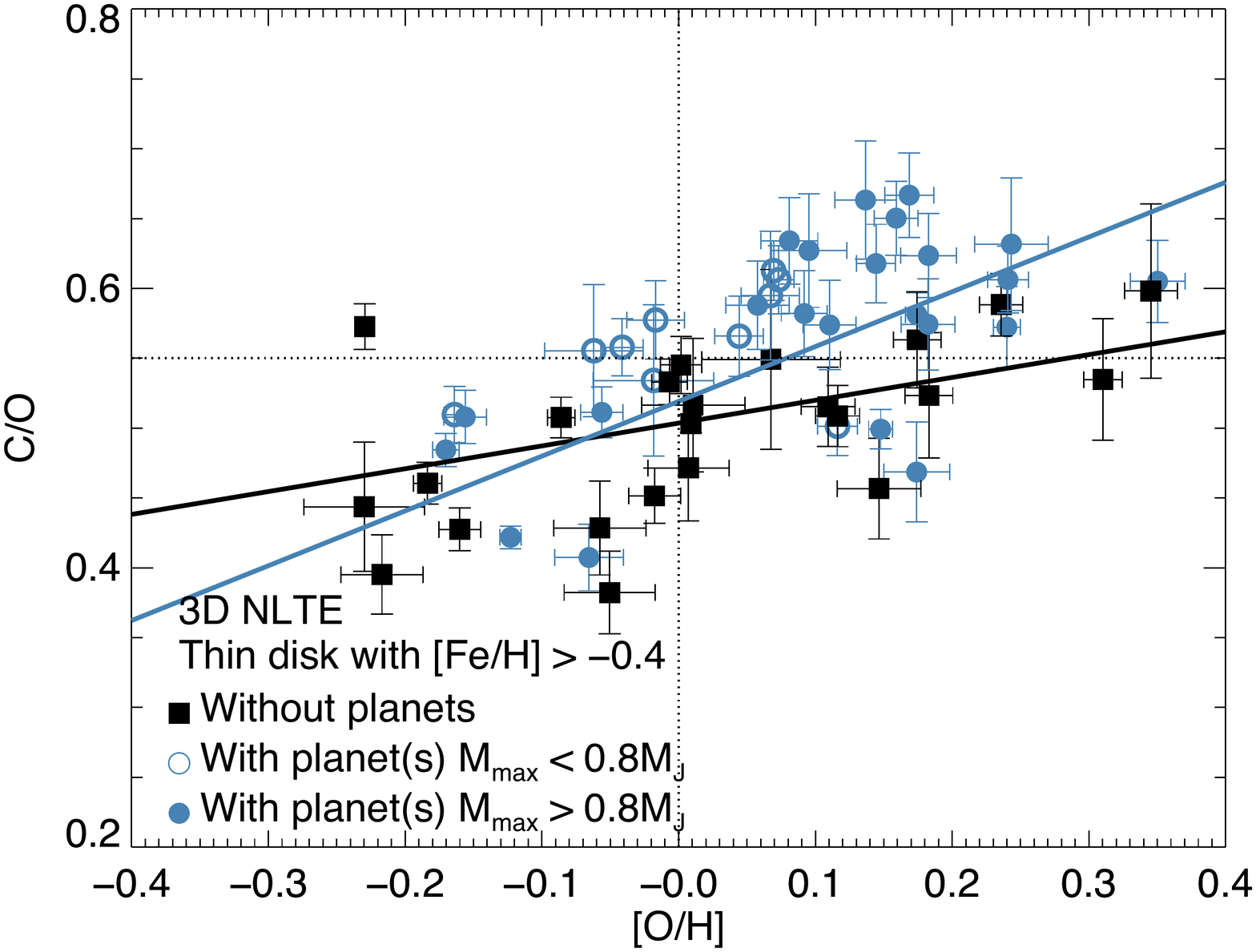}\includegraphics[scale=0.33]{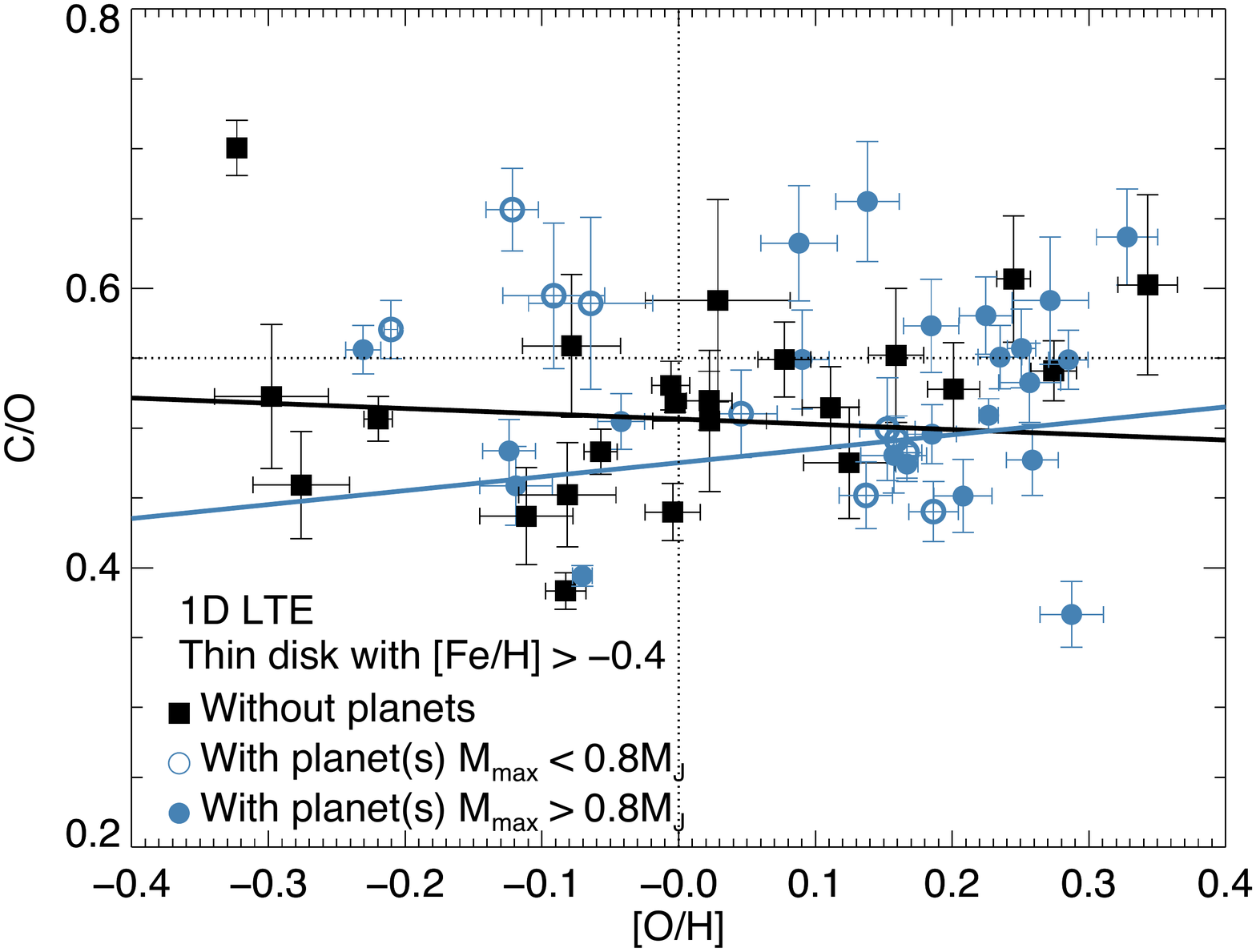}
        \caption{Carbon to oxygen abundance ratios,
        for thin-disk stars without and with confirmed planet detections.
        The latter are further separated according to the maximum planet
        mass in the system.
        The panels show results based on different 
        line formation models, the 3D non-LTE model being preferred. 
        The nominal solar value 
        $\mathrm{C/O}=0.55$~is shown as a horizontal dashed line.
        Also plotted are error-weighted lines of best fit 
        for stars without and with confirmed planet detections
        (irrespective of maximum planet mass).
        The error bars do not account for the systematic 
        uncertainty in the solar carbon and
        oxygen abundances.}
        \label{fig:planets1}
    \end{center}
\end{figure*}

\subsection{$\abrat{C}{Fe}$~and $\abrat{O}{Fe}$~in 
the disks and high-$\upalpha$~halo population}
\label{discussion_gce_cfeofe}

We verify the main results of \citet{2014A&amp;A...568A..25N}
concerning the different stellar populations 
in Figs~\ref{fig:abundances1}---\ref{fig:abundances3}.
There is a gap between the thin and thick disks
of around $0.1\,\dex$~in
$\abrat{C}{Fe}$~and $\abrat{O}{Fe}$~at $\feh\approx0.0$.
The thick-disk stars and high-$\upalpha$~halo population stars 
overlap in the $\abrat{C}{Fe}$~and $\abrat{O}{Fe}$~versus $\feh$~planes:
this is consistent with the interpretation that
the high-$\upalpha$~halo population is composed of ancient
thick-disk stars that have been heated by past accretion events
to obtain halo-like kinematics \citep{2018ApJ...863..113H}.

\fig{fig:abundances1} and \fig{fig:abundances2} show that
compared to the thin disk, the thick disk has higher 
$\abrat{C}{Fe}$~and $\abrat{O}{Fe}$~over the entire observed metallicity range,
with the possible exception of at the highest $\feh$:
the two populations may meet at $\abrat{C}{Fe}\approx-0.10$,
and~$\abrat{O}{Fe}\approx-0.15$, when $\feh\approx0.4$, although the sample
would need to be extended to confirm this.  
The thin and thick disks increase
linearly in both $\abrat{C}{Fe}$~and $\abrat{O}{Fe}$~with decreasing
$\feh$, the thick disk seemingly rising more steeply than the thin disk.
At lower metallicities, a plateau is reached in the thick disk
(and high-$\upalpha$~halo population)
of $\abrat{O}{Fe}\approx0.6$~at $\feh\approx-1.0$.
These results can be interpreted by considering the cosmic
origins of carbon, oxygen, and iron (\sect{introduction}).
As oxygen is an $\upalpha$-element, the $\abrat{O}{Fe}$~plateau reflects
a steady rate of oxygen and iron enrichment of the 
cosmos by core-collapse
supernova.  The gradual decrease of
$\abrat{O}{Fe}$, and also of $\abrat{C}{Fe}$, towards
higher $\feh$~reflects the increasing iron 
enrichment from Type-Ia supernovae, which have
delayed time distribution relative to core-collapse supernovae
(e.g.~\citealt{2012MNRAS.426.3282M}).

There is a hint that 
$\abrat{C}{Fe}$~has a slight 
increase with increasing metallicity in
the region $-1.5\lesssim\feh\lesssim-1.0$,
reaching a maximum of $\abrat{C}{Fe}\approx0.2$~at
$\feh\approx-1.0$.
This can be interpreted as early pollution of
(presumably) the most massive AGB stars which have relatively short
lifetimes ($10^7$~to $10^8$~years). This causes an increase in
$\abrat{C}{Fe}$ with $\feh$, until the contribution of 
iron from Type-Ia supernova
becomes more dominant. These results are  in agreement with measurements of
$\abrat{Ba}{Fe}$~in Galactic halo stars, 
which show evidence of pollution of barium (and of
other neutron-capture elements) by AGB stars, already at these low
$\feh\approx-1.5$ \citep{2007A&A...476..935F,2012A&A...545A..31H}.

The observed linear decrease in 
$\abrat{C}{Fe}$~(\fig{fig:abundances1}) 
and $\abrat{O}{Fe}$~(\fig{fig:abundances2}) with increasing
metallicity above
$\feh\gtrsim0.0$~is in contrast to some recent results.
For example \citet{2018ApJ...852...49H} find 
instead a gently rising trend of $\abrat{C}{Fe}$~with 
increasing $\feh$~at super-solar metallicities,
and a plateau in $\abrat{O}{Fe}$.
The data of \citet{2018ApJ...852...49H} is drawn from APOGEE DR13,
where the stars are all red giants, and where
the abundances of carbon and oxygen were 
mainly inferred from a 1D LTE analysis of infra-red CH, CO, and OH lines.
In contrast, the present stellar sample is composed of F~and G~dwarfs,
where the abundances of carbon in the high-metallicity stars
were obtained from the \ion{C}{I} $505.2\,\nm$~and $538.0\,\nm$~lines,
and the abundances of oxygen
were obtained from the \ion{O}{I} $777\,\nm$~multiplet.
Concerning the $\abrat{O}{Fe}$~trend, our results agree well
with earlier 1D non-LTE studies of  
the \ion{O}{I} $777\,\nm$~multiplet in dwarfs
\citep[e.g.][]{2013ApJ...764...78R,
2014A&amp;A...562A..71B,2018MNRAS.478.4513B}.
We speculate that this discrepancy could be alleviated
if the APOGEE abundances were corrected for the severe
3D effects that are expected for molecular lines in red giants
\citep[e.g.][]{2007A&amp;A...469..687C,2011A&amp;A...529A.158H}.

\subsection{$\abrat{C}{O}$~in the disks
and high-$\upalpha$~halo population}
\label{discussion_gce_co}

The thin disk is systematically higher than the thick disk
in the $\abrat{C}{O}$~versus $\abrat{O}{H}$~plane.  
\fig{fig:abundances3} shows
that this amounts to $\approx0.15\,\dex$~in $\abrat{C}{O}$
at~$\abrat{O}{H}\approx0.0$. 
Furthermore, there is a clear underdensity of stars
in the $\abrat{C}{O}$---$\abrat{O}{H}$~plane, in the region 
$\abrat{C}{O}\approx-0.2$~and
$\abrat{O}{H}\approx-0.2$.

This separation of the thin disk
from the thick disk (together with the 
high-$\upalpha$~halo
population, which could be interpreted 
as the heated thick disk --- \sect{discussion_gce_cfeofe})
appears to be consistent with the Milky Way having had
two main infall episodes, the latter responsible for 
forming the thin disk
\citep{1997ApJ...477..765C,2001ApJ...554.1044C}.
The location of this underdensity corresponds to the
discontinuity in several of the 
two-infall Galactic chemical evolution models 
recently presented in \citet{2019arXiv190709476R},
that marks the onset of the second infall episode
(see their Fig.~1) and thus separates the younger thin disk
from the older thick disk (e.g.~\citealt{2018MNRAS.475.5487S}).

In both the thin disk and the thick disk
(and high-$\upalpha$~halo population), there 
are trends of increasing $\abrat{C}{O}$~with
increasing $\abrat{O}{H}$.
This could reflect an increasing rate of
carbon enrichment from AGB stars at later epochs,
or an increasing rate of carbon enrichment from metallicity-dependent winds
from massive stars at later epochs,
the two phenomena having degenerate elemental abundance
signatures (see \sect{introduction}).

\subsection{The low-metallicity stars}
\label{discussion_lowalpha}

The low- and high-$\upalpha$~halo populations clearly separate into low and
high groups in both the $\abrat{C}{Fe}$~and $\abrat{O}{Fe}$~versus 
$\feh$~planes in Figs~\ref{fig:abundances1}---\ref{fig:abundances2}.
The high-$\upalpha$~halo population can perhaps be interpreted 
as the heated thick disk, and we thus discuss the results for those
stars in the previous subsections, 
Sects~\ref{discussion_gce_cfeofe}---\ref{discussion_gce_co}.

If the unclassified stars from the VLT/UVES sample
are in fact primarily belonging to the low-$\upalpha$~halo population
(\sect{obs_sample}), then it appears that
the trends of $\abrat{C}{Fe}$~and $\abrat{O}{Fe}$~versus $\feh$~for
this combined set of low-metallicity stars are 
qualitatively similar to that of
the high-$\upalpha$~halo population, albeit
with a larger scatter, and offset towards lower metallicities.
Namely, $\abrat{C}{Fe}$~has a mild rise with decreasing
$\feh$, while $\abrat{O}{Fe}$~has a steeper rise with decreasing $\feh$
(see \sect{discussion_gce_cfeofe}).

The trends inferred here for oxygen 
in the low-$\upalpha$~halo population
are similar to those found in dwarf satellite galaxies
via red giant stars \citep[e.g.][]{2019A&A...626A..15H}. The same is true for
carbon \citep[e.g.][]{2015ApJ...801..125K,
2015A&A...574A.129S,2016A&A...585A..70L}, after taking into account mixing in
the red giant branch stars which results in an offset in
$\abrat{C}{Fe}$~\citep[e.g.][]{2000A&A...354..169G,2006A&A...455..291S}.
One interpretation of these findings, combined with information from other
elemental abundance ratios, as well as kinematics and ages
\citep{2010A&amp;A...511L..10N,2011A&A...530A..15N,2012A&A...543A..28N,      2012A&A...538A..21S},
is that the low-$\upalpha$~halo is a younger population, composed of stars
accreted in a past merger event with a dwarf satellite galaxy \citep[Gaia
Enceladus;][]{2018Natur.563...85H}.

If the low-$\upalpha$~halo population is to
be interpreted as an accreted dwarf galaxy, one would expect that the knee in
the $\abrat{\upalpha}{Fe}$~versus $\feh$~plane is shifted to lower
metallicities, reflecting the weaker star formation rate in less-massive
systems \citep[e.g.][]{2009ARA&A..47..371T}. Recent Galactic chemical        evolution modelling suggest that this knee could be at around
$\feh\approx-2.5$~\citep{2019arXiv190303465V}.
The corresponding plateau value of $\abrat{O}{Fe}\approx0.7$~at 
the lowest metallicities $\feh\lesssim-2.5$~in the unclassified stars, 
is roughly $0.1\,\dex$~higher than that
of the high-$\upalpha$~halo population. This is consistent with the
metallicity dependence of theoretical core-collapse supernovae yields, which
predict $\abrat{O}{Fe}$ to be $\approx0.15$~dex higher at 
$\feh\approx-3$ compared to
at solar metallicity \citep{2006ApJ...653.1145K}. Similarly, theoretical
predictions indicate that 
$\abrat{C}{Fe}$~in the yields of core-collapse supernovae
might increase towards lower metallicities
\citep{2006ApJ...653.1145K,2010ApJ...724..341H}.

Interestingly, in the $\abrat{C}{O}$~versus $\abrat{O}{H}$~plane
(\fig{fig:abundances3}), there is not clear evidence for an offset between the
low-$\upalpha$~halo population and the thick disk/high-$\upalpha$~halo
population. Rather, the low-$\upalpha$~halo population continues the trend of
decreasing $\abrat{C}{O}$~with decreasing $\abrat{O}{H}$.
The unclassified stars, which
are possibly dominated by the low-$\upalpha$ halo population
(\sect{obs_sample}), form a plateau of $\abrat{C}{O}\approx-0.6$~below
$\abrat{O}{H}\approx-1.0$.  This plateau could be interpreted as reflecting
primary carbon and oxygen nucleosynthesis from the cores of massive
stars at earlier epochs, with a negligible contribution from AGB stars or
metallicity-dependent winds from massive stars (see \sect{introduction}).

The most oxygen-poor star 
in \fig{fig:abundances3} is CD~-24~17504.  
Given the uncertainties,
it is not clear if this star is in fact an outlier to the mean trend: 
the large error bars reflect the difficulty in reliably 
measuring the equivalent width of the \ion{O}{I} $777\,\nm$~multiplet
(see Fig.~4 of \citealt{2009A&amp;A...500.1143F}).
If the star is indeed an outlier, this may reflect its
uncertain status as a
so-called carbon-enhanced metal-poor (CEMP) star
(having $\abrat{C}{Fe}>0.7$; \citealt{2007ApJ...655..492A}).
A 1D LTE analysis of CH lines in this star suggests that it 
does belong to this category,
with $\abrat{C}{Fe}\approx1.1$~\citep{2015ApJ...808...53J}.
However our 3D non-LTE analysis of \ion{C}{I}~lines
implies $\abrat{C}{Fe}\approx0.3$: strictly speaking
not a CEMP star, however still with a moderate enhancement of carbon.
More discussion of this star, and two other stars in this sample
that were previously reported as CEMP stars (G~64-12 and G~64-36; 
\citealt{2016ApJ...829L..24P}), can be found in
\citet{2019A&A...622L...4A} and
\citet{2019ApJ...879...37N}.

\subsection{$\mathrm{C/O}$~in stars with and
without confirmed planet detections}
\label{discussion_planets}

Given the importance of $\mathrm{C/O}$~on 
planet formation and characterisation
\citep[e.g.][]{2016ApJ...831...20B}, we
briefly investigate the implications of our results on
our understanding of exoplanets.
In \fig{fig:planets1} we illustrate
$\mathrm{C/O}$\footnote{$\mathrm{C/O}\equiv
N_{\mathrm{C}}/N_{\mathrm{O}}$}~versus $\abrat{O}{H}$~for
thin-disk stars
with and without confirmed planet detections,
drawing on data from the NASA exoplanet archive
\citep{2013PASP..125..989A}.
By plotting $\mathrm{C/O}$~versus $\abrat{O}{H}$,
rather than histograms of $\mathrm{C/O}$~or even
$\mathrm{C/O}$~versus $\feh$, it is easier to 
disentangle the effects of Galactic chemical evolution;
while restricting the
analysis to a single population (here thin-disk stars) 
partially removes systematics arising from age differences.

The 3D non-LTE results shown in \fig{fig:planets1} indicate that, at least
for $\abrat{O}{H}\gtrsim0.0$, thin-disk stars with confirmed planet
detections have larger values of $\mathrm{C/O}$~than stars without confirmed
planet detections at given $\abrat{O}{H}$. A similar result was found by
\citet[][]{2019A&A...621A.112P}, however the signature is much stronger here.
There is no apparent bias with planetary mass, apart from that stars with
higher metallicities (here traced by $\abrat{O}{H}$)~tend 
to host more massive planets
\citep[e.g.][]{2005ApJ...622.1102F,
2010PASP..122..905J,2019Geosc...9..105A}. The
mean rising trend in the plot is a result of the increasing production of
carbon at later cosmic times (\sect{discussion_gce_co}).

The result that $\mathrm{C/O}$~is higher in stars with confirmed
planet detections than in stars without confirmed
planet detections, could possibly be extrapolated to say 
that protoplanetary disks with higher values of $\mathrm{C/O}$~give
rise to more planets, or at least to more planets that are massive.
The high binding energy of the $\mathrm{CO}$~molecule makes
the value of $\mathrm{C/O}$~a sensitive parameter of the
protoplanetary disk chemistry, and it is possible that 
$\mathrm{C/O}$~could play some intrinsic role
in the formation efficiency of planets.

\subsection{Implications of 3D non-LTE spectral line formation}
\label{discussion_3n}

It is clear from comparing
the 3D non-LTE and 1D LTE results in
in Figs~\ref{fig:abundances1}---\ref{fig:planets1}
that using improved line formation models
reduces the scatter about the mean trends in the elemental abundances. 
The reduction in scatter is most apparent 
in the $\abrat{C}{O}$~and 
$\mathrm{C/O}$~versus $\abrat{O}{H}$~diagrams.
In particular, the 3D non-LTE results suggest that
thin-disk stars with confirmed planet detections have 
larger values of $\mathrm{C/O}$~than stars without
confirmed planet detections at given
$\abrat{O}{H}$~(\sect{discussion_planets}),
while this signature is not apparent in the 
analogous 1D LTE results
(\fig{fig:planets1}).

Using 3D non-LTE line formation models not only
impacts the scatter in the abundance ratios, but also the mean trends.
For carbon, although the 1D LTE 
$\abrat{C}{Fe}$~versus $\feh$~trend in \fig{fig:abundances1}
rises steeply towards lower metallicities, the 
$\abrat{C}{Fe}$~versus $\feh$~trend is much flatter.
This is because for \ion{C}{I}, the (absolute and differential) 
3D non-LTE versus 1D LTE abundance corrections 
are negative and are more severe 
towards lower metallicities (reaching $-0.3\,\dex$ for
the near infra-red \ion{C}{I} lines;
\sect{results_abcor_c1}).

It is important to note that other carbon abundance diagnostics
are also susceptible to large systematic errors:
in particular, negative and severe 3D LTE versus 1D LTE 
abundance corrections are expected 
for CH lines \citep[reaching as much as
$-1.0\,\dex$; e.g.][]{2006ApJ...644L.121C,
2016A&A...593A..48G,2017A&amp;A...598L..10G}.
Thus 1D LTE analyses significantly overestimate
carbon abundances in metal-poor stars and thus
the fraction of CEMP stars in our Galaxy
\citep{2019A&A...622L...4A,2019ApJ...879...37N}.

For oxygen, the 3D non-LTE $\abrat{O}{Fe}$~versus $\feh$~trend in
\fig{fig:abundances2} is steeper in both the metal-rich regime,
and in the metal-poor regime, than the corresponding 1D LTE trend.
This is because for \ion{O}{I}
there are severe negative absolute 3D non-LTE versus
1D LTE abundance corrections at high metallicities
(reaching $-0.6\,\dex$ for
the \ion{O}{I} $777\,\nm$~multiplet; 
\sect{results_abcor_o1}). This corresponds
to moderate negative differential abundance corrections 
with respect to the Sun
(the solar absolute abundance corrections are
around $-0.2\,\dex$ for
the \ion{O}{I} $777\,\nm$~multiplet.)
At low metallicities however, the absolute abundance
corrections for \ion{O}{I} lines
are closer to zero, and consequently the differential abundance 
corrections with respect to the Sun are significant, and positive
(\sect{results_abcor_o1}).

The 3D non-LTE
$\abrat{C}{O}$~versus $\abrat{O}{H}$~trend in \fig{fig:abundances3}
is strikingly different to the analogous 1D LTE trend,
as a result of the differential 3D non-LTE abundance corrections
going in opposite directions for \ion{C}{I} and \ion{O}{I}
in the low metallicity regime.
While the 3D non-LTE trend is monotonic,
the 1D LTE trend turns over at 
$\feh\approx-1.0$, which has been interpreted 
as a possible nucleosynthesis signature of 
Population III stars \citep{2004A&amp;A...414..931A,
2009A&amp;A...500.1143F}
or rapidly-rotating massive Population II stars
\citep{2006A&A...449L..27C}.
This means that there is no longer a need to introduce exotic
nucleosynthesis channels to explain the observations
of carbon and oxygen abundances in metal-poor stars,
as discussed in \citet{2019A&A...622L...4A},
at least down to $\abrat{O}{H}\approx-2.5$.

As explained in \sect{results_abcor},
in the absence of 3D non-LTE models,
1D non-LTE models should be used instead,
at least for the high-excitation
\ion{C}{I}~and \ion{O}{I} lines discussed here.
A comparison of 1D LTE, 3D LTE, 1D non-LTE,
and 3D non-LTE results for the metal-poor stars can be found in 
\citep{2019A&A...622L...4A}, where it is clear
that the 1D non-LTE results better resemble the 3D non-LTE ones,
at least compared to 1D LTE, or 3D LTE.

\section{Conclusion}
\label{conclusion}

We have presented extensive grids of
3D non-LTE versus 1D LTE abundance corrections 
for \ion{C}{I} and \ion{O}{I} lines
for FGK-type dwarfs and sub-giants with
$5000\lesssim\teff/\mathrm{K}\lesssim6500$,
$3.0\leq\log\left(g / \mathrm{cm\,s^{-2}}\right)\leq5.0$,
and $-3.0\leq\feh\leq0.5$.  We have also presented
grids of 1D non-LTE versus 1D LTE abundance corrections 
for \ion{C}{I} and \ion{O}{I} lines that extend
to even hotter and cooler late-type stars.
The (absolute) 3D non-LTE versus 1D LTE abundance corrections
can be as severe as $-0.3\,\dex$~for \ion{C}{I} lines,
and $-0.6\,\dex$~for \ion{O}{I} lines.

In addition, we have presented 3D LTE versus
1D LTE abundance corrections for \ion{Fe}{II} lines
(which are expected to suffer negligible non-LTE effects)
for late-type FGK-type dwarfs, sub-giants and giants with
$4000\lesssim\teff/\mathrm{K}\lesssim6500$,
$1.5\leq\log\left(g / \mathrm{cm\,s^{-2}}\right)\leq5.0$
and $-4.0\leq\feh\leq0.5$.
The 3D LTE versus 1D LTE abundance corrections
for \ion{Fe}{II} lines are usually only of the order
$-0.05$~to $+0.15\,\dex$.

We used the abundance corrections
to re-analyse the carbon, oxygen, and iron abundances
in $187$~F~and G~dwarfs previously presented in the literature.
It is clear that (differential) 3D non-LTE effects 
significantly reduce the scatter in the results and change the
mean Galactic chemical evolution trends. This is most apparent in
the $\abrat{C}{O}$~versus $\abrat{O}{H}$~plane, where severe 
3D non-LTE versus 1D LTE differential abundance corrections 
in both \ion{C}{I} and \ion{O}{I} work in the same direction such that,
while 1D LTE models predict a sharp minimum of
$\abrat{C}{O}\approx-0.5$~at $\abrat{O}{H}\approx-1.0$,
3D non-LTE models predict $\abrat{C}{O}$~to 
monotonically decrease with decreasing $\abrat{O}{H}$,
to a plateau of $\abrat{C}{O}\approx-0.6$~below
$\abrat{O}{H}\approx-1.0$.
Some discussion of this result in the context of
proposed exotic nucleosynthetic signatures
can be found in \citealt{2019A&A...622L...4A}.

The reduction in scatter due to the
3D non-LTE abundance corrections also has implications on studies
of exoplanets.  We found evidence that, at a given
value of $\abrat{O}{H}$, thin-disk stars with higher values of 
$\mathrm{C/O}$~tend to be more likely to host planets.
This result could be used to constrain planet formation models,
the physics of which are still not perfectly understood.

Although beyond the scope of this work,
the improved carbon and (in particular) oxygen 
abundances presented here have implications on stellar ages
as determined from stellar evolutionary models.
They have a direct influence on the rate of burning via the CNO cycle,
and are important sources of opacity in stellar interiors
\citep[e.g.][]{1993ApJ...414..580S};
these factors can make the shapes and locations of stellar isochrones
in the Hertzsprung-Russell diagram sensitive to the
assumed abundances \citep[e.g.][]{2013ApJ...765L..12B}.
Given the significant differential abundance corrections
for both \ion{C}{I} and \ion{O}{I} at low metallicity,
this should be relevant not only in the context of 
absolute ages of the oldest stars \citep[e.g.][]{2014ApJ...792..110V}, 
but also in the context of differential ages of
low-$\upalpha$~and high-$\upalpha$~halo population stars
\citep{2016ApJ...833..161G}.

The grids of line-by-line abundance corrections 
are presented in the online Tables via the CDS.
It is extremely cheap and simple to apply these abundance 
corrections
to 1D LTE abundances, to improve the accuracy of 
spectroscopic analyses of late-type stars.
We demonstrated this here for $187$~stars, but in principle,
as long as line-by-line 1D LTE abundances are available,
this approach can easily be applied to the 
extremely large spectroscopic surveys 
of over $10^{6}$~stars, which are planned
or currently underway
\citep[e.g.][]{2016SPIE.9908E..1GD,
2018MNRAS.478.4513B,2018AJ....156..126J,2019Msngr.175....3D}.

\begin{acknowledgements}
The authors thank the referee, Matthias Steffen,
for providing valuable feedback that improved the quality 
of this study. The authors also thank Bertram Bitsch 
for helpful discussions and comments concerning these results.
AMA and \'AS acknowledge
funds from the Alexander von Humboldt Foundation in the
framework of the Sofja Kovalevskaja Award endowed by the Federal Ministry of
Education and Research.
Funding for the Stellar Astrophysics Centre is provided by The Danish
National Research Foundation (grant DNRF106).
This work has made use of data from the European Space Agency (ESA) mission
{\it Gaia} (\url{https://www.cosmos.esa.int/gaia}), processed by the {\it Gaia}
Data Processing and Analysis Consortium (DPAC,
\url{https://www.cosmos.esa.int/web/gaia/dpac/consortium}). Funding for the DPAC
has been provided by national institutions, in particular the institutions
participating in the {\it Gaia} Multilateral Agreement.
This research has made use of the NASA Exoplanet Archive, which is
operated by the California Institute of Technology, under contract with the
National Aeronautics and Space Administration under the Exoplanet
Exploration Program.
This work was supported by computational resources provided by the Australian
Government through the National Computational Infrastructure (NCI)
under the National Computational Merit Allocation Scheme.
\end{acknowledgements}


\bibliographystyle{aa} 
\bibliography{bibl.bib}


\label{lastpage}
\end{document}